\documentclass[11pt]{article}

\usepackage[T1]{fontenc}

\usepackage{lineno,hyperref}
\modulolinenumbers[5]
\usepackage{amsmath,amsfonts, amsthm}
\usepackage{mathtools}
\usepackage{algorithm}
\usepackage{algpseudocode}
\usepackage{graphicx}
\usepackage{hyperref}
\usepackage{tikz}
\usepackage{pgfplots}
\usepackage{caption}
\usepackage{subcaption}
\usepackage{balance}
\usepackage{textcomp}
\usepackage{xcolor}
\usepackage{titlesec}
\usepackage[multiple]{footmisc}
\usepackage{bigfoot}
\usepackage{textcomp}
\usepackage{url}
\usepackage{hyperref}
\usepackage{array}
\usepackage{enumitem}
\usepackage{array}
\usepackage{authblk}
\usepackage{floatflt}
\usepackage{wrapfig}
\usepackage[top=1in, bottom=1in, left=1in, right=1in]{geometry}

\allowdisplaybreaks

\date{\vspace{-5ex}}

\newtheorem{definition}{Definition} 
\newtheorem{theorem}{Theorem} 
\newtheorem{assumption}{Assumption} 
\newtheorem{problem}{Problem} 
\newtheorem{remark}{Remark}
\newtheorem{lemma}{Lemma}
\newtheorem{corollary}{Corollary}

\newtheorem{example}{Example}

\newcommand{\mypara}[1]{\vspace{0.3em} \noindent {\bf #1}}

\begin{document}
\title{Distributionally Robust Predictive Runtime Verification under Spatio-Temporal Logic Specifications}
\author[1]{Yiqi Zhao}
\author[1]{Emily Zhu}
\author[2]{Bardh Hoxha}
\author[2]{Georgios Fainekos}
\author[1]{Jyotirmoy V. Deshmukh}
\author[1]{Lars Lindemann}
\affil[1]{Thomas Lord Department of Computer Science, University of Southern California}
\affil[2]{Toyota NA R\&D}

\maketitle
\begin{abstract}
Cyber-physical systems (CPS) designed in simulators, often consisting of multiple interacting agents (e.g. in multi-agent formations), behave differently in the real-world. We would like to verify these systems during runtime when they are deployed. Thus, we propose robust predictive runtime verification (RPRV) algorithms for: (1) general stochastic CPS under signal temporal logic (STL) tasks, and (2) stochastic multi-agent systems (MAS) under spatio-temporal logic tasks. The RPRV problem presents the following challenges: (1) there may not be sufficient data on the behavior of the deployed CPS, (2) predictive models based on design phase system trajectories may encounter distribution shift during real-world deployment, and (3) the algorithms need to scale to the complexity of MAS and be applicable to spatio-temporal logic tasks. To address these challenges, we assume knowledge of an upper bound on the statistical distance (in terms of an f-divergence) between the trajectory distributions of the system at deployment and design time. We are motivated by our prior work \cite{zhao2024robust,lindemann2023conformal} where we proposed an accurate and an interpretable  RPRV algorithm for general CPS, which we here extend to the MAS setting and spatio-temporal logic tasks. Specifically, we use a learned predictive model to estimate the system behavior at runtime and {\em robust conformal prediction} to obtain probabilistic guarantees by accounting for distribution shifts. Building on \cite{zhao2024robust}, we perform robust conformal prediction over the robust semantics of spatio-temporal reach and escape logic (STREL) to obtain centralized RPRV algorithms for MAS. We empirically validate our results in a drone swarm simulator, where we show the scalability of our RPRV algorithms to MAS and analyze the impact of different trajectory predictors on the verification result. To the best of our knowledge, these are the first statistically valid algorithms for MAS under distribution shift.
\end{abstract}
\section{Introduction}
\label{sec:introduction}
Cyber-physical Systems (CPS) are often stochastic in that they operate in non-deterministic environments and are subject to uncertain dynamics controlled by learning-enabled components. In many applications, stochastic systems consist of multiple agents that interact with each other in often uncertain ways, e.g., in smart cities \cite{roscia2013smart} and autonomous systems \cite{chen2020autonomous}. The safe design of CPS relies on formal verification and has received attention for both general CPS and MAS. Stochastic MAS are challenging to design and verify due to inherent uncertainty and scalability challenges.
\begin{floatingfigure}[r]{0.6\textwidth}
    \vspace{-0.2cm}
    \centering
    \includegraphics[width=0.6\textwidth]{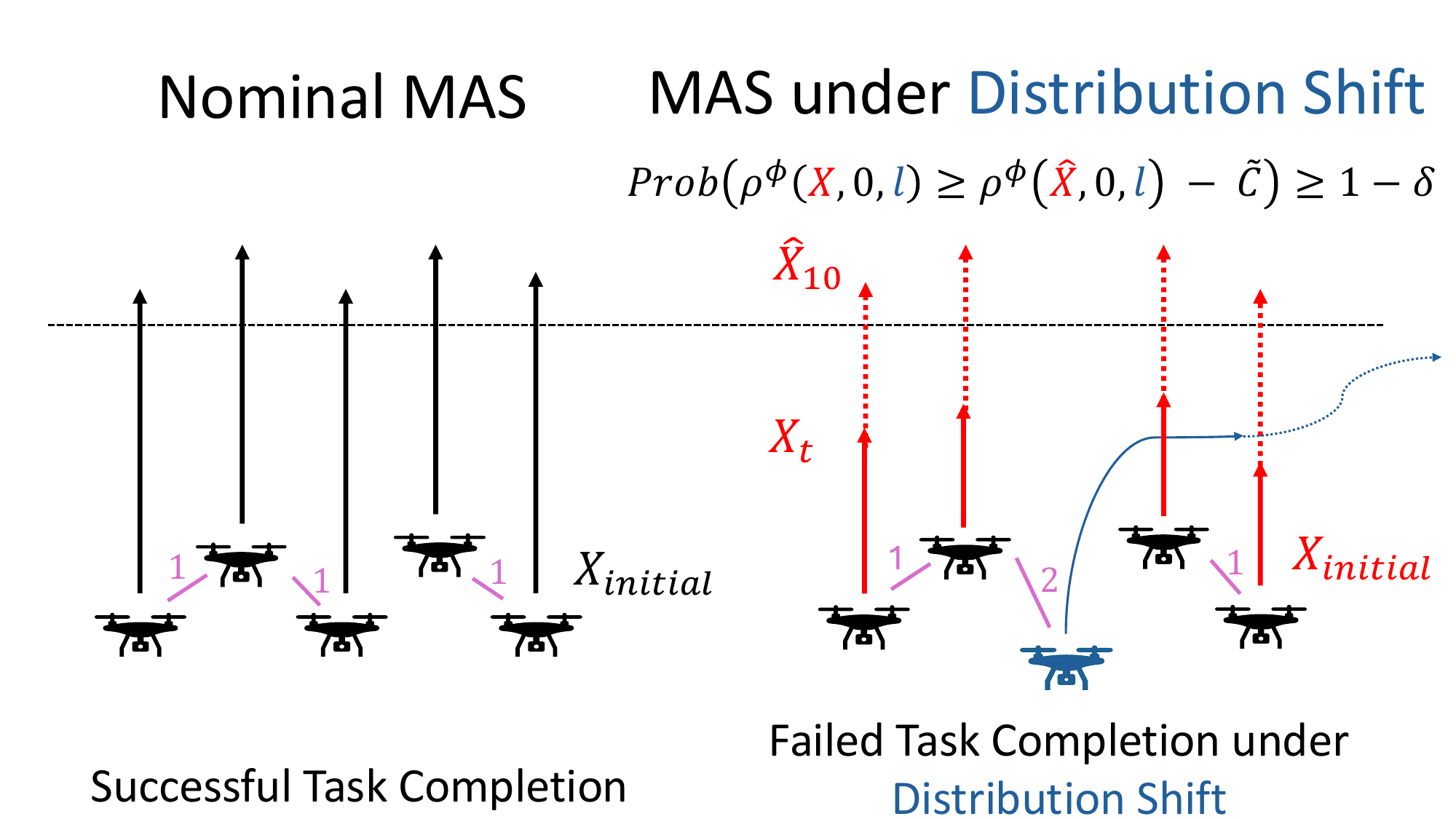}
    \caption{Left: A drone swarm reaching a goal configuration with communication cost (in terms of communication time in seconds) in pink. Right: The swarm with an adversarial agent (an agent, here in blue, purposely not following the nominal control strategy) performing the same task inducing a distribution shift.}
    \label{fig:multiagent_example}
\end{floatingfigure}
When designing CPS, one typically relies on simulators, which differ from the actually deployed system. Simulators model the CPS as a distribution $\mathcal{D}_0$ over the space of system trajectories, while the actual trajectory distribution $\mathcal{D}$ of the deployed system may differ from $\mathcal{D}_0$. For MAS, this is further exaggerated by the possible presence of adversarial agents (see Figure \ref{fig:multiagent_example}) and  inter-agent interactions. In \cite{zhao2024robust}, we proposed robust predictive runtime verification (RPRV) algorithms for general CPSs against a specification $\phi$ expressed in signal temporal logic (STL). Specifically, at runtime, we compute  the probability that a system trajectory will satisfy (or violate) the specification $\phi$ using the already observed part of the system trajectory. The presented algorithms are robust in that they account for all test distributions $\mathcal{D}$  that are contained within a set of possible distributions $P(\mathcal{D}_0)$ (e.g., from the deployed system) which are close to the training distribution $\mathcal{D}_0$ (e.g., describing a simulator) under a suitable distance measure. We present both an accurate algorithm where the verification results are statistically tight and an interpretable method where the reason for the violation or satisfaction of a logical formula can be specified in terms of the violation or satisfaction of atomic predicates in the formula.

In this work, we build up on \cite{zhao2024robust} and design RPRV algorithms for MAS. Specifically, we argue for the use of spatio-temporal reach and escape logic (STREL) \cite{bartocci2017monitoring, nenzi2022logic} to capture complex spatio-temporal logic tasks to reason over complex multi-agent behaviors. While previous research has proposed algorithms for online verification of MAS \cite{micalizio2007line, ferrando2022towards} (a more detailed discussion is presented later), the methods are not designed to offer statistical guarantees that generalize to settings where the trajectory distributions differ from test and design time (a phenomenon called  {\em distribution shift}).  Our focus here is thus to develop {\em robust predictive runtime verification (RPRV) algorithms} with guarantees even when the test distribution $\mathcal{D}$ differs from the training distribution $\mathcal{D}_0$, with a specific focus on stochastic MAS under STREL specifications. In this regard, the designed algorithms should be scalable (in terms of the number of agents) and be able to deal with the complexity of spatio-temporal logic tasks. We make the following contributions.
\begin{itemize}[wide, labelwidth=!, labelindent=0pt]
    \item To set the stage, we follow \cite{zhao2024robust} and present an accurate and an interpretable RPRV algorithm for general CPS under STL specifications that: (1) uses trajectory predictors to predict future system behavior, and (2) leverages robust conformal prediction \cite{cauchois2024robust} to quantify prediction uncertainty under the STL robust semantics using calibration data from $\mathcal{D}_0$. We show that both the accurate and interpretable algorithms are valid statistically for all test distributions $\mathcal{D} \in P(\mathcal{D}_0)$.
    \item By describing the network topology of an MAS as a function of the individual agent states, we propose centralized RPRV algorithms for MAS under STREL specifications. Our RPRV algorithms predict the future behavior of the MAS and use robust conformal prediction to quantify prediction uncertainty under the STREL robust semantics. We provide a complexity analysis on the algorithms.
    \item We provide an analysis following \cite{zhao2024robust} for the relationship between confidence, number of calibration data, and amount of distribution shift in the CPS and the MAS setting.
    \item We empirically validate our RPRV algorithms within a drone swarm simulator. We (1) show the scalability of our RPRV algorithms to MAS and STREL, and (2) analyze the impact of different trajectory predictors on the verification result.
\end{itemize}

\subsection{Related Work.} 

\mypara{(Offline) Verification of Stochastic Systems.} 
Formal verification of general CPS models is known to be an undecidable problem \cite{asarin1998achilles}.
To address the computational cost and inefficiency of abstraction-based verification algorithms, techniques such as statistical model checking have been used to give statistical guarantees \cite{agha2018survey,legay2019statistical,zarei2020statistical}. 
In \cite{sen2004statistical}, statistical hypothesis testing is performed with executions from a black-box system to verify continuous stochastic logic specifications. Statistical verification under STL specifications was first considered in \cite{bartocci2013robustness}. The works in \cite{zarei2020statistical,qin2022statistical} consider the problem of statistical verification of learning-enabled CPS with respect to STL specifications. Recent work has also sought to combine model-based techniques and statistical, data-driven techniques \cite{salamati2021data}. For instance, in \cite{pedrielli2023part} surrogate Gaussian process models of the system are learned and used to obtain probabilistic guarantees on satisfying STL specifications. These works assume that the distribution from which data is sampled is fixed. The authors in \cite{dutta2023distributionally} propose an active sampling approach using imprecise neural networks to promote distributional robustness in statistical verification of stochastic autonomous systems. Prediction intervals that are valid under distribution shift are designed in \cite{huang2024adaptive}  using conformal prediction. In \cite{schon2024data}, barrier certificates are constructed with guarantees under distribution shift. Robust prediction under distribution shift is proposed in autonomous driving \cite{stoler2024safeshift} and epistemic uncertainty-aware planning is considered in \cite{filos2020can}. 

\mypara{Runtime Verification of Stochastic Systems.} In runtime verification (RV), we are instead interested in verifying system properties during the operation of the system solely based on the currently observed system trajectory \cite{bartocci2018introduction,jaeger2020statistical}. RV techniques complement offline verification techniques, e.g., used for verifying STL specifications \cite{ma2021predictive, deshmukh2017robust}. Predictive RV is a special class of RV where we use a system model to predict the future system behavior from the currently observed system trajectory to either check if (hidden) system states satisfy a given specification \cite{sistla2011runtime} or if the system may violate system specifications in the future \cite{bortolussi2019neural,qin2020clairvoyant,cairoli2023learning,yu2024modeljournal}. In our prior work \cite{lindemann2023conformal}, we present predictive RV algorithms for general CPS using conformal prediction, a statistical tool for uncertainty quantification \cite{vovk2005algorithmic},  by calibrating prediction errors of a trajectory predictor to obtain valid probabilistic verification guarantees on the satisfaction of STL specifications. Similar in spirit, the authors in \cite{cairoli2023conformal} use a technique known as conformalized quantile regression \cite{romano2019conformalized} to design predictive RV algorithms that also provide probabilistic verification guarantees on the satisfaction of STL formulas. The authors in \cite{mao2023safe} take a first step in the direction of robust safety verification using conformal prediction and robust evaluators that minimize distribution shifts from high-dimensional measurements. However, to the best of our knowledge,  only our prior work \cite{zhao2024robust}, which we extend here, provides statistically valid RV guarantees under distribution shift. 

\mypara{Verification of Multi-agent Systems.} There exists numerous literature on specification languages for  MAS, see e.g.,  \cite{bartocci2017monitoring, ma2020sastl, nenzi2022logic}. Distributed runtime verification is considered in \cite{francalanza2018runtime, cooper1991consistent, chauhan2013distributed}. STL monitoring of distributed partially synchronous CPS is considered in \cite{momtaz2023monitoring} via solving a satisfiability modulo theory encoding. We, however, consider centralized runtime verification with a synchronized global clock, but with focus on stochastic MAS. Statistical model checking of MAS is considered in \cite{haghighi2015spatel, herd2015quantitative}.     Conformal prediction is used in \cite{muthali2023multi} for probabilistic reachability analysis of MAS. In this paper, we are interested in RPRV of STREL \cite{bartocci2017monitoring, nenzi2022logic} due to its high expressitivity and growing application in MAS, e.g., STREL is used to formulate requirements for autononomous aircraft systems 
\cite{schirmer2023hierarchy} and is accompanied by a tool for offline monitoring \cite{nenzi2023moonlight}. Online monitoring with STREL using imprecise signals is considered in \cite{visconti2021online}. However, to the best of our knowledge, no existing work in MAS runtime verification handles the distribution shift as we do in this paper.

%
%


\section{Problem Formulation}
\label{sec:rnn}

To describe general stochastic CPS, we consider an unknown test distribution $\mathcal{D}$ over finite-length system trajectories $X:=(X_0,X_1\hdots)\sim \mathcal{D}$ where\footnote{We use $\mathbb{R}^m$ with $m \in \mathbb{N}$ to represent the $m$-dimensional Euclidean space. We use $\mathbb{R}^\infty$ to represent the extended real line and any subscript to limit the scope of the defined space. For instance, $\mathbb{R}^\infty_{\ge 0}$ represents the non-negative extended real line.} $X_\tau \in \mathbb{R}^N$ is the state of the system at time $\tau$. We make no assumption about $\mathcal{D}$ other than the finite-length of the trajectory, e.g.,  $\mathcal{D}$ can describe distributions over finite executions of Markov decision processes or hybrid stochastic systems.

Now, consider an MAS with $L$ agents, where each agent is labeled with a distinct index $l \in \{1, \hdots, L\}$. Consider again the random vector $X \sim \mathcal{D}$. We express the state information of the multi-agent system via instantiating $X_\tau$ as $X_\tau := (X_\tau[1], \hdots, X_\tau[L])\in \mathbb{R}^{N}$ with $N := nL$ where $X_\tau[l]\in \mathbb{R}^{n}$ represents the state of agent $l$ at time $\tau$. We further define $X[l] := (X_0[l], X_1[l], \hdots)$.  While the distribution $\mathcal{D}$ is completely unknown, we assume that we have access to $K$ calibration trajectories $(X^{(1)},\hdots,X^{(K)})$ from a training distribution $\mathcal{D}_0$ that is close to $\mathcal{D}$ (as specified later).\footnote{The distributions $\mathcal{D}$ and $\mathcal{D}_0$ are  defined over the same  probability space $(\Omega,\mathcal{F},\mathbb{P})$ where $\Omega$ is
the sample space, $\mathcal{F}$ is a $\sigma$-algebra of $\Omega$, and $\mathbb{P}:\mathcal{F}\to[0,1]$ is a probability 
measure. For simplicity, we will mostly use the notation $\text{Prob}$ to be independent of the underlying probability space.}

\begin{assumption}\label{ass1}
 	We have access to a calibration dataset $S:=(X^{(1)},\hdots,X^{(K)})$ in which each of the $K$ trajectories $X^{(i)}:=(X_0^{(i)},X_1^{(i)},\hdots)$ is independently drawn from a training distribution $\mathcal{D}_0$, i.e., $X^{(i)}\sim\mathcal{D}_0$.\footnote{For instance, we can obtain such i.i.d. trajectories from a simulator that we can query repeatedly with a fixed distribution over simulation parameters.} We additionally have access to a training dataset $S_{tr} \coloneq \{X^{(K + 1)}, \hdots, X^{(K + \Gamma)}\}$, independent from $S$, in which each of the $\Gamma$ trajectories is again independently drawn from $\mathcal{D}_0$.
 \end{assumption}
We show shortly in Section \ref{sec:robust_cp} that the training dataset $S_{tr}$ will be used to train a trajectory predictor and the calibration dataset $S$ will be used to compute statistically valid prediction sets for runtime verification. Assumption \ref{ass1} holds in many applications. For a general CPS, an example is a setting in which $\mathcal{D}_0$ describes the motion of a robot within a high-fidelity simulator, while $\mathcal{D}$ describes a real robot operating in a lab environment. For a MAS, $\mathcal{D}_0$ describes nominal trajectories of a swarm of drones, while $\mathcal{D}$ describes the presence of adversarial agents, see Figure \ref{fig:multiagent_example}.
 
 To measure closeness of the distributions $\mathcal{D}_0$ and $\mathcal{D}$, we use the f-divergence, a statistical distance, that quantifies the similarity between $\mathcal{D}_0$ and $\mathcal{D}$ and thereby the distribution shift. Specifically, the f-divergence $D_f(\mathcal{D},\mathcal{D}_0)$ is defined as $D_f(\mathcal{D},\mathcal{D}_0):=\int_\mathcal{X} f\Big(\frac{d \mathcal{D}}{d \mathcal{D}_0}\Big) d \mathcal{D}_0$  where $\mathcal{X}$ is the support of $\mathcal{D}_0$, $\mathcal{D}$ is absolutely continuous with respect to $\mathcal{D}_0$, and $\frac{d \mathcal{D}}{d \mathcal{D}_0}$ is the Radon-Nikodym derivative of $\mathcal{D}$ with respect to $\mathcal{D}_0$. The function $f:[0,\infty)\to\mathbb{R}$ is convex and satisfies $f(1)=0$. If we set $f(z) := \frac{1}{2}|z - 1|$, we obtain the total variation distance  $TV(\mathcal{D}, \mathcal{D}_0) := \frac{1}{2}\int_\mathcal{X}|P(x) - Q(x)|dx$ where $P$ and $Q$ are probability density functions to $\mathcal{D}$ and $\mathcal{D}_0$.
\begin{assumption}\label{ass2}
 	The test and training distributions $\mathcal{D}$ and $\mathcal{D}_0$ are such that $D_f(\mathcal{D},\mathcal{D}_0)\le \epsilon$ where $\epsilon>0$. We hence assume that $\mathcal{D}\in P(\mathcal{D}_0):=\{\mathcal{D}' \mid D_f(\mathcal{D}',\mathcal{D}_0)\le \epsilon\}$.
 \end{assumption}
We emphasize that the parameter $\epsilon$ is a measure of the permissible distribution shift in terms of the f-divergence $D_f$. We provide more discussion on estimating $\epsilon$ in our experiments later. A detailed discussion on how $\epsilon$ can be estimated is also provided in \cite{cauchois2024robust}. In practice,  $\epsilon$ is often not exactly known beforehand and acts as a hyperparameter that we can use to robustify our predictive runtime verification algorithms. This is common practice in other areas such as robust control \cite{zhou1996robust}.

\mypara{Challenges in Runtime Verification.} Given a specification $\phi$ for a general CPS (e.g., formulated in STL) or a specification $\psi$ for a MAS (e.g., formulated in STREL) and a partial observation $(X_0, \hdots, X_t)$ from the test trajectory $X\sim \mathcal{D}$ at runtime $t$, we want to compute the probability that $X$ satisfies $\phi$ or $\psi$, respectively. We introduce the syntax and semantics of STL and STREL later in the section. The challenges are that we only have knowledge about the training distribution $\mathcal{D}_0$ as per Assumption \ref{ass1}, and our knowledge about the test distribution $\mathcal{D}$ is limited to the fact that $\mathcal{D}_0$ and $\mathcal{D}$ are $\epsilon$-close. For MAS, we specifically face scalability challenges. We design RPRV algorithms for general CPS and  MAS that are predictive in that we use predictions $\hat{X}_{\tau|t}$ of future states  $X_\tau$ for $\tau>t$ and  robust as we provide valid probabilistic guarantees as long as  $\mathcal{D}\in P(\mathcal{D}_0)$.



\subsection{Signal Temporal Logic for General CPS}
\label{sec:STL}
We use signal temporal logic (STL) to express system specifications over general CPS and define STL over discrete-time trajectories $x:=(x_0,x_1,\hdots)$, e.g., $x$ can be a realization of the stochastic trajectory $X$. We note that readers with limited background in temporal logics can, if they like to, skip the following formal definitions of syntax and semantics of an STL formula $\phi$ and instead think of $\phi$ as a high-level system specification that is imposed on the system at time $\tau_0$. We let $(x,\tau_0)\models \phi$ indicate that $x$ satisfies $\phi$ and we assume that bounded trajectories $x$ of length $L^\phi$ are sufficient to compute $(x,\tau_0)\models \phi$. The notation $\rho^\phi(x,\tau_0)\in\mathbb{R}^\infty$ will indicate how well $\phi$ is satisfied by $x$ at time $\tau_0$ with larger values indicating better satisfaction. 

\mypara{Syntax.} The atomic elements of STL are predicates that are functions $\pi:\mathbb{R}^N\to\{\text{True},\text{False}\}$. The predicate $\pi$ is defined via a predicate function $h:\mathbb{R}^N\to\mathbb{R}$ as $\pi(x_\tau):=\text{True}$ if $h(x_\tau)\ge 0$ and $\pi(x_\tau):=\text{False}$ otherwise, where $h$ is Borel-measurable. The syntax of STL is recursively defined as 
\begin{align}\label{eq:full_STL}
\phi \; ::= \; \text{True} \; | \; \pi \; | \;  \neg \phi' \; | \; \phi' \wedge \phi'' \; | \; \phi'  U_I \phi'' \; \color{blue}
\end{align}
where $\phi'$ and $\phi''$ are STL formulas. The Boolean operators $\neg$ and $\wedge$ encode negations (``not'') and conjunctions (``and''), respectively. The until operator $\phi' {U}_I \phi''$ encodes that $\phi'$ is true from now on until $\phi''$ becomes true at some future time within the time interval $I\subseteq \mathbb{R}_{\ge 0}$. We can further derive disjunction ($\phi' \vee \phi'':=\neg(\neg\phi' \wedge \neg\phi'')$), eventually ($F_I\phi:=\text{True} U_I\phi$), and always ($G_I\phi:=\neg F_I\neg \phi$).

\mypara{Semantics.} To determine if a trajectory $x$ satisfies an STL formula $\phi$ that is enabled at time $\tau_0$, we can define the semantics as a relation $\models$, i.e.,  $(x,\tau_0) \models\phi$ means that $\phi$ is satisfied. While the STL semantics are fairly standard \cite{maler2004monitoring}, we recall them in Appendix \ref{app:STL} in the supplementary material. Additionally, we can define robust semantics $\rho^{\phi}(x,\tau_0)\in\mathbb{R}^\infty$ that indicate how robustly the formula $\phi$ is satisfied or violated \cite{donze2,fainekos2009robustness}, see Appendix \ref{app:STL}. Larger and positive values of $\rho^{\phi}(x,\tau_0)$ hence indicate that the specification is satisfied more robustly. Importantly, it holds that $(x,\tau_0)\models \phi$ if $\rho^\phi(x,\tau_0)>0$ due to \cite[Proposition 16]{fainekos2009robustness}. We emphasize that STL can be used to express temporal properties of a MAS or of single agents within a MAS (as we do in Section \ref{sec:stl_rprv}), but that STL lacks the ability to specify spatial agent interactions, which are well-captured by STREL, shown in the next section. We emphasize the assumption that we consider bounded STL formulas $\phi$, i.e., all time intervals $I$ within the formula $\phi$ are bounded. Satisfaction of bounded STL formulas can be decided by finite length trajectories \cite{sadraddini2015robust}. The minimum length is given by the formula length $L^\phi$, i.e., with knowledge of $(x_{\tau_0},\hdots,x_{\tau_0+L^\phi})$ we can compute $(x,\tau_0) \models\phi$, see again Appendix \ref{app:STL} for more details.

\subsection{Spatio-Temporal Reach and Escape Logic for MAS}
\label{sec:STREL}
MAS can be described by continuous states and discrete graphs modeling inter-agent communication or interactions \cite{rahmani2009controllability}. Often, the edges are weighted to denote the communication cost (e.g. communication time) or inter-agent distance. Two agents are connected (via an edge) if they can communicate with each other (or more generally if they can exchange information \cite{rahmani2009controllability}). In this paper, we assume that  edges are undirected with non-negative weights, which is natural in applications that involve communication requirements.
\begin{figure}[H]
    \centering
    \vspace{-0.7cm}
    \includegraphics[width=0.7\textwidth]{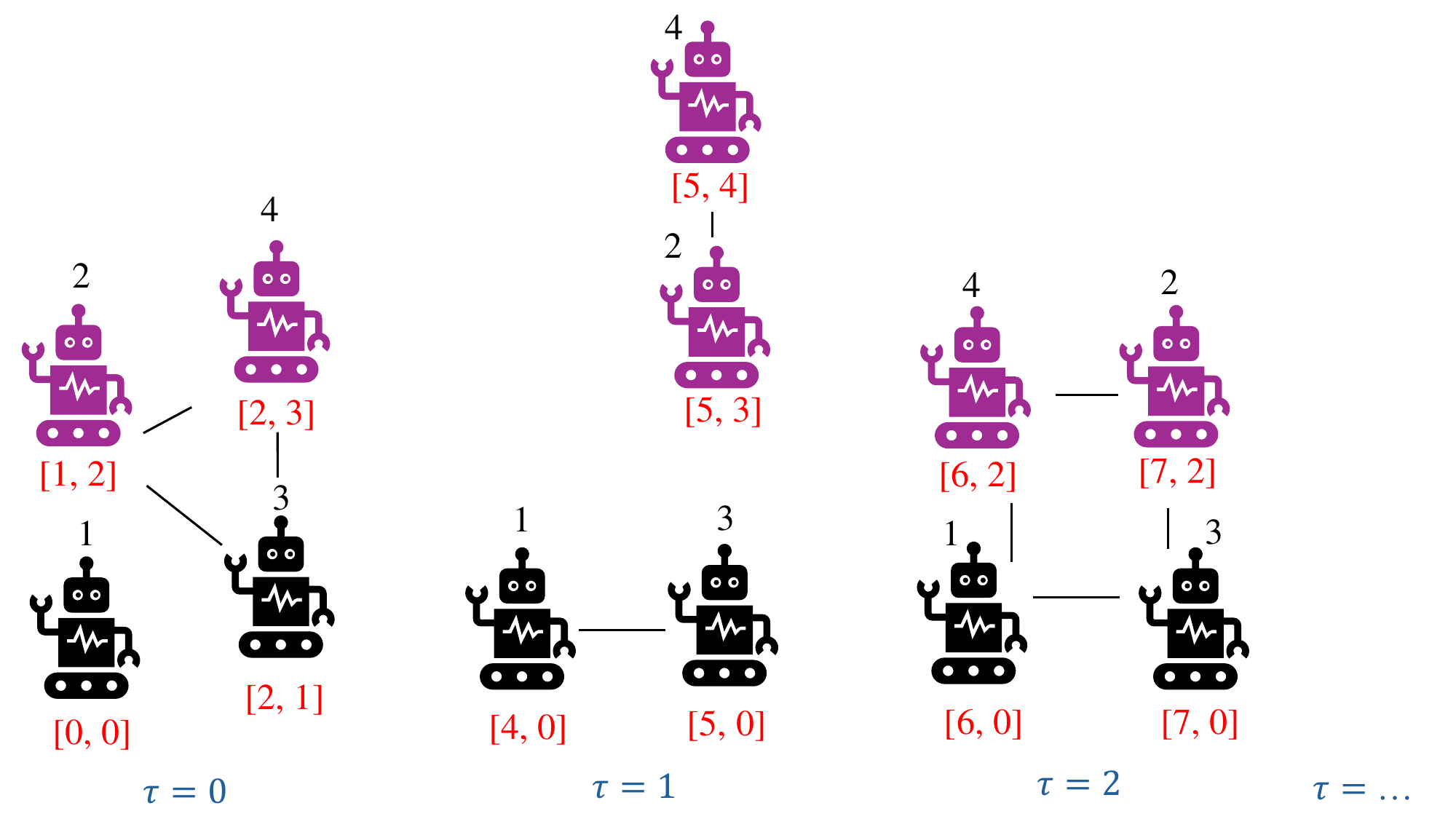}
    \caption{Example of a Multi-agent System moving from left to right at different time instances $\tau$. Agents in purple and in black satisfy two different predicates.}
    \label{fig:multiagent_running}
\end{figure}

\mypara{Modeling graphs as state dependent weights.} To reason about logical specifications over the graph structure of an MAS, we rely on a function describing the relationship between any two agents based on their state information. We thus assume to have access to a Borel-measurable weight function $w: \{1, \hdots, L\}^2 \times \mathbb{T} \times \mathbb{R}^{N \times \mathbb{T}} \rightarrow \mathbb{R}_{\ge 0}^\infty$, where $\mathbb{T} \coloneqq \{0, \hdots, \tau_0 + L^\psi\}$ with $\psi$ being a STREL specification and $L^\psi$ being its  formula length (which we both define  shortly). For two labels $l_1$ and $l_2$, any time instance $\tau \in \mathbb{T}$, and the state trajectory $X$ over the multi-agent system, we let $w(l_1, l_2, \tau, X)$ denote the weight between the two agents denoted by $l_1$ and $l_2$ at $\tau$. We require that $w(l_1, l_2, \tau, X) = w(l_2, l_1, \tau, X)$ so that the graph is undirected. We define that two agents labeled $l_1$ and $l_2$ are disconnected at time $\tau$ if and only if $w(l_1, l_2, \tau, X) := \infty$. We assume the underlying graph is non-reflexive (i.e., each label is disconnected from itself at all times) so that $w(l, l, \tau, X) := \infty$ for any agent $l$ at any time $\tau$. Further, if $w(l_1, l_2, \tau, X) = w'$ where $w'$ is finite valued, we equivalently write $l_1 \xrightarrow{w'} l_2$ at $\tau$. By the above definition, $w$ and $X$ together describe graph relations between the agents as functions over its states and labels. This will enable our RPRV algorithms under STREL specifications by directly accessing and predicting the random vector $X$.\footnote{One could instead model the weights between two agents more generally as states by considering the joint state $X := (X[1], \hdots, X[L], W_{1, 1}, \hdots, W_{L, L})$, where $W_{l_i, l_j}$ describes the weight between agents $l_i$ and $l_j$. Here, the weights $W_{1, 1}, \hdots, W_{L, L}$ can be independent of the individual agent states $X[1], \hdots, X[L]$. However, such a modeling choice would require us later to predict the future behavior of $W_{1, 1}, \hdots, W_{L, L}$ within our runtime monitoring approach which may be challenging in practice, especially when connections randomly disappear.} The assumption that the weight is a function of the agents' states and labels is not restrictive and arises in many applications in robotics and CPS. Examples of weight functions include $w(l_1, l_2, \tau, X) := \|X_\tau[l_1] - X_\tau[l_2]\|_2$ for specifying the Euclidean distance between two agents. The weight function can also be $w(l_1, l_2, \tau, X) := c\|X_\tau[l_1] - X_\tau[l_2]\|_2$ with $c \in \mathbb{R}_{> 0}$ to reflect that the communication time is proportional to the distance between two agents.

\mypara{Preliminaries for STREL. }  For any time $\tau$, we define a route $r_\tau$ as an infinite sequence of agent labels $l_0l_1\hdots \in \mathcal{L}^\omega$ where $w(l_i, l_{i + 1}, \tau, X) \neq \infty$ for any $i \ge 0$. We drop the subscript $\tau$ in $r$ when clear from the context. In the route $r$, $l_0$ is the starting agent of the route. Let $r[i] := l_i$ and $r[i \hdots]$ be the suffix route $l_il_{i + 1}\hdots$. Let $Routes(\tau, l_0)$ be the set of all routes starting from $l_0$. We also define $r(l') \coloneqq \min\{i \mid r[i] = l'\} \text{ if } l' \in r$ and $r(l') \coloneqq \infty \text{ otherwise}$ (as the first agent with label $l'$ within the route $r$).  Finally, we define the weight accumulation function $d: \mathbb{N}^\infty \times \mathcal{L}^\omega \times \mathbb{T} \rightarrow \mathbb{R}^\infty$ as $d(i, r, \tau) \coloneqq 0 \text{ if $i = 0$}$, $d(i, r, \tau) \coloneqq \infty \text{ if $i = \infty$}$, and $d(i, r, \tau) \coloneqq w(r[0], r[1], \tau, X) + d(i - 1, r[1\hdots], \tau) \text{ if $i > 0$}$. Intuitively, $d(i, r, \tau)$ recursively computes the distance (in terms of edge weights) along the route $r$ from the starting agent to the $i$th agent in $r$. We define the distance to agent $l'$ from the starting agent along the route $r$ to be $\tilde{d}(l', r, \tau) := d(r(l'), r, \tau)$. The minimum distance between two agents $l'$ and $l''$ at time $\tau$ is therefore defined as $\tilde{d}_{\min}(l', l'', \tau) := \min \{\tilde{d}(l'', r, \tau) \mid r \in Routes(\tau, l')\}$.

\begin{example}
    \label{example:graph_assumption}
    Consider a group of robots moving from left to right in Figure \ref{fig:multiagent_running} with their labels in black and states (locations) in red. In this example, $X_0 \coloneqq (X_0[1], X_0[2], X_0[3], X_0[4]) \coloneqq (0, 0, 1, 2, 2, 1, 2, 3)$. Suppose that two robots are connected when their Euclidean distance is at most 2 (as also indicated in Figure \ref{fig:multiagent_running}). For instance, note that robots 1 and 3 are disconnected at time $0$ since $\|X_0[1] - X_0[3]\|_2 \coloneq \sqrt{5} > 2$. To allow computation of the number of edges between any two agents, we let the weight of each connection be $1$ so that each edge lying on a route connecting two agents is counted exactly once when computing the distance along the route. Then, $w(1, 2, 0, X) := \infty$ but $w(1, 3, 1, X) = 1$. At $\tau = 2$, consider the route $r := 24132...$ where $l_0 = 2$; then $r(3) := 3$, $\tilde{d}(3, r, \tau) = 3$, and $\tilde{d}_{\min}(2, 3, \tau) = 1$.
\end{example}
\mypara{Syntax.} STREL \cite{bartocci2017monitoring, nenzi2022logic} extends STL by spatial operators reach and escape. The atomic elements of STREL are agent-dependent predicates $\pi:\mathbb{R}^n \times \{1, \hdots, L\} \to\{\text{True},\text{False}\}$. The predicate $\pi$ is defined via a Borel-measurable predicate function $h:\mathbb{R}^n\to\mathbb{R}$ as $\pi(x_\tau, l):= \text{True}$ if $h(x_\tau[l])\ge 0$ and $\pi(x_\tau, l) := \text{False}$ otherwise. The syntax of STREL is defined as $\psi::= \text{True} | \pi |  \neg \psi' | \psi' \wedge \psi'' | \psi'  U_I \psi'' | \psi' R_{[d_1, d_2]}\psi'' | \mathcal{E}_{[d_1, d_2]}\psi'$. Note that the first five operators are inherited from STL. In addition, the reach operator $\psi' R_{[d_1, d_2]}\psi''$ is an analogy of the until operator in the space domain and encodes the behavior of reaching a location $l'$ satisfying $\psi''$ within a distance of $[d_1, d_2] \subseteq \mathbb{R}^\infty_{\ge 0}$ along a path in which all nodes starting from $l$ satisfy $\psi'$. The escape operator $\mathcal{E}_{[d_1, d_2]}\psi'$ encodes the existence of a node $l'$ satisfying $\psi'$ with a minimum distance within $[d_1, d_2]$ from the central agent $l$ to be monitored such that there exists a path from the central agent $l$ to $l'$ in which all nodes on that path leading to $l'$ satisfy $\psi'$. Similar to the temporal operators, the spatial operators induce new operators for somewhere ($\mathcal{M}_{[d_1, d_2]} \psi := \text{True} R_{[d_1, d_2]}\psi$), everywhere ($\mathcal{N}_{[d_1, d_2]}\psi := \neg \mathcal{M}_{[d_1, d_2]} \neg \psi$), and surround ($\psi' \mathcal{S}_{\le d}\psi'' := \psi' \wedge \neg (\psi'R_{[0, d]}\neg(\psi' \vee \psi''))\wedge \neg(\mathcal{E}_{[d, \infty]}\psi')$). The somewhere and everywhere operators are the space equivalents of eventually and always operators, respectively. The surround operator denotes if an agent is in a region satisfying $\psi'$ and is surrounded by agents satisfying $\psi''$. As opposed to STL, STREL permits reasoning over inter-agent relationships in MAS.

\mypara{Semantics.} For STREL, we define the semantics $(x, \tau_0, l) \models \psi$ which indicate that the agent $l$ satisfies $\psi$ at time $\tau_0$.  We define robust semantics for STREL, which we denote by $\rho^\psi(x, \tau_0, l)$. We recall the STREL semantics, closely following \cite{nenzi2022logic}, along with the formula length $L^\psi$ in Appendix \ref{app:STREL} in the supplementary material.   Our RPRV algorithms rely on computation of the robust semantics, which we also summarize from \cite{nenzi2022logic} in Appendix \ref{app:STREL}. Importantly, the semantics are sound.
\begin{theorem}
\label{thm:soundness_strel}
     It holds that $(x,\tau_0,l)\models \psi$ if $\rho^\psi(x,\tau_0,l)>0$ and $(x,\tau_0,l)\models \neg \psi$ if $\rho^\psi(x,\tau_0,l) < 0$.
\end{theorem}

This soundness property is mentioned in \cite{nenzi2022logic}, but not proven. For imprecise signals (see its definition in \cite{nenzi2022logic}), the soundness property is shown in \cite{visconti2021online}. We show the proof in our setting in Appendix \ref{app:proof} in the supplementary material.  Similar as for STL, we assume that the STREL formula $\psi$ is bounded in time (i.e., $L^\psi \neq \infty$).
\begin{example}
    \label{example:strel}
    Consider again Example \ref{example:graph_assumption}, now with $X_\tau[l] = (X_\tau[l][0], X_\tau[l][1])$ denoting the position of agent $l$. Consider the predicate $\pi(X_\tau, l) := X_\tau[l][1] \ge 1.5$. In Figure \ref{fig:multiagent_running}, we mark all agents satisfying $\pi$ at time $\tau$ in purple and all agents that satisfy $\neg\pi$ in black. Consider $\psi := G_{[0, 2]}(\mathcal{M}_{[0, 2]}\pi)$. Clearly, $(X, 0, 2) \models \psi$ since at all times between $0$ and $2$, robot $2$ is connected to an agent that is purple. In fact it is itself purple at all time. Correspondingly, $\rho^{\psi}(X, 0, 2) = 0.5$. On the contrary, $(X, 0, 3) \not\models \psi$ as demonstrated by the counter-example at $\tau := 1$, and it holds that $\rho^{\psi}(X, 0, 3) = -1.5$.
\end{example}
\begin{remark}We remark on differences compared to the original definition of STREL from \cite{bartocci2017monitoring, nenzi2022logic}: (1) we do not rely on the definition of an explicit graph structure and instead rely on the weight function $w$. (2) although the definition of the signal domain (in the form of a semi-ring) from \cite{bartocci2017monitoring, nenzi2022logic} is more general, we follow \cite{visconti2021online} and only define the qualitative semantics induced by the boolean interpretation and the robust semantics induced by the $R^\infty$ interpretation both on real-valued signals. Different from \cite{visconti2021online}, we do not consider  interval semantics. (3) Similarly to \cite{visconti2021online}, we simplify the definition of distance from \cite{bartocci2017monitoring, nenzi2022logic} with the weight accumulation function.
\end{remark}

\subsection{Robust Predictive Runtime Verification}\label{sec:robust_cp}

Assume that we have observed the states $X_{\text{obs}}:=(X_0,\hdots,X_t)$ at runtime $t$, i.e., all states up until time $t$ are known, while future states $X_{\text{un}}:=(X_{t+1},X_{t+2},\hdots)$ from ${X}=(X_{\text{obs}},X_{\text{un}})\sim \mathcal{D}$ are not known. In this paper, we are interested in Problem \ref{prob1} for general CPS and Problem \ref{prob2} for MAS.
  
 \begin{problem}\label{prob1}
Let $\mathcal{D}_0$ be a training distribution, $\mathcal{D}$ be a test distribution from $P(\mathcal{D}_0)$ that satisfies Assumption \ref{ass2}, and $\phi$ be a bounded STL formula imposed at time $\tau_0$. Given the current time $t$, observations $X_{\text{obs}}$ from $X\sim \mathcal{D}$, and a failure probability $\delta\in(0,1)$, compute a lower bound  $\rho^*$ such that $\text{Prob}(\rho^\phi(X,\tau_0)\ge \rho^*)\ge 1-\delta$.  
\end{problem} 

\begin{problem}
    \label{prob2}
    Let $\mathcal{D}_0$ be a training distribution, $\mathcal{D}$ be a test distribution from $P(\mathcal{D}_0)$ satisfying Assumption \ref{ass2}, and $\psi$ be a temporally bounded STREL formula imposed at time $\tau_0$ and agent $l$. Given the current time $t$, observations $X_{\text{obs}}$ from $X\sim \mathcal{D}$, and a failure probability $\delta\in(0,1)$, compute a lower bound  $\rho^*$ such that $\text{Prob}(\rho^\psi(X,\tau_0, l)\ge \rho^*)\ge 1-\delta$.  
\end{problem}

For each problem, once we have computed the lower bound $\rho^*$, we remark that $\text{Prob}\big((X,\tau_0)\models \phi\big)\ge 1-\delta$ if $\rho^*>0$ (where $\phi$ is replaced by $\psi$ in the case of an MAS) due to the soundness of the robust semantics, see \cite[Proposition 16]{fainekos2009robustness} for STL and Theorem \ref{thm:soundness_strel} for STREL.  

\mypara{Trajectory Predictor.} Our RPRV algorithms for both general CPS (Problem \ref{prob1}) and for MAS (Problem \ref{prob2}) use trajectory predictors $\mu$ that map observations $X_\text{obs}$ at time $t$ into predictions $\hat{X}_{t + 1 | t} , \hat{X}_{t + 2 | t} \hdots$ of future states $X_\text{un}$. Therefore, we train a trajectory predictor $\mu$ on $S_{tr}$, the training trajectory dataset. Commonly used trajectory predictors range from recurrent neural networks (RNN) and long short-term memory (LSTM) networks \cite{hochreiter1997long, fjellstrom2022long} to support vector machines \cite{cortes1995support, ristanoski2013time} and autoregressive integrated moving average models \cite{box2015time, mehrmolaei2016time}. Since we consider temporally bounded STL and STREL formulas, only a finite prediction horizon is needed. Therefore, we let $H := \tau_0 + L^\phi - t$ be the prediction horizon needed for the computation of satisfaction of $\phi$ in case of STL (where $H$ is defined similarly for $\psi$ in case of STREL) imposed at $\tau_0$. To facilitate our discussion, we define the predicted trajectory $\hat{X} := (X_\text{obs}, \hat{X}_{t + 1 | t} , \hdots, \hat{X}_{t + H | t})$ with predictions $(\hat{X}_{t + 1 | t} , \hdots, \hat{X}_{t + H | t}):=\mu(X_\text{obs})$. We use the same notation for trajectories $X^{(i)}:=(X_\text{obs}^{(i)},X_\text{un}^{(i)})$ from the calibration dataset $S$. We also define the predicted calibration trajectory $\hat{X}^{(i)} := (X_\text{obs}^{(i)}, \hat{X}^{(i)}_{t + 1 | t} , \hdots, \hat{X}^{(i)}_{t + H | t})$ where $(\hat{X}^{(i)}_{t + 1 | t} , \hdots, \hat{X}^{(i)}_{t + H | t}):=\mu(X_\text{obs}^{(i)})$.



\subsection{Robust Conformal Prediction}
\label{sec:intro_conf}

Our solutions to Problem \ref{prob1} and \ref{prob2} rely on quantifying uncertainty of trajectory predictors. We thus use the calibration dataset $S$ from $\mathcal{D}_0$ along with robust conformal prediction as presented in \cite{cauchois2024robust} to account for the distribution shift between $\mathcal{D}_0$ and $\mathcal{D}$. Robust conformal prediction is an extension of conformal prediction which is a  statistical tool for uncertainty quantification \cite{vovk2005algorithmic,shafer2008tutorial, angelopoulos2021gentle,lei2018distribution}.

\mypara{Conformal Prediction (CP).} Let $R^{(0)},\hdots,R^{(K)}\sim \mathcal{R}_0$ be $K+1$ i.i.d. random variables following a training distribution $\mathcal{R}_0$.\footnote{In fact, exchangeability of $R^{(0)},\hdots,R^{(K)}$ would be sufficient which is a weaker requirement than being i.i.d.} The variable $R^{(i)}$ can be freely defined and is referred to as the nonconformity score. In regression, a common choice for $R^{(i)}$ is the prediction error $|Z^{(i)}-\mu(U^{(i)})|$ where the predictor $\mu$ attempts to predict   $Z^{(i)}$ based on an input $U^{(i)}$. We note that a large nonconformity score indicates a large prediction error.  Our goal is thus to obtain an upper bound for $R^{(0)}$ (our test data) from $R^{(1)},\hdots,R^{(K)}$ (our calibration data).  Formally, given a failure probability $\delta\in (0,1)$, we want to compute a constant $C$ (which depends on $R^{(1)},\hdots,R^{(K)}$) such that $\text{Prob}(R^{(0)}\le C)\ge 1-\delta$.

The probability $\text{Prob}(\cdot)$ is defined over the product measure of $R^{(0)},\hdots,R^{(K)}$.
By a simple statistical argument, one can obtain $C$ to be the $1/K$ corrected $(1-\delta)$th quantile of the empirical distribution of the values $R^{(1)},\hdots,R^{(K)}$, i.e., 
\setlength{\abovedisplayskip}{2pt}
\setlength{\belowdisplayskip}{2pt}
\begin{equation}\label{eq:vanilla_quantile}
C:=\text{Quantile}_{(1+1/K)(1-\delta)}(R^{(1)},\hdots,R^{(K)}).
\end{equation}

Formally, for $\beta\in [0,1]$,  the quantile function is defined as $\text{Quantile}_{\beta}(R^{(1)}, \dots, R^{(K)}):=\text{inf}\{z\in \mathbb{R}|\text{Prob}(Z\le z)\ge \beta\}$ where the random variable $Z\sim\sum_i \delta_{R^{(i)}}/K$ where $\delta_{R^{(i)}}$ is a dirac distribution centered at $R^{(i)}$. Equation \eqref{eq:vanilla_quantile} thus requires  $0\le (1+1/K)(1-\delta)\le 1$ and imposes the implicit lower bound $(K+1)(1-\delta) \le K$ on the  number of data $K$.  If this bound is satisfied and $R^{(1)},\hdots,R^{(K)}$ are sorted in non-decreasing order, we obtain $C:=R^{(p)}$ with $p:=\lceil (K+1)(1-\delta)\rceil$, i.e., $C$ is the $p$th smallest nonconformity score. We remark that we trivially have $C := \infty$ if $(K+1)(1-\delta) > K$.

\mypara{Robust CP.} In this paper, our test data is different from the training data. We thus use a robust version of conformal prediction based on \cite{cauchois2024robust}. Assume that $R^{(0)},\hdots,R^{(K)}$ are again independent, but not identically distributed in the sense that $R^{(0)}\sim\mathcal{R}$ while $R^{(1)},\hdots,R^{(K)}\sim\mathcal{R}_0$ where $\mathcal{R}$ is a test distribution. Under the assumption that test and training distributions $\mathcal{R}$ and $\mathcal{R}_0$ are close, the calibration data from the training distribution can still be used to bound $R^{(0)}$.

\begin{lemma}[Corollary 2.2 in \cite{cauchois2024robust}] \label{lemma:1}
    Let $R^{(0)},\hdots,R^{(K)}$ be independent random variables with $R^{(0)}\sim\mathcal{R}$ and $R^{(1)},\hdots,R^{(K)}\sim\mathcal{R}_0$ where the distributions $\mathcal{R}$ and $\mathcal{R}_0$ are such that $D_f(\mathcal{R},\mathcal{R}_0)\le \epsilon$. For a failure probability $\delta\in (0,1)$, it holds that
    \begin{align}\label{eq:guarantee_robust}
        \text{Prob}(R^{(0)}\le \tilde{C})\ge 1-\delta
    \end{align}
    \begin{align}\label{eq:C_tilde}
        \text{where } \tilde{C}:= \text{Quantile}_{1-\tilde{\delta}}(R^{(1)},\hdots,R^{(K)}) 
    \end{align}
    with $\tilde{\delta}:=1-g^{-1}(1-\delta_K)$ being obtained by solving a series of convex optimization problems as $\delta_K:=1-g\big((1+1/K)g^{-1}(1-\delta)\big)$ where $g(\beta):=\inf \{z\in[0,1]|\beta f(\frac{z}{\beta})+ (1-\beta)f(\frac{1 - z}{1 - \beta}) \le \epsilon\}$ and $g^{-1}(\tau):=\sup \{\beta\in[0,1]|g(\beta)\le \tau\}$.
\end{lemma}

We remark that equation \eqref{eq:C_tilde} requires $0\le 1-\tilde{\delta}\le 1$ and poses restrictions on the number of data $K$, the failure probability $\delta$, and the distribution shift $\epsilon$ as we elaborate on later in the paper. Note that $g$ and $g^{-1}$ are both solutions to convex programs. The solution to $g^{-1}(\tau)$ can thus be computed efficiently, e.g., using line search over $\beta\in(0,1)$. See \cite{cauchois2024robust} for details of computation and the role of each component in Lemma \ref{lemma:1}. In special cases, we can even obtain closed-form solutions for $g$, e.g., for $f(z):= \frac{1}{2}|z-1|$ (for the total variation distance), $g(\beta)=\max(0,\beta-\epsilon)$.

\mypara{Induced Distribution Shift.} Note that in runtime verification we assume an $\epsilon$-bounded distribution shift in terms of the $f$-divergence on the trajectory level, as described by $\mathcal{D}$ and $\mathcal{D}_0$. In the runtime verification algorithms, it will be necessary to quantify the induced distribution shift of functions that are defined over $X\sim \mathcal{D}$ and $X_0\sim \mathcal{D}_0$. The following result follows trivially.
\begin{lemma}[Data processing inequality]\label{lemma:2}
Let $\mathcal{D}$ and $\mathcal{D}_0$ be distributions such that $D_f(\mathcal{D},\mathcal{D}_0)\le \epsilon$ and let $R:\mathcal{X}\to \mathbb{R}$ be a measurable function. For $X\sim\mathcal{D}$ and $X_0\sim\mathcal{D}_0$, let $\mathcal{R}$ and $\mathcal{R}_0$ denote the induced distributions of $R(X)$ and $R(X_0)$, respectively. Then, it holds that $D_f(\mathcal{D},\mathcal{D}_0)\le \epsilon \Rightarrow D_f(\mathcal{R},\mathcal{R}_0)\le \epsilon.$
\setlength{\abovedisplayskip}{2pt}
\setlength{\belowdisplayskip}{2pt}
\end{lemma}
We remark that the robust semantics of STL and of STREL over discrete-time finite-length trajectories are either (1) function compositions of maximum or minimum operators over a finite set of $\mathbb{R}$-valued measureable mappings for STL and over countable sets for STREL, or (2) infinity (or negative infinity) valued for a measurable set of states (as the weight functions are measurable). The robust semantics of STL and STREL are both measurable, allowing distributions over $\mathcal{R}$, the test nonconformity distribution, and $\mathcal{R}_0$, the training nonconformity distribution, and invocations of the Data Processing Inequality in Lemma \ref{lemma:2}.

\section{RPRV Algorithms for General CPS with STL Specifications}\label{sec:general_cps_rprv}
Our  RPRV algorithms for general CPS follow our work \cite{zhao2024robust} which were  inspired by \cite{lindemann2023conformal}, but can deal with  distribution shifts $D_f(\mathcal{D},\mathcal{D}_0)\le \epsilon$. The first algorithm directly uses robust conformal prediction from Lemma \ref{lemma:1} to obtain a probabilistic lower bound $\rho^*$ for the robust semantics $\rho^\phi(X,\tau_0)$. This algorithm provides a tight verification result, but lacks interpretability, i.e., if $\phi$ is violated no explanation is provided. The second algorithm  uses robust conformal prediction to obtain a probabilistic lower bound for the robust semantics $\rho^\pi(X,\tau)$ of each predicate $\pi$ at each time $\tau$, then used to obtain a probabilistic lower bound $\rho^*$ for $\rho^{\phi}(X,\tau_0)$.  We use the following running example.


\begin{example}
\label{example:1}
We consider the F-16 aircraft  from \cite{heidlauf2018verification} with a hybrid controller modelled by $16$ states. We  only consider the height $h$ (given in ft) as the state to verify the STL specification $\phi:=G_{[0, 105]} h \ge 60$ and to find $\rho^*$ from Problem \ref{prob1} for $\delta:=0.2$. To construct a simple academic example, we collect a single trajectory $x_c$ from the simulator and then add independent noise to $x_c$ at each time, i.e., we let $\mathcal{N}(x_c(t), 3^2)$ and $\mathcal{N}(x_c(t), 3.5^2)$ describe $\mathcal{D}_0$ and $\mathcal{D}$, respectively. We assume that we have $K:=2000$ calibration trajectories $X^{(i)}$ with $i\in\{1,\hdots,K\}$ from $\mathcal{D}_0$ as per Assumption \ref{ass1}.  
\end{example}

\subsection{Accurate Robust STL Predictive Runtime Verification}
\label{subsec:robust_direct}
We first apply robust conformal prediction directly to the robust semantics $\rho^\phi$. To do so, we need to account for prediction errors in $\hat{X}$, defined in Section \ref{sec:robust_cp}, and the distribution shift between the calibration trajectories $X^{(i)}\sim \mathcal{D}_0$ and the test trajectory $X\sim \mathcal{D}$. Specifically, consider the nonconformity score
\setlength{\abovedisplayskip}{2pt}
\setlength{\belowdisplayskip}{2pt}
\begin{equation}
\label{eq:R_direct}
    R^{(i)} = \rho^\phi(\hat{X}^{(i)}, \tau_0) - \rho^\phi(X^{(i)}, \tau_0)
\end{equation}
\noindent
for each calibration trajectory $X^{(i)}\in S$. Intuitively, this nonconformity score measures the difference between the predicted robust semantics $\rho^\phi(\hat{X}^{(i)}, \tau_0)$ and the true robust semantics $\rho^\phi(X^{(i)}, \tau_0)$. For the test trajectory $X$, we analogously define the test nonconformity score $R := \rho^\phi(\hat{X}, \tau_0) - \rho^\phi(X, \tau_0)$ which we cannot compute during runtime as $X$ is unknown. Let the induced distribution of $R^{(i)}$ and $R$ be $\mathcal{R}_0$ and $\mathcal{R}$, i.e., $R^{(i)} \sim \mathcal{R}_0$ and $R \sim \mathcal{R}$ where $\mathcal{R}$ is a shifted version of the distribution $\mathcal{R}_0$.

By this construction of $R^{(i)}$ and $R$, it is easy to see from Lemmas \ref{lemma:1} and \ref{lemma:2} that $\text{Prob}((\rho^\phi(\hat{X}, \tau_0) - \rho^\phi(X, \tau_0) \le \tilde{C})\ge 1-\delta$ where $\tilde{C} := \text{Quantile}_{1 - \tilde{\delta}}(R^{(1)}, \hdots, R^{(K)})$ and $\tilde{\delta} := 1 - g^{-1}(1 - \delta_K)$. We summarize these results in Theorem \ref{theorem:1}, and provide a brief and formal proof in Appendix \ref{app:proof}.

\begin{theorem}\label{theorem:1}
Let the conditions from Problem \ref{prob1} hold. Then, it holds that $Prob(\rho^\phi(X, \tau_0) \ge \rho^{\phi}(\hat{X},\tau_0)-\tilde{C}) \ge 1 - \delta$ where $\tilde{C}$ is computed as in \eqref{eq:C_tilde} with the nonconformity score $R^{(i)}$ in \eqref{eq:R_direct} defined for all calibration trajectories $X^{(i)}\in S$.
\end{theorem}
\begin{figure*}
    \centering
    \begin{subfigure}[t]{0.3\textwidth}
        \includegraphics[width=\textwidth]{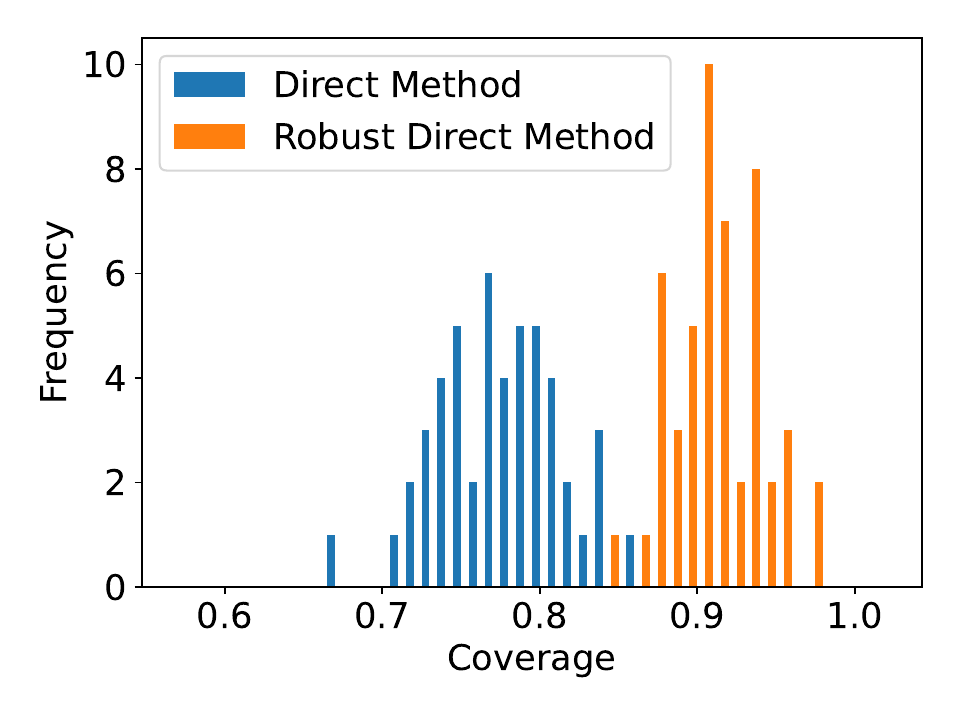}
        \caption{Histogram of coverage: accurate (direct) methods.}
        \label{fig:direct_coverages}
    \end{subfigure}
    \begin{subfigure}[t]{0.3\textwidth} 
        \includegraphics[width=\textwidth]{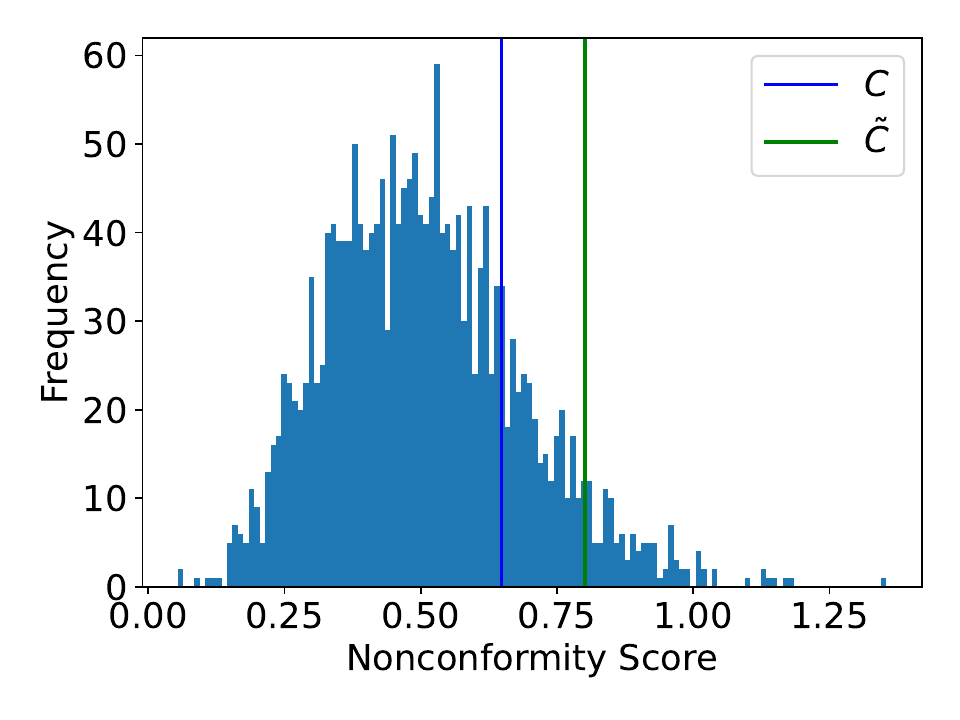}
        \caption{Histogram of $R^{(i)}$ from \eqref{eq:R_indirect} with $\tilde{C}$ and $C$.}
        \label{fig:indirect_1_nonconformities}
    \end{subfigure}
    \begin{subfigure}[t]{0.3\textwidth} 
        \includegraphics[width=\textwidth]{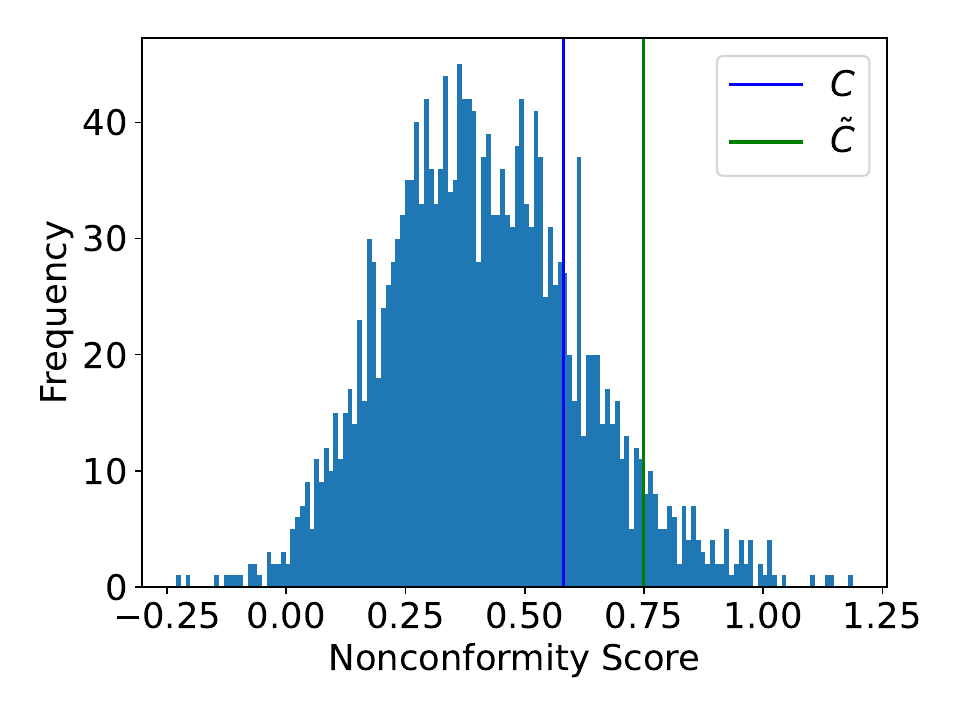}
        \caption{Histogram of $R^{(i)}$ from \eqref{eq:R_hybrid} with $\tilde{C}$ and $C$.}
        \label{fig:indirect_2_nonconformities}
    \end{subfigure}
    \begin{subfigure}[t]{0.3\textwidth}
    \includegraphics[width=\textwidth]{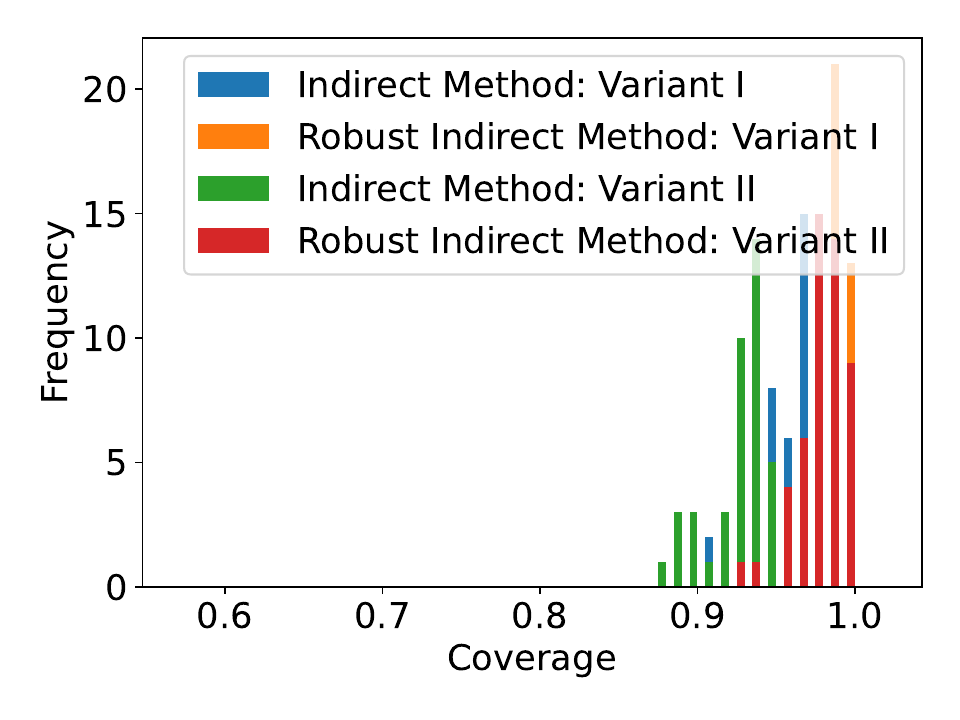}
    \caption{Histogram of coverage: interpretable (indirect) methods.}
    \label{fig:f_16_indirect_coverages}
    \end{subfigure}
    \caption{Running example: Histograms of nonconformity scores \eqref{eq:R_indirect}, and \eqref{eq:R_hybrid},  and empirical coverage plots of $\rho^\phi(X, \tau_0) \ge \rho^*$ for the accurate and interpretable (Variant I and II) methods.}
    \label{fig:combined}
\end{figure*}
Intuitively, this result says that the robust semantics $\rho^\phi(X, \tau_0)$ is lower bounded by the predicted robust semantics $\rho^\phi(\hat{X},\tau_0)$ adjusted by the value $\tilde{C}$, and that this holds with high probability. 
\begin{example}
\label{example:2}
    Recall Example \ref{example:1}. We set $t:=100$ and train an LSTM on $500$ trajectories from $\mathcal{D}_0$ to predict the next $H:=5$ time steps. We follow standard procedure to evaluate statistical guarantees from conformal prediction, see \cite[Section 2]{lindemann2024formal}. Specifically, we  perform the following experiment $50$ times: we sample $2000$ calibration trajectories from $\mathcal{D}_0$ and $100$ test trajectories from $\mathcal{D}$. In this case, we computed $\tilde{C}$ for the total variation distance with $\epsilon:=0.142$, which is such that $TV(\mathcal{R},\mathcal{R}_0)\le \epsilon$. In Section \ref{sec:simulations}, we explain in more detail how we estimate $TV(\mathcal{R},\mathcal{R}_0)$ in practice. It becomes evident in Figure \ref{fig:direct_coverages} where we plot the empirical coverage over all $50$ experiments for both algorithms. In other words, for each experiment we compute the ratio of how many of the $100$ test trajectories satisfy $\rho^\phi(X^{(i)}, \tau_0) \ge \rho^{\phi}(\hat{X}^{(i)},\tau_0)-\tilde{C}$ (where $\tilde{C}$ is replaced by $C$ for the non-robust version from \cite{lindemann2023conformal}), and plot the histogram over these ratios. As we aim for $1-\delta=0.8$ coverage, we can observe that only the robust algorithm from Theorem \ref{theorem:1} achieves the desired coverage.
\end{example}

\subsection{Interpretable STL Robust Predictive Runtime Verification}
\label{subsec:robust_indirect}
The accurate algorithm provides a precise verification result, but lacks interpretability. Specifically, if the satisfaction of a formula is not guaranteed by the direct algorithm, the reason for possible violation is unknown. Therefore, it is crucial to examine the robust semantics of each individual predicate in the STL formula. Consider the following example with a deterministic signal.
\begin{wrapfigure}{r}{6cm} 
    \centering
    \vspace{3pt}
    \begin{tikzpicture}
        \begin{axis}[
            width=6cm,
            height=4cm,
            xlabel={\small Time},
            ylabel={\small Value},
            xmin=-0.5,
            xmax=8.5,
            ymin=-1.5,
            ymax=3.5,
            xtick={0,1,2,3,4,5,6,7,8},
            ytick={-1,0,1,2,3},
            grid=major,
            legend pos=north west,
            legend style = {font=\tiny},
            title={\small Signal $x$ in Example \ref{example:interp}.}
        ]
        \addplot[
            color=blue,
            mark=*,
            thick,
            mark options={fill=blue}
        ] coordinates {
            (0, -1) (1, -1) (2, -1) (3, -1) 
            (4, 1) (5, 1) (6, 2) (7, 2) (8, 2)
        };
        \addplot[
            color=red,
            mark=square*,
            thick,
            mark options={fill=red}
        ] coordinates {
            (0, 1) (1, 1) (2,1) (3, 2) 
            (4, 2) (5, 2) (6, 3) (7, 3) (8, 3)
        };
        \legend{$x[0]$, $x[1]$}
        \end{axis}
    \end{tikzpicture}
\end{wrapfigure}
\begin{example}
\label{example:interp}
    Consider an STL formula $\phi := G_{[0, 5]}(F_{[0, 3]}(x[0] \ge 0 \wedge x[1] \ge 0))$ and the example signals $x[0]$ and $x[1]$ in the figure on the right. We can compute $\rho(x, 0) \coloneqq -1$, but the reason of violation is unclear unless we investigate the signal further. We notice that $x_0[0] < 0, x_1[0] < 0, x_2[0] < 0$ and $x_3[0] < 0$, which suggests that $\phi$ is violated at time $0$ particularly by the state $x[0]$. Our interpretable runtime verification algorithm therefore seeks to provide information on the satisfaction of each predicate in all time.
\end{example}
To provide theoretical guarantee (which we state later in Theorem \ref{theorem:2}), we assume that the STL formula $\phi$ is in positive normal form, i.e., that $\phi$ contains no negations. Note that every STL formula $\phi$ can be re-written in positive normal form, see e.g., \cite{sadraddini2015robust}. We remark that a formula with negation in front of predicates can be written in positive normal form by introducing equivalent negation-free predicates. Let the formula $\phi$ consists of $m$ predicates $\pi_i$, and define $\mathcal{P} := \big\{(\pi_i, \tau)| i\in\{1,\hdots,m\}, \tau\in \{t+1,\hdots,t+H\}\big\}$ as the set of all predicates and times. We define interpretability formally below.
\begin{definition}
\label{def:interpretability_stl}
    Given an STL formula $\phi$, a runtime verification algorithm is \textbf{interpretable} if it finds probabilistic lower bounds $\rho^*_{\pi, \tau}$ of the robust semantics $\rho^{\pi}(X,\tau)$ for all predicates and times $(\pi,\tau)\in \mathcal{P}$, i.e., such that
    \begin{align}\label{eq:guarantee_prediate}
\text{Prob}(\rho^{\pi}(X, \tau) \ge \rho_{\pi,\tau}^*, \forall (\pi, \tau) \in \mathcal{P}) \ge 1 - \delta.
\end{align}
\end{definition}

Intuitively, $\rho_{\pi,\tau}^* \ge 0$ certifies that $\pi$ is satisfied at time $\tau$ with probability $1-\delta$, and $\rho_{\pi,\tau}^* < 0$ represents possible sources of violation. Alternatively, one can monitor $\neg \phi$ in which case $\rho_{\pi,\tau}^* \ge 0$ indicates sources of violation. We now design an interpretable robust predictive runtime verification algorithm, referred to as the robust interpretable method. Before we propose two ways of computing $\rho_{\pi,\tau}^*$ from the predicted trajectory $\hat{X}$, we state our main results upfront. We recursively define the probabilistic robust semantics $\bar{\rho}^\phi$, which provide the desired probabilistic lower bound $\rho^*$ in Problem \ref{prob1}, starting from predicates as
$\bar{\rho}^{\pi}(\hat{X},\tau) := 
     h(X_\tau)$ if $\tau\le t$ and 
     $\bar{\rho}^{\pi}(\hat{X},\tau) :=\rho_{\pi,\tau}^*$ otherwise,  while the other Boolean and temporal operators follow standard semantics, as summarized in Appendix \ref{app:STL}. We next state our main results, proven in Appendix \ref{app:proof}.
\begin{theorem}\label{theorem:2}
Let the conditions from Problem 1 hold, and let $\phi$ further be in positive normal form. If the lower bounds $\rho^*_{\pi, \tau}$  satisfy  equation \eqref{eq:guarantee_prediate}, then it holds that  $\text{Prob}(\rho^{\phi}(X, \tau_0) \ge \bar{\rho}^\phi(\hat{X},\tau_0)) \ge 1 - \delta$ where $\bar{\rho}^\phi(\hat{X},\tau_0)$ is recursively constructed from $\rho^*_{\pi, \tau}$ as previously described.
\end{theorem}
\mypara{Computing $\rho_{\pi,\tau}^*$ on the state level (Variant I).} We now present two ways to compute $\rho_{\pi,\tau}^*$ that satisfy equation \eqref{eq:guarantee_prediate}. In the first method (Variant I), we compute prediction regions for trajectory predictions via robust conformal prediction. Therefore, we define the nonconformity score 
\begin{align}\label{eq:R_indirect}
    R^{(i)} := \max_{\tau \in \{t+1, \hdots, t+H\}} \|X^{(i)}_\tau - \hat{X}^{(i)}_{\tau|t}\|/\alpha_\tau
\end{align}
where $\alpha_\tau>0$ are constants that normalize the prediction errors at times $\tau$, following a similar idea to \cite{cleaveland2023lcp}. In this work, however, we simply propose to compute $\alpha_\tau:=\max_i \|X^{(i)}_\tau - \hat{X}^{(i)}_{\tau|t}\|$ over an additional set of trajectories $X^{(i)}$ from $\mathcal{D}_0$ that is independent from the dataset $S$, such as the set of training trajectories used to train the predictor $\mu$ on. Next, we define $\mathcal{B}_\tau := \{\zeta\in\mathbb{R}^N | \|\zeta - \hat{X}_{\tau|t}\| \le \tilde{C}\alpha_\tau\}$ which is a norm ball of radius $\tilde{C}\alpha_\tau$ with center at $\hat{X}_{\tau|t}$. We then compute the worst case value of $\rho^\phi(\zeta,\tau)$ over all $\zeta\in \mathcal{B}_{\tau}$, i.e., we let 
\begin{align}\label{eq:worst_case_robustness}
    \rho^*_{\pi, \tau} = \inf_{\zeta\in \mathcal{B}_{\tau}} h(\zeta).
\end{align}

Finally, we relate this construction to equation \eqref{eq:guarantee_prediate} and prove the following result in Appendix \ref{app:proof} using results from \cite{lindemann2019robust}.

\begin{lemma}\label{lemma:variant1}
    Let the conditions from Problem \ref{prob1} hold. If $\alpha_\tau>0$ for all $\tau \in \{t+1, \hdots, t+H\}$, then
    \begin{align} \label{eq:prob_indirect_1}
    \text{Prob}(\|X_\tau - \hat{X}_{\tau|t}\|\le \tilde{C}\alpha_\tau,  \forall \tau \in \{t+1, \hdots, t+H\}) \ge 1 - \delta
\end{align}
where $\tilde{C}$ is computed as in \eqref{eq:C_tilde} with the nonconformity score $R^{(i)}$ in \eqref{eq:R_indirect} defined for all calibration trajectories $X^{(i)}\in S$. Under the same conditions, it holds that $\rho^*_{\pi, \tau}$ in \eqref{eq:worst_case_robustness} satisfy~\eqref{eq:guarantee_prediate}.
\end{lemma}

Theorem \ref{theorem:2} and Lemma \ref{lemma:variant1} together present an interpretable predictive runtime monitor that can account for distribution shifts between $\mathcal{D}_0$ and $\mathcal{D}$ via the values of $\rho^*_{\pi, \tau}$. 

\mypara{Computing $\rho_{\pi,\tau}^*$ on the predicate level (Variant II).}  While Variant I provides interpretability, it may be the case that taking the infimum in equation \eqref{eq:worst_case_robustness} is conservative, e.g., as we show in the case study in \cite{zhao2024robust} and later in Example \ref{example:3}. We thus present a second method (called Variant II) where we compute prediction regions for each predicate $\pi$. Therefore, consider the nonconformity score
\begin{align}\label{eq:R_hybrid}
    R^{(i)} := \max_{(\pi, \tau) \in \mathcal{P}}(\rho^{\pi}(\hat{X}^{(i)}, \tau) - \rho^{\pi}(X^{(i)}, \tau))/\alpha_{\pi, \tau}
\end{align}
where $\alpha_{\pi, \tau}>0$ are again normalization constants. In this work, we use $\alpha_{\pi, \tau}:=\max_i |\rho^{\pi}(\hat{X}^{(i)}, \tau) - \rho^{\pi}(X^{(i)}, \tau)|$ over an additional set of trajectories $X^{(i)}$ from $\mathcal{D}_0$ that is independent from $S$. We conclude with the following result for which we provide a proof in Appendix \ref{app:proof}.
\begin{lemma}\label{lemma:variant2}
Let the conditions from Problem \ref{prob1} hold. If $\alpha_{\pi, \tau}>0$ for all $(\pi, \tau) \in \mathcal{P}$, then 
    \begin{align} \label{eq:prob_indirect_2}
    \text{Prob}(\rho^{\pi}(\hat{X}, \tau) - \rho^{\pi}(X, \tau) \le \tilde{C}\alpha_{\pi, \tau}, \forall (\pi, \tau) \in \mathcal{P}) \ge 1 - \delta
\end{align}
where $\tilde{C}$ is computed as in \eqref{eq:C_tilde} with the nonconformity score $R^{(i)}$ in \eqref{eq:R_hybrid} defined for all calibration trajectories $X^{(i)}\in S$. Under the same conditions, it holds that $\rho^*_{\pi, \tau}:=\rho^{\pi}(\hat{X}, \tau)-\tilde{C}\alpha_{\pi, \tau}$  satisfies~\eqref{eq:guarantee_prediate}.
\end{lemma}


\begin{example}
    \label{example:3}
    Recall Example \ref{example:1}. Similar as for the accurate algorithm in Example \ref{example:2}, we perform the same experiment of sampling $2000$ calibration trajectories and $100$ test trajectories $50$ times. For one of these experiments, we show in Figures \ref{fig:indirect_1_nonconformities} and \ref{fig:indirect_2_nonconformities} the histograms of the nonconformity scores $R^{(i)}$ among the calibration set from \eqref{eq:R_indirect} and \eqref{eq:R_hybrid} along with the robust prediction region $\tilde{C}$. As in Example \ref{example:2}, we use $\epsilon:=0.142$ which is such that $TV(\mathcal{R},\mathcal{R}_0)\le \epsilon$  where the induced distributions $\mathcal{R}$ and $\mathcal{R}_0$ are now with respect to equations \eqref{eq:R_indirect} and \eqref{eq:R_hybrid}. For comparison, we also plot the non-robust prediction region $C$ from \cite{lindemann2023conformal} which corresponds to the the case where $\epsilon=0$. In Figure \ref{fig:f_16_indirect_coverages},  we plot the histogram over all $50$ experiments of the empirical coverage for $\rho^{\phi}(X^{(i)}, \tau_0) \ge \rho^*$ on test trajectories $X^{(i)}$  for Variants I and II. We achieve the desired coverage of $1-\delta=0.8$ and compare to the non-robust versions (using $C$ instead of $\tilde{C}$). In this case, these also achieve an empirical coverage of $0.8$ as the interpretable algorithms are more conservative than the accurate algorithm. We also notice that the Variant II achieves lower coverage in Figure \ref{fig:f_16_indirect_coverages}, suggesting a reduction in conservatism from Variant I.
\end{example}

\section{RPRV Algorithms for MAS with STREL Specifications}\label{sec:spatio_temporal}
We now present RPRV algorithms for MAS under STREL specifications $\psi$. We focus on centralized monitoring where we have access to the full state information $X_t$ at the current time $t$.

\subsection{Accurate Robust STREL Predictive Runtime Verification}
We define  the predicted trajectory $\hat{X} := (X_\text{obs}, \hat{X}_{t + 1 | t} , \hdots, \hat{X}_{t + H | t})$ where $\hat{X}_{\tau}$  are predictions of the MAS state $X_\tau=(X_\tau[1],\hdots,X_\tau[L])$ at times $\tau\in\{t+1,\hdots,t+H\}$. Since we model graphs as state dependent weights (recall Section \ref{sec:STREL}), we can compute the robust semantics $\rho^\psi$ of a STREL specification $\psi$ directly over the predicted trajectory $\hat{X}$. Hence,  consider the nonconformity score
\begin{equation}
\label{eq:nonc_multi_direct}
\begin{aligned}
    R^{(i)} = \rho^\psi(\hat{X}^{(i)}, \tau_0, l) -  \rho^\psi(X^{(i)}, \tau_0, l).
\end{aligned}
\end{equation}
We can now again apply robust conformal prediction  to compute $\tilde{C}$ according to equation \eqref{eq:C_tilde}. The result below follows in the same way as Theorem \ref{theorem:1} by invoking Lemma \ref{lemma:1}, and is thus omitted. 
\begin{theorem}\label{theorem:multi_direct}
Let the conditions from Problem \ref{prob2} hold. Then, it holds that $Prob(\rho^\psi(X, \tau_0, l) \ge \rho^{\psi}(\hat{X},\tau_0, l)-\tilde{C}) \ge 1 - \delta$ where $\tilde{C}$ is computed as in \eqref{eq:C_tilde} with the nonconformity score $R^{(i)}$ in \eqref{eq:nonc_multi_direct} defined for all calibration trajectories $X^{(i)}\in S$.
\end{theorem}
We note that the soundness property of STREL, as shown in Theorem \ref{thm:soundness_strel}, means that $\rho^{\psi}(\hat{X},\tau_0, l)-\tilde{C}>0$ ensures that $(X,\tau_0, l)\models \psi$ with probability no less than $1-\delta$.

\subsection{Interpretable Robust STREL Predictive Runtime Verification}
Similar to Section \ref{subsec:robust_indirect}, we assume that the STREL formula is in positive normal form, i.e., the formula $\psi$ contains no negations. Let the STREL formula $\psi$  consist of $m$ predicates $\pi_i$, and define $\mathcal{P} := \{(\pi_i, \tau, l') \mid i \in \{1, \hdots, m\}, \tau \in \{t + 1, \hdots, t + H\}, l' \in \{1, \hdots, L\}\}$. In this case, for any two trajectories $x, x': \mathbb{N} \rightarrow \mathbb{R}^N$ with the same prefix $x_\tau = x'_\tau$ for times $\tau \le t$, it holds that $\rho^\psi(x, \tau_0, l) \ge \rho^\psi(x', \tau_0, l)$ for any $l \in \{1, \hdots, L\}$ if $\rho^\pi(x, \tau, l) \ge \rho^\pi(x', \tau, l)$ for all $(\pi, \tau, l) \in \mathcal{P}$. This follows a similar reasoning as in the STL case as the robust semantics are defined with only max/min operators if negations are excluded. In the remainder, motivated by Definition \ref{def:interpretability_stl}, we use probabilistic lower bounds $\rho^*_{\pi, \tau, l}$ of the robust semantics $\rho^{\pi}(X,\tau,l)$ for all predicates, times, and agents $(\pi, \tau, l)\in \mathcal{P}$, i.e., such that
\begin{align}\label{eq:guarantee_predicate_strel}
    \text{Prob}(\rho^\pi(X, \tau, l') \ge \rho^*_{\pi, \tau, l}, \forall (\pi, \tau, l) \in \mathcal{P}) \ge 1 - \delta
\end{align}

We then recursively define the probabilistic robust semantics $\bar{\rho}^\psi$, which provide the desired probabilistic lower bound $\rho^*$ in Problem \ref{prob2}, starting from predicates as
$\bar{\rho}^{\psi}(\hat{X},\tau,l) := 
     h(X_\tau[l])$ if $\tau\le t$ and 
     $\bar{\rho}^{\pi}(\hat{X},\tau,l) :=\rho_{\pi,\tau,l}^*$ otherwise,  while the other Boolean, temporal, and spatial operators follow standard semantics, as summarized in Appendix \ref{app:STREL}. We next state our main results, which follows similar reasoning as Theorem \ref{theorem:2}, so that the proof is omitted.
\begin{theorem}\label{theorem:ind_strel}
Let the conditions from Problem \ref{prob2} hold, and let $\psi$ further be in positive normal form. If the lower bounds $\rho^*_{\pi, \tau, l'}$  satisfy  equation \eqref{eq:guarantee_predicate_strel}, then it holds that  $\text{Prob}(\rho^{\psi}(X, \tau_0, l) \ge \bar{\rho}^\psi(\hat{X},\tau_0, l)) \ge 1 - \delta$ where $\bar{\rho}^\psi(\hat{X}, \tau_0, l)$ is recursively constructed from $\rho^*_{\pi, \tau, l}$ as previously described.
\end{theorem}
\mypara{Computing $\rho^*_{\pi, \tau, l}$ on the state level (Variant I).} We compute $\rho^*_{\pi, \tau, l}$ in equation \eqref{eq:guarantee_predicate_strel} similar to  $\rho^*_{\pi, \tau}$ in equation \eqref{eq:guarantee_prediate} for Variant I. However, the difference here is that we also have to take agents $l\in \{1, \hdots, L\}$ into account similar to \cite{yu2023signal}. We thus consider the nonconformity score
\begin{align}\label{eq:r_indirect_strel}
     R^{(i)} := \max_{\tau \in \{t+1, \hdots, t+H\}, l \in \{1, \hdots, L\}} \|X^{(i)}_\tau[l] - \hat{X}^{(i)}_{\tau|t}[l]\|/\alpha_{\tau, l'}
\end{align}
where $\alpha_{\tau, l} := \max_i\|X_\tau^{(i)}[l] - \hat{X}_{\tau|t}^{(i)}[l]\| > 0$ over an additional set of trajectories $X^{(i)}$ from $\mathcal{D}_0$ separate from the calibration set $S$. We let $\rho^*_{\pi, \tau, l} = \inf_{\zeta \in \mathcal{B}_{\tau, l}}h(\zeta)$ where $B_{\tau, l} := \{\zeta \in \mathbb{R}^{n_l} \mid \|\zeta - \hat{X}_{\tau|t}[l]\| \le \tilde{C}\alpha_{\tau, l}\}$ with $\tilde{C}$ computed as in \eqref{eq:C_tilde} with the nonconformity score of \eqref{eq:r_indirect_strel}. We can then conclude Corollary \ref{corollary:variant_1_strel}, where we again omit the proof is as it follows similarly to Lemma \ref{lemma:variant1}.
\begin{corollary}
    \label{corollary:variant_1_strel} Let the conditions from Problem \ref{prob2} hold. If $\alpha_{\tau, l'} > 0$ for all $(\tau, l) \in \{t+1, \hdots, t+H\} \times \{1, \hdots, L\}$, then
    \begin{align}
        \text{Prob}(\|X_\tau[l] - \hat{X}_{\tau|t}[l]\|\le \tilde{C}\alpha_{\tau, l'},  \forall (\tau, l) \in \{t+1, \hdots, t+H\} \times \{1, \hdots, L\}) \ge 1 - \delta
    \end{align}
    where $\tilde{C}$ is computed as in \eqref{eq:C_tilde} with the nonconformity score $R^{(i)}$ in \eqref{eq:r_indirect_strel} defined for all calibration trajectories $X^{(i)}\in S$. Under the same conditions, it holds that $\rho^*_{\pi, \tau, l}$ satisfies~\eqref{eq:guarantee_predicate_strel}.
\end{corollary}

\mypara{Computing $\rho^*_{\pi, \tau, l}$ on the predicate level (Variant II).} We compute $\rho^*_{\pi, \tau, l}$ in equation \eqref{eq:guarantee_predicate_strel} similar to  $\rho^*_{\pi, \tau}$ in Lemma \ref{lemma:variant2} for Variant II. However, we again have to take agents $l\in \{1, \hdots, L\}$ into account and consider the nonconformity score
\begin{align}\label{eq:r_hybrid_strel}
    R^{(i)} := \max_{(\pi, \tau, l) \in \mathcal{P}}(\rho^\pi(\hat{X}^{(i)}, \tau, l) - \rho^\pi(X^{(i)}, \tau, l'))/\alpha_{\pi, \tau, l}
\end{align}
where $\alpha_{\pi, \tau, l} := \max_i|\rho^\pi(\hat{X}^{(i)}, \tau, l) - \rho^\pi(X^{(i)}, \tau, l)| > 0$ is again computed over an additional dataset of $X^{(i)}$ from $\mathcal{D}_0$ independent from $S$. We obtain the following result similar to Lemma \ref{lemma:variant2}.

\begin{corollary}\label{corollary:variant_2_strel}
Let the conditions from Problem \ref{prob2} hold. If $\alpha_{\pi, \tau, l}>0$ for all $(\pi, \tau, l) \in \mathcal{P}$, then 
    \begin{align}
    \text{Prob}(\rho^{\pi}(\hat{X}, \tau, l) - \rho^{\pi}(X, \tau, l) \le \tilde{C}\alpha_{\pi, \tau, l}, \forall (\pi, \tau, l) \in \mathcal{P}) \ge 1 - \delta
\end{align}
where $\tilde{C}$ is computed as in \eqref{eq:C_tilde} with the nonconformity score $R^{(i)}$ in \eqref{eq:r_hybrid_strel} defined for all calibration trajectories $X^{(i)}\in S$. Under the same conditions, it holds that $\rho^*_{\pi, \tau, l}:=\rho^{\pi}(\hat{X}, \tau, l)-\tilde{C}\alpha_{\pi, \tau, l}$ satisfies~\eqref{eq:guarantee_predicate_strel}.
\end{corollary}

\section{Data Requirements, Distribution Shift, and Algorithm Complexity}
\label{sec:data_complexity}

\mypara{Data Requirements and Distribution Shift.} For the computation of  $\tilde{C}$ in Lemma \ref{lemma:1}, we require that $1-\tilde{\delta}\in [0,1]$ which is equivalent to  $1-\delta_K=(1+1/K)g^{-1}(1-\delta)\in [0,1]$ as the function $g$ and its inverse $g^{-1}$ have domains $[0,1]$. We note that the lower bound $0\le (1+1/K)g^{-1}(1-\delta)$ is  satisfied for any $K>0$. The upper bound $(1+1/K)g^{-1}(1-\delta)\le 1$, on the other hand, poses a lower bound on the number $K$ of calibration trajectories as $K \ge \Big\lceil \frac{g^{-1}(1 - \delta)}{1 - g^{-1}(1 - \delta)} \Big\rceil$.
This condition can be seen as a requirement on the number of calibration data $K$ if $g^{-1}(1 - \delta)<1$. However, it is important to observe the case where $g^{-1}(1 - \delta)=1$. It is this condition that imposes additional conditions on the confidence $1-\delta$ and the distribution shift $\epsilon$ for a given $f$-divergence. For instance, for $f(t) = \frac{1}{2}|t - 1|$, associated with the total variation distance, we know that $g^{-1}(1 - \delta)=\text{argsup}_{\beta\in[0,1]} \max(0,\beta-\epsilon)\le 1-\delta$ which is equivalent to $1$ if $\epsilon\ge \delta$. More generally, the condition $g^{-1}(1 - \delta)=1$ will constrain the permissible distribution shift $\epsilon$ for a confidence of $1-\delta$. 
\begin{corollary}
    Let the conditions from Problem \ref{prob1} (or Problem \ref{prob2}) hold. If $K \ge \Big\lceil \frac{g^{-1}(1 - \delta)}{1 - g^{-1}(1 - \delta)} \Big\rceil$ with $g^{-1}(1 - \delta)<1$, then the algorithms presented in Theorems \ref{theorem:1} and \ref{theorem:2} (or in Corollaries \ref{theorem:multi_direct} and \ref{theorem:ind_strel}) provide nontrivial verification results in the sense that $\tilde{C}<\infty$.
\end{corollary}

\mypara{Complexity of the MAS RPRV Algorithms.} We next  discuss the time complexity of our RPRV algorithms. We focus our discussion on a STREL formula $\psi$ (recall that STREL is strictly more expressive than STL). We discuss the difference in online and offline procedures of our algorithms. Assuming access to a trained predictor, we remark that \textbf{the offline procedure} includes the computation of the statistical bounds $\tilde{C}$ from Theorem \ref{theorem:multi_direct} for the accurate method from Corollary \ref{corollary:variant_1_strel} for the interpretable method (Variant I) and from Corollary \ref{corollary:variant_2_strel} for the interpretable method (Variant II). The procedure is offline since no test data from the test distribution $\mathcal{D}$ is required. In the \textbf{online procedure}, we apply the bound $\tilde{C}$ computed in the offline procedure to attain the lower bound robust semantics $\rho^*$. The online procedure requires a partially realized test trajectory and is conducted in runtime at time $t$. 

We show that the accurate method is faster during runtime than the interpretable method, but may be slower during offline calibration for complex specifications. We denote the time for computing the robust semantics $\rho^\psi$ for a given trajectory $X$ by $T_{\text{comp},\psi}$. We discuss the time complexity of computing the STREL robust semantics in Appendix \ref{app:STREL}. 

The \textbf{offline} complexity of the accurate method, i.e., of computing $\tilde{C}$ in Theorem \ref{theorem:multi_direct}, is $O(\max(KT_{\text{comp},\psi}, \\K\log(K)))$ where $K\log(K)$ denotes the time complexity for sorting the $K$ nonconformity scores. The offline complexity of the interpretable method (Variant I), i.e., of computing $\tilde{C}$ in Corollary \ref{corollary:variant_1_strel}, however, is $O(\max(HLK, K\log(K)))$ where we assume that computation of  $\alpha_{\tau, l'}$ is negligible since it is $O(HLK)$. The offline complexity of the interpretable method (Variant II), i.e., of computing $\tilde{C}$ in Corollary \ref{corollary:variant_2_strel},  is $O(\max(HLK\Pi,  K\log(K)))$, where $\Pi$ is the number of predicates in the formula $\psi$. 

The  \textbf{online} complexity of the accurate method, i.e., of computing $\rho^*$, is $T_{\text{comp},\psi}$. For the interpretable method (variant I) it is $O(\max(T_{opt}|\mathcal{P}|, T_{\text{comp},\psi}))$ where $T_{opt}$ denotes the worst case time complexity for computing $\rho^*_{\pi, \tau, l}$ among all $(\pi, \tau, l) \in \mathcal{P}$. For the interpretable method (Variant II), it is $O(\max(|\mathcal{P}|, T_{\text{comp},\psi}))$, which is significantly faster than Variant I when non-convex predicates are present in $\psi$. We evaluate the complexity empirically in our case study in Section \ref{sec:simulations}.

\begin{figure}[t] 
    \centering

    \begin{subfigure}{0.4\textwidth} 
        \includegraphics[width=\textwidth]{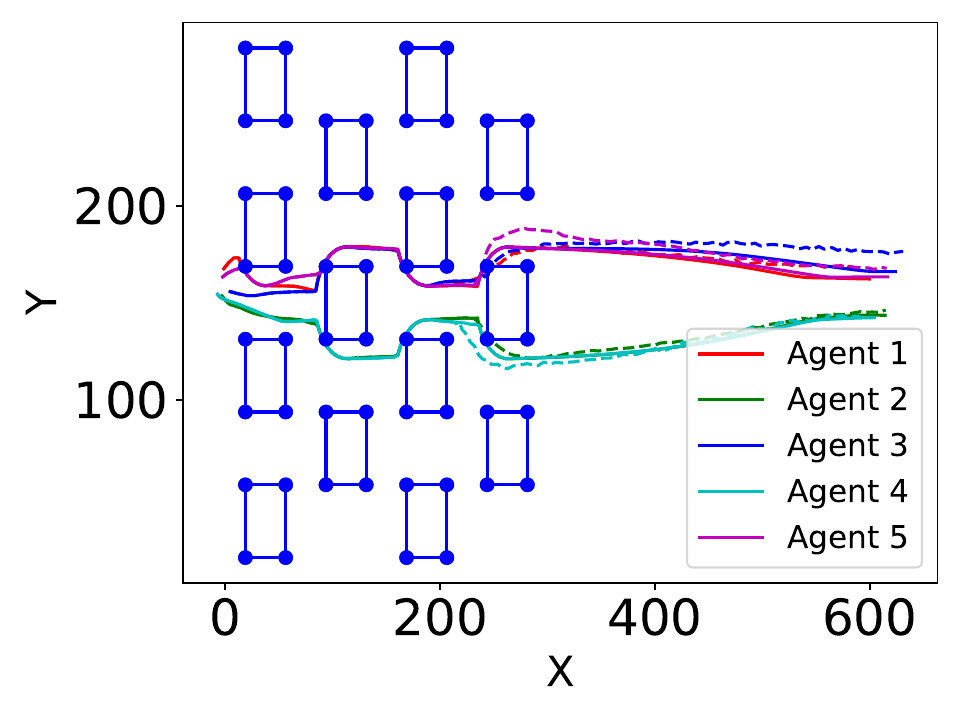}
        \caption{Example trajectory (in solid) from $\mathcal{D}_0$ 
          with prediction (in dashed)}
        \label{fig:example_trajectory_nominal}
    \end{subfigure}
    \begin{subfigure}{0.4\textwidth}
        \includegraphics[width=\textwidth]{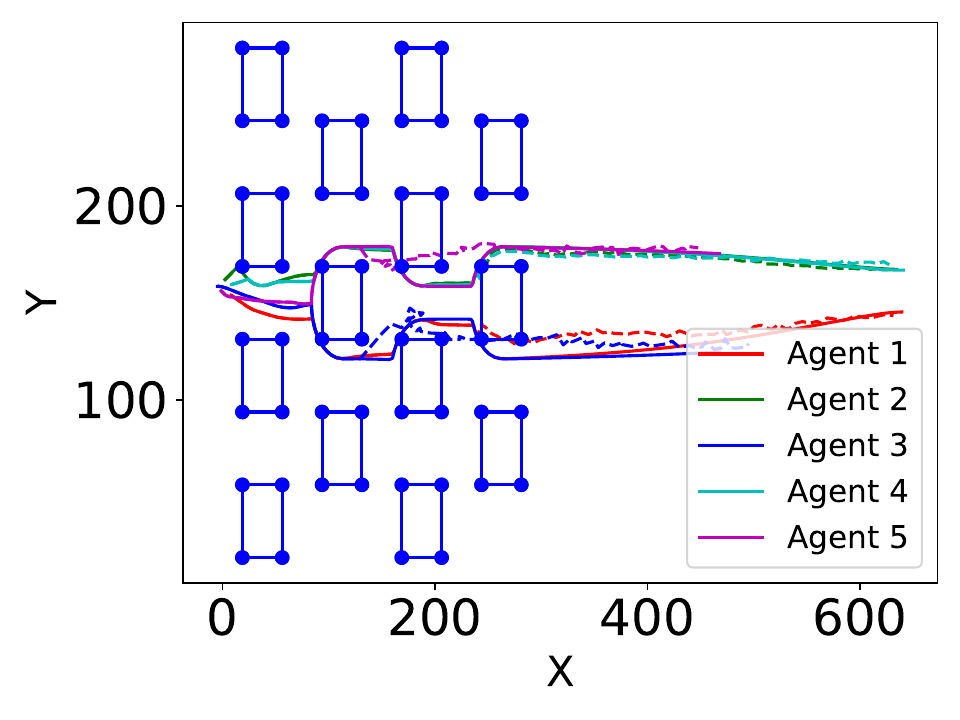}
        \caption{Example trajectory (in solid) from $\mathcal{D}$ 
          with prediction (in dashed)}
        \label{fig:example_trajectory_shifted}
    \end{subfigure}
    
    \caption{Example Trajectories for the Case Study}
    \label{fig:multiagent_example_trajectories}
\end{figure}
\section{Case Study: Drone-swarm Simulation}\label{sec:simulations} 
Consider Swarmlab \cite{soria2020swarmlab}, a Matlab-based drone swarm simulator, where we consider a group of drones navigating through obstacles. To validate the proposed RPRV algorithms, we present a case study for
runtime verification of both a single-agent CPS and  
an MAS. We show scalability of the MAS RPRV algorithms and analyze the effect of different predictors on the verification results\footnote{The code for the case study can be found at: \url{https://github.com/SAIDS-Lab/Robust_Spatio-Temporal_Predictive_Runtime_Verification}}.

\mypara{Estimation of Distribution Shifts.}
For validation purpose, we estimate the distribution shift $\mathcal{D}_f(\mathcal{D}, \mathcal{D}_0)$ from the training and test datasets. To do so, we randomly sample test trajectories from $\mathcal{D}$ (denoted by $T_{\mathcal{D}}$), and training trajectories from $\mathcal{D}_0$ (denoted by $T_{\mathcal{D}_0}$). For STL verification with single-agent CPS, for each of the nonconformity scores $R^{(i)}$ from \eqref{eq:R_direct}, \eqref{eq:R_indirect}, and \eqref{eq:R_hybrid}, we perform the following procedure: we calculate $R^{(i)}$ for the trajectories from $T_{\mathcal{D}_0}$ and $T_{\mathcal{D}}$ to obtain the empirical distributions of $\mathcal{R}_0$ and $\mathcal{R}$, respectively, where $\mathcal{R}_0$ and $\mathcal{R}$ are the induced distributions using the nonconformity scores over $\mathcal{D}_0$ and $\mathcal{D}$. Specifically, we use kernel density estimators with Gaussian kernels to estimate the empirical probability density functions (PDFs) of each distribution. We then numerically evaluate the distribution shift by computing $TV(\mathcal{R}, \mathcal{R}_0) = \frac{1}{2}\int_\mathcal{X}|q(x) - p(x)|dx$ where $p(x)$ and $q(x)$ are the estimated PDFs associated with $\mathcal{R}$ and $\mathcal{R}_0$. We obtain the values $\epsilon_1$, $\epsilon_2$, and $\epsilon_3$ that indicate the estimated distribution shifts on \eqref{eq:R_direct}, \eqref{eq:R_indirect}, and \eqref{eq:R_hybrid}. Finally, we take $\epsilon := \max(\epsilon_1, \epsilon_2, \epsilon_3)$ so that $\epsilon$ is greater than the estimated distribution shift of $ D_f(\mathcal{R}, \mathcal{R}_0)$ for all $\mathcal{R}$ and $\mathcal{R}_0$ in \eqref{eq:R_direct}, \eqref{eq:R_indirect}, and \eqref{eq:R_hybrid}. For STREL verification with MAS, we perform the same procedure but with the nonconformity scores $R^{(i)}$ from \eqref{eq:nonc_multi_direct}, \eqref{eq:r_indirect_strel}, and \eqref{eq:r_hybrid_strel}. As $\epsilon$ is often not exactly known and has to be estimated in practice, we note that $\epsilon$ can be thought of as a parameter  that robustifies our predictive runtime verification algorithms, as it is common practice in other areas such as robust control \cite{zhou1996robust}. The purpose of the estimation of $\epsilon$ here is for validation of our algorithms.

\mypara{System Description.} Consider $L$ drones with three-dimensional state $X_\tau[l] \in \mathbb{R}^3$ describing the location of drone $l\in\{1,\hdots,L\}$ at time $\tau$. The initial position of each drone is uniformly sampled as $X_0[l] \sim P_0 + 20[U(0, 1), U(0, 1), U(0, 1)]^T$, where $P_0 := [-10, 150, 50]^T$ and  where $U(a, b)$ describes the uniform distribution with a range of $(a, b)$. The task of the swarm is to navigate through a cluttered environment with 14 fixed parallelepiped obstacles (see Figure \ref{fig:multiagent_example_trajectories}). Each drone is assumed to be a point-mass, and the swarm adapts the Olfati-Saber algorithm for navigation, which is detailed in \cite{soria2020swarmlab} and requires the agents to establish consensus to accomplish a constant distance from their neighbors and to move in a constant speed towards the equilibrium. For validation, we generate trajectories by a swarm controller with a reference speed, which the drones track using the Olfati-Saber algorithm, of 6 and 5.9  for the training distribution $\mathcal{D}_0$ and the test distribution $\mathcal{D}$, respectively. We assume to have access to 1000 trajectories from $D_0$, which we denote by $Z_0$, and 500 test trajectories from $D$, which we denote by $Z$. We assume the current time is $t := 50$ and train an LSTM predictor on 200 trajectories, which we denote by $Z_{train}\subset Z_0$. For illustration, we show in Figure \ref{fig:multiagent_example_trajectories} one example trajectory (with solid lines) with their predictions (with dashed lines) separately from $Z_0$ and $Z$ with their 2D view from the perpective of vertical axis up to time $T := 120$ with $L := 5$. We also show in Figure \ref{fig:multiagent_example_trajectories} the locations of the obstacles. Note that the purple and blue trajectories in the right figure of Figure \ref{fig:multiagent_example_trajectories} are shorter, denoting a decrease in the reference speed, which results in worse predictions, as denoted by the dashed lines.

\subsection{Validation of STL RPRV Methods}\label{sec:stl_rprv}
\textit{We consider the RPRV of an STL formula over a single drone in a swarm of $5$ drones}, i.e., $L \coloneqq 5$. To compute $\epsilon$, we let $T_{\mathcal{D}_0} \coloneqq Z_0 \setminus Z_{train}$ and $T_{\mathcal{D}} \coloneqq Z$. Using the aforementioned procedure, we set $\epsilon := 0.172$. Since the specification is only considered over agent 1 (shown shortly), it is sufficient to consider $\|X_\tau[1]^{(i)} - \hat{X}[1]^{(i)}\|$ in \eqref{eq:R_indirect} and in computing $\alpha_\tau$, as we do in computing $\epsilon$ and in implementing the interpretable method (Variant I) in the experiments in this subsection. Similarly, we consider $\mathcal{B}_\tau := \{\zeta\in\mathbb{R}^N | \|\zeta[1] - \hat{X}_{\tau|t}[1]\| \le \tilde{C}\alpha_\tau\}$ for the interpretable method (Variant I).

\mypara{System Requirement.} In terms of the requirement, we want to verify the safety and the reachability of a single agent within the general CPS by considering the formula
\begin{align*}
    \phi &\coloneqq G_{[0, T]}(X[1][2] \ge 10 \wedge \min_{oc \in \text{obstacle centers}}\|X[1][0, 1] - oc[0, 1]\|_\infty \ge 18.75) \wedge F_{[0, T]}(X[1][0] \ge 600);
\end{align*}
where $X[1][n]$ extracts the $n$-th dimension of $X[1]$ and $X[1][0, 1]$ extracts the 1st and 2nd dimension of $X[1]$. We seek to monitor $\phi$ with $\tau_0 := 0$. The formula $\min_{oc \in \text{obstacle centers}}\|X[1][0, 1] - oc[0, 1]\|_\infty \ge 18.75$ makes sure that agent does not collide with the obstacles. The formula $G_{[0, T]}(X[1][2] \ge 10 \wedge \min_{oc \in \text{obstacle centers}}\|X[1][0, 1] - oc[0, 1]\|_\infty \ge 18.75)$ is therefore a safety requirement specifying that the agent does not collide to the ground nor to the obstacles at all time up to $T := 120$. The formula $F_{[0, T]}(X[1][0] \ge 600)$ is a task completion requirement specifying that the agent promptly reaches the goal configuration.

\begin{figure*}
    \centering
    \begin{subfigure}[t]{0.35\textwidth}
    \includegraphics[width=\textwidth]{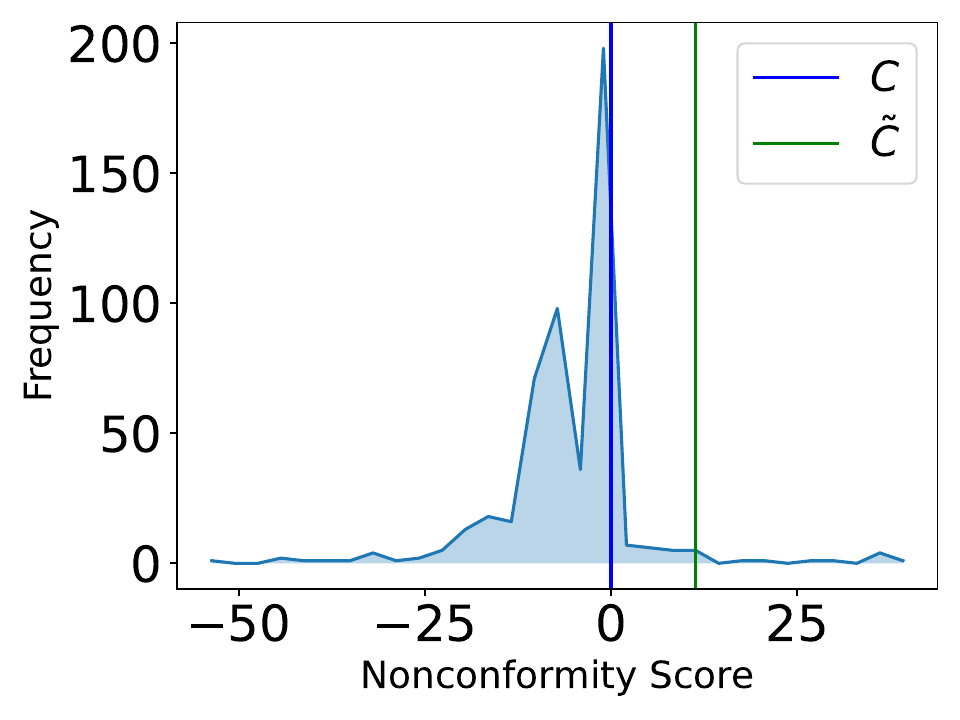}
    \caption{Histogram of $R^{(i)}$ from \eqref{eq:R_direct} with $\tilde{C}$ and $C$.}
    \label{fig:stl_direct_nonconformities}
    \end{subfigure}
     \hspace{1mm}
    \begin{subfigure}[t]{0.35\textwidth}
    \includegraphics[width=\textwidth]{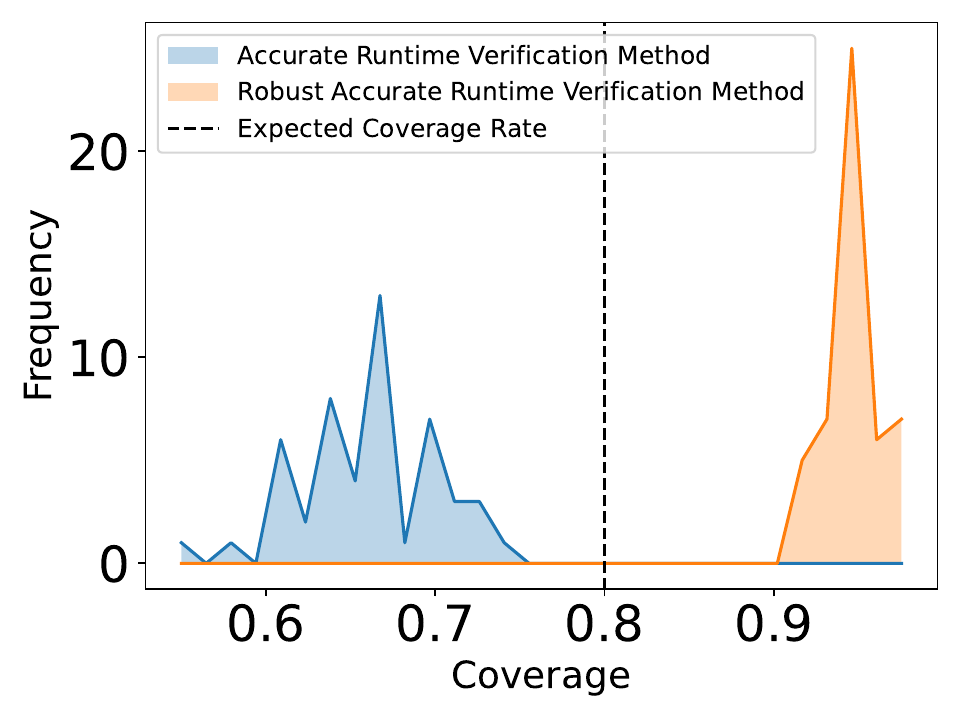}
    \caption{Histogram of coverage: accurate methods.}
    \label{fig:stl_direct_coverage}
    \end{subfigure}
     \hspace{1mm}
    \begin{subfigure}[t]{0.32\textwidth}
    \includegraphics[width=\textwidth]{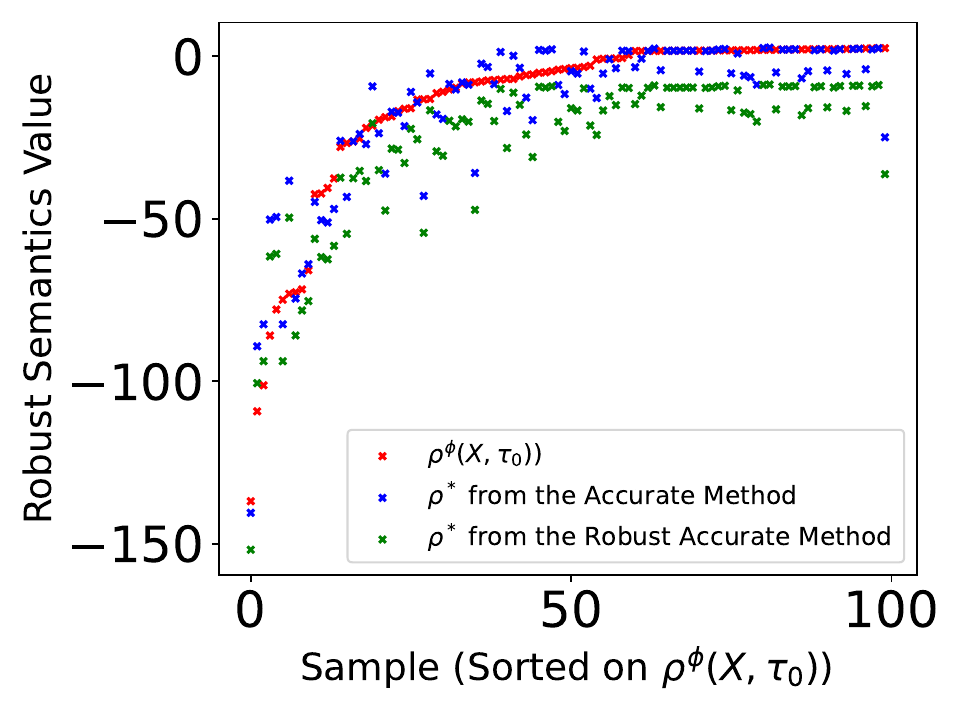}
    \caption{$\rho^\phi(X, \tau_0)$ and $\rho^*$ for the accurate methods.}
    \label{fig:stl_direct_comparison}
    \end{subfigure}
    \hspace{1mm}
    \begin{subfigure}[t]{0.32\textwidth}
    \includegraphics[width=\textwidth]{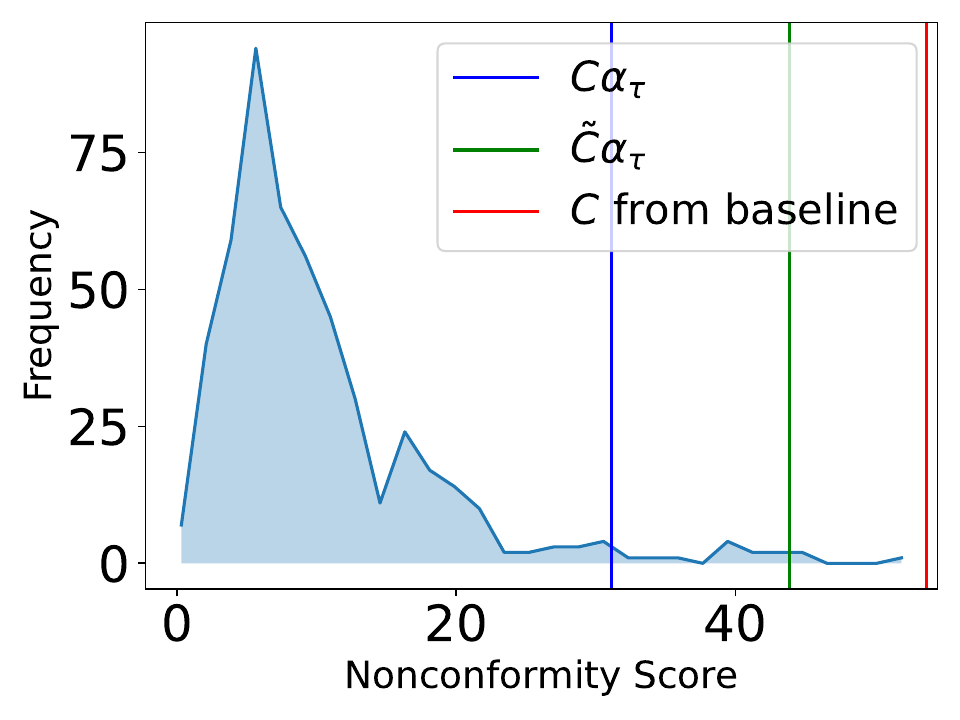}
    \caption{Histogram of residuals with $\tilde{C}\alpha_\tau$, $C\alpha_\tau$, and $C$ from \cite{lindemann2023conformal}.}
    \label{fig:stl_indirect_nonconformities}
    \end{subfigure}
    \hspace{1mm}
    \begin{subfigure}[t]{0.32\textwidth}
    \includegraphics[width=\textwidth]{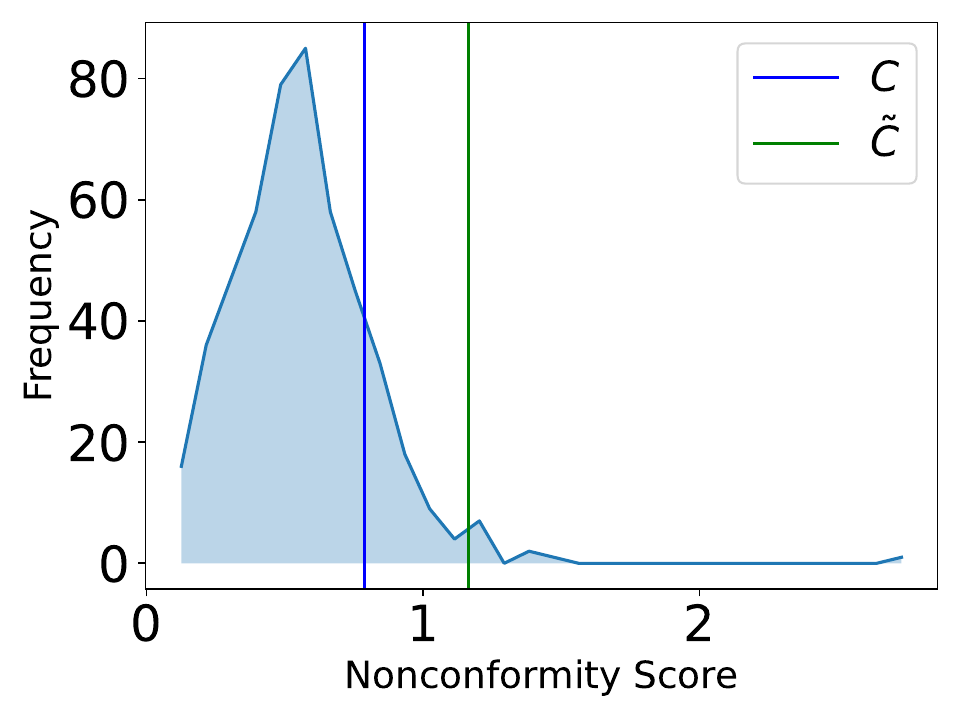}
    \caption{Histogram of $R^{(i)}$ from \eqref{eq:R_hybrid} with $\tilde{C}$ and $C$.}
    \label{fig:stl_hybrid_nonconformities}
    \end{subfigure}
    \hspace{1mm}
    \begin{subfigure}[t]{0.35\textwidth}
    \includegraphics[width=\textwidth]{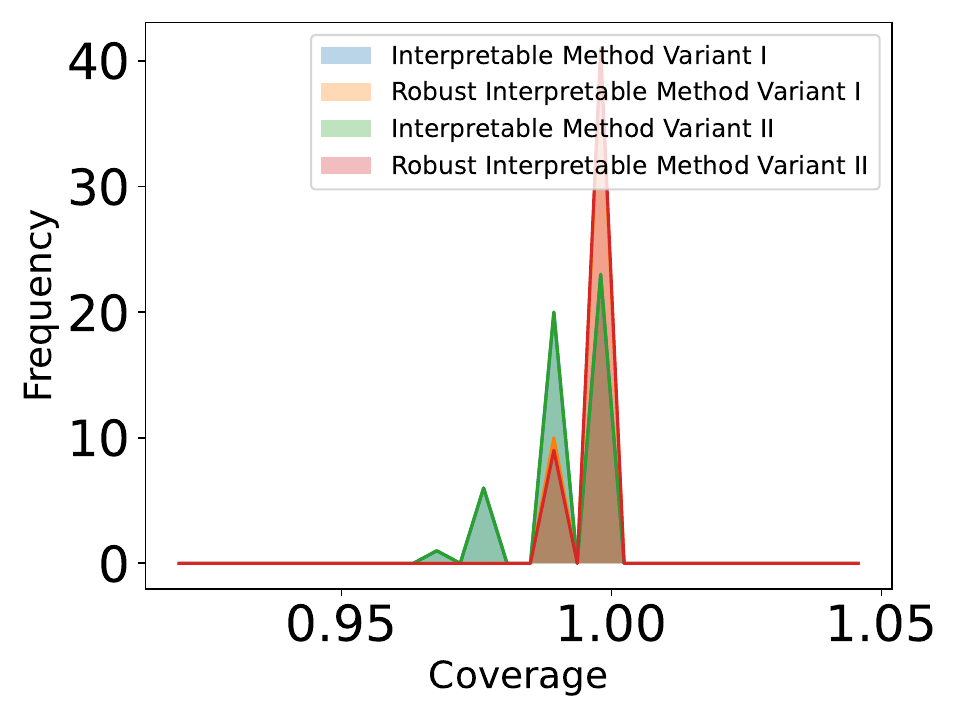}
    \caption{Histogram of coverage: interpretable methods.}
    \label{fig:stl_indirect_coverages}
    \end{subfigure}
    \hspace{1mm}
    \begin{subfigure}[t]{0.35\textwidth}
    \includegraphics[width=\textwidth]{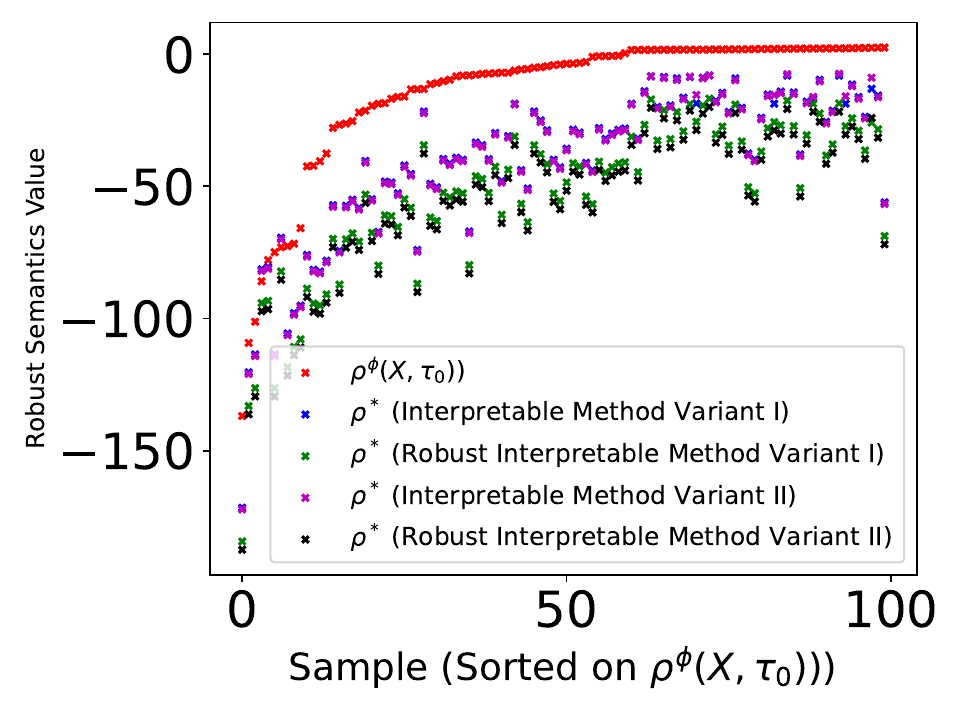}
    \caption{$\rho^\phi(X, \tau_0)$ and $\rho^*$ for the interpretable methods.}
    \label{fig:stl_indirect_comparison}
    \end{subfigure}
    \caption{Results for the STL RPRV case study}
    \label{fig:results_rprv_stl}
\end{figure*}

\mypara{Validation and Comparison of Accurate Method.} For this case study, we seek to find $\rho^*$ from Problem \ref{prob1} for a failure probability $\delta := 0.2$. To illustrate the validity, we compare to a baseline method from \cite{lindemann2023conformal}, where we use non-robust conformal prediction (i.e. $\epsilon \coloneqq 0$) from \eqref{eq:vanilla_quantile} instead of robust conformal prediction from \eqref{eq:C_tilde}. We run the following experiments 50 times: we sample $K \coloneqq 500$ calibration trajectories from $Z_0$ and $100$ test trajectories from $Z$. For one of these experiments, we show the histogram of nonconformity scores $R^{(i)}$ from \eqref{eq:R_direct} for the calibration data and the robust prediction region $\tilde{C}$ (in green) from \eqref{eq:C_tilde} in Figure \ref{fig:stl_direct_nonconformities}. For comparison, we also show the prediction region $C$ (in blue) from the non-robust accurate method in \cite{lindemann2023conformal} (where $\epsilon = 0$), which is smaller than $\tilde{C}$ and cannot deal with the distribution shifts. In Figure \ref{fig:stl_direct_coverage}, we plot the empirical coverage over the $50$ experiments: for each experiment, we compute the percentage of the test trajectories satisfying $\rho^\phi(X^{(i)}) \ge \rho^\phi(\hat{X}^{(i)}) - C$ and $\rho^\phi(X^{(i)}) \ge \rho^\phi(\hat{X}^{(i)}) - \tilde{C}$ for the non-robust and robust methods. We remark that comparing to the non-robust method, the robust method achieves an empirical coverage that center above the expected success rate of $0.8$ (i.e., $1 - \delta$), which we denote by the dashed line. For one experiment, we show the true robust semantics $\rho^\phi(X, \tau_0)$ for the ground truth test data and the predicted worst-case robust semantics $\rho^*$ in Figure \ref{fig:stl_direct_comparison} for the non-robust and robust RPRV algorithms. As we see, the robust method is more conservative than the non-robust counterpart (as the robust method achieves lower $\rho^*$) and hence accounts for the distribution shift.

\mypara{Validation and Comparison of Interpretable Methods.} We perform the same validation experiments as for the accurate method again for $50$ times with $\delta \coloneqq 0.2$. We remark that since the interpretable method (Variant I) is inspired by the interpretable (indirect) method from \cite{lindemann2023conformal}, we compare our interpretable RPRV algorithms with the interpretable method from \cite{lindemann2023conformal} to illustrate the reduction of conservatism in our proposed RPRV methods. We also compare to the non-robust interpretable methods, where we consider the prediction region $C$ from \eqref{eq:vanilla_quantile} instead of $\tilde{C}$. We run the following experiments $50$ times: we again sample $K \coloneqq 500$ calibration trajectories from $Z_0$ and $100$ test trajectories from $Z$. \emph{Variant I.} For one of these experiments, we show the histogram of $\|X_\tau^{(i)}[1] - \hat{X}^{(i)}_\tau[1]\|$ in Figure \ref{fig:stl_indirect_nonconformities} over the calibration set where $\tau := 120$. For comparison, we also show $C\alpha_\tau$ and $\tilde{C}\alpha_\tau$ following our result in Lemma \ref{lemma:variant1} where $C$ and $\tilde{C}$ are computed from equations \eqref{eq:vanilla_quantile} and \eqref{eq:C_tilde} with $\alpha_\tau$ and $R^{(i)}$ in \eqref{eq:R_indirect} accompanying the aforementioned changes where we focus on agent $1$. We also plot the prediction region from \cite{lindemann2023conformal}, which we here denote by $C$ from baseline. As we see, the prediction $C$ from the baseline model in \cite{lindemann2023conformal} is more conservative. \emph{Variant II.} In Figure \ref{fig:stl_hybrid_nonconformities}, we show the histogram of $R^{(i)}$ in equation \eqref{eq:R_hybrid} over calibration data along with $\tilde{C}$ from \eqref{eq:C_tilde} and $C$ from \eqref{eq:vanilla_quantile}. As expected, the interpretable algorithms are more conservative than the accurate algorithm and all achieve an empirical coverage for $\rho^\phi(X^{(i)}, \tau_0) \ge \rho^*$ greater than the desired coverage of $1 - \delta = 0.8$. This is demonstrated in Figure \ref{fig:stl_indirect_coverages} where we plot the coverage over the $50$ experiments. For one experiment, we show the true robust semantics $\rho^\phi(X, \tau_0)$ for the 100 ground truth test data  and the predicted worst-case robust semantics $\rho^*$ for Variants I and II in Figure \ref{fig:stl_indirect_comparison}.

\subsection{Validation of STREL RPRV Methods}\label{sec:validation_rprv}
\textit{We now consider the RPRV of a STREL formula for an MAS, where we consider $L \coloneqq 5, 7, 10$ to showcase the scalability of the algorithms.} To compute $\epsilon$, we again select $T_{\mathcal{D}_0} \coloneqq Z_0 \setminus Z_{train}$ and $T_\mathcal{D} = Z$, and we set $\epsilon \coloneqq 0.140, 0.077, 0.145$ respectively for $L = 5, 7, 10$.

\mypara{System Requirement.} With STL RPRV, we cannot monitor the communication between the agents within an MAS, which is possible with STREL. To illustrate STREL RPRV, we consider the following setting with a central observer located at the ground level that can communicate with any drones operating below the height of 50m. Each drone, even if located above the height of 50m, can forward its information to a fixed set of other drones. Namely, each drone can communicate with drone $2$ and drone $2$ can communicate with all other drones. One can think of the connection topology as a communication protocol set prior to the swarm operation. We again emphasize that if two drones $l_1$ and $l_2$ can communicate with each other at time $\tau$, $w(l_1, l_2, \tau, X)$ is finitely valued. As a natural choice of communication cost, we consider the weight to be the communication time, which we assume is proportional to the distance between two drones. Formally, we let $w(l_1, l_2, \tau, X) := 0.2\|X[l_1] - X[l_2]\|_2$ if the drones  $l_1$ and $l_2$ can communicate (according to the aforementioned protocol) and $w(l_1, l_2, \tau, X) := \infty$ otherwise. Therefore, if the central observer desires to gather information of the behavior of a specific drone $l$, it is required that at all times, either $l$ is below the height of $50$m or $l$ can communicate with a drone below the height of $50$.
\begin{figure*}
    \centering
    \begin{subfigure}[t]{0.32\textwidth}
    \includegraphics[width=\textwidth]{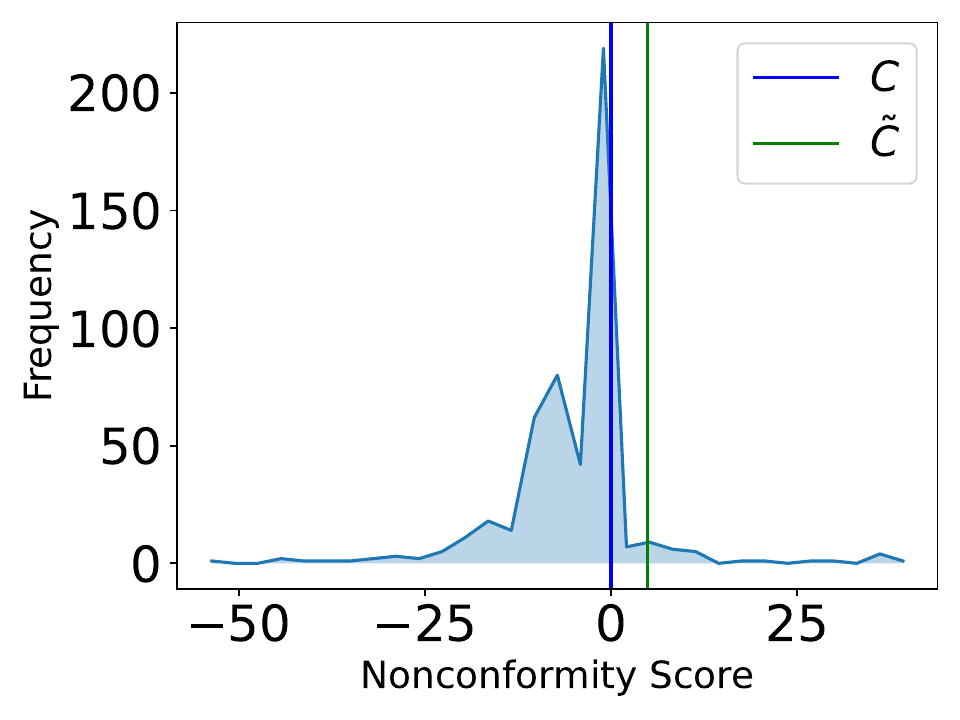}
    \caption{Histogram of $R^{(i)}$ from \eqref{eq:nonc_multi_direct} with $L = 5$.}
    \label{fig:direct_nonconformities_5}
    \end{subfigure}
    \hspace{1mm}
    \begin{subfigure}[t]{0.32\textwidth}
    \includegraphics[width=\textwidth]{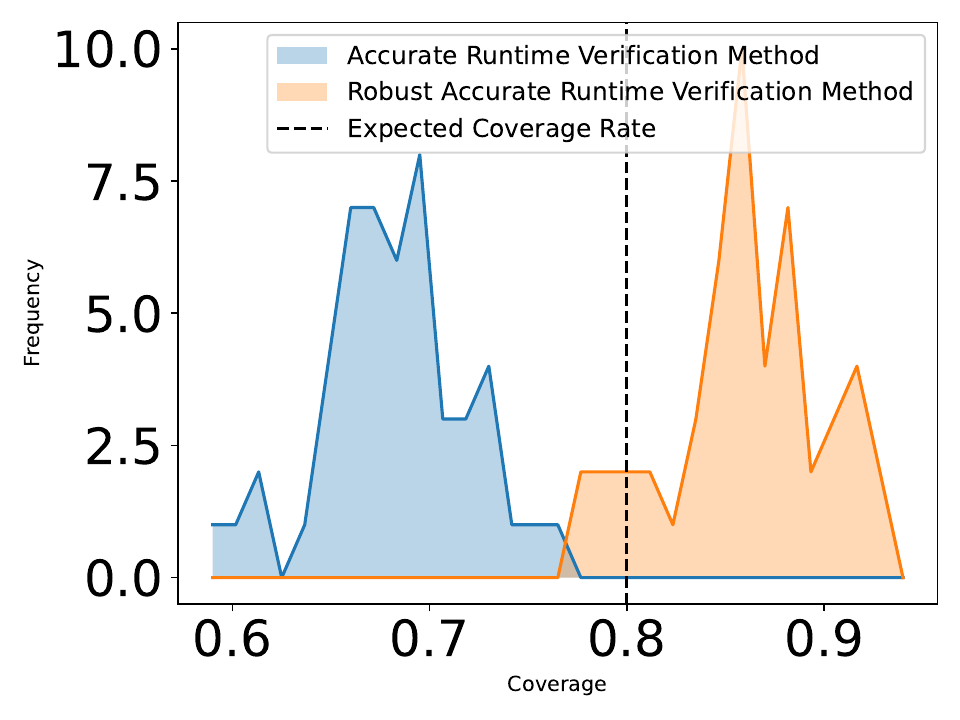}
    \caption{Histogram of coverage: accurate methods with $L = 5$.}
    \label{fig:direct_coverages_5}
    \end{subfigure}
    \hspace{1mm}
    \begin{subfigure}[t]{0.32\textwidth}
    \includegraphics[width=\textwidth]{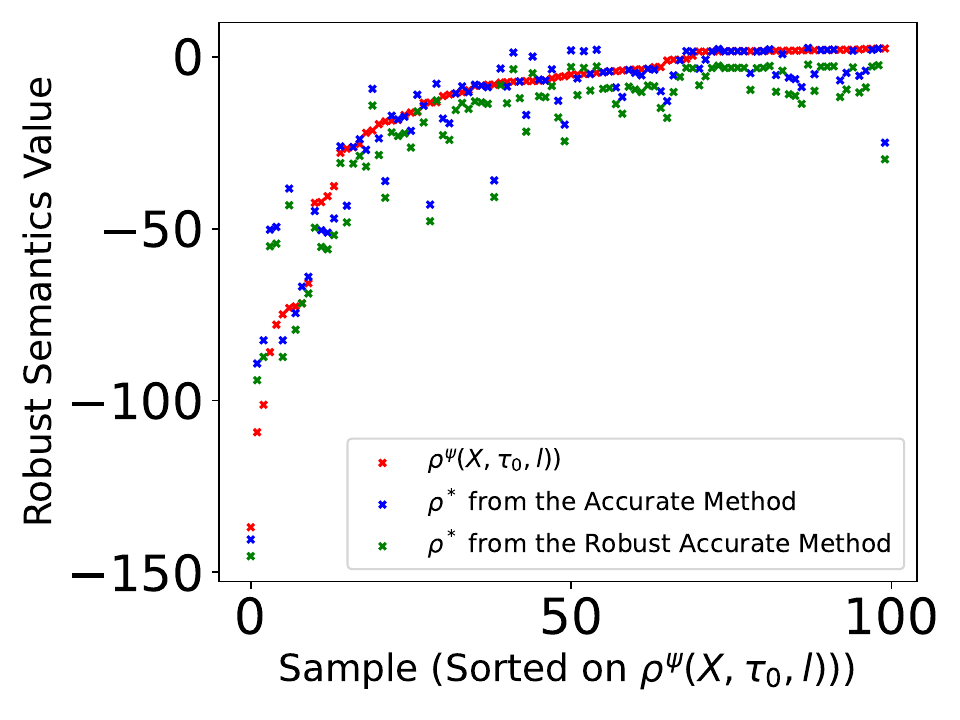}
    \caption{$\rho^\psi(X, \tau_0, l)$ and $\rho^*$ with $L = 5$ for the accurate methods.}
    \label{fig:direct_robustness_5}
    \end{subfigure}
    \hspace{1mm}
     \begin{subfigure}[t]{0.32\textwidth}
    \includegraphics[width=\textwidth]{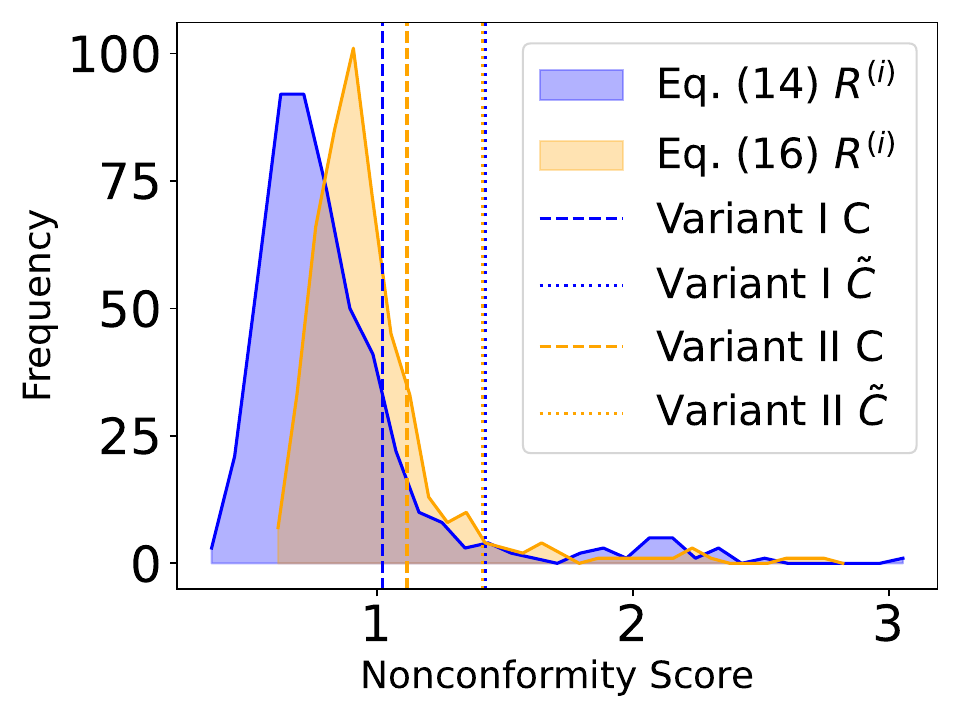}
    \caption{Histogram of $R^{(i)}$ from \eqref{eq:r_indirect_strel} and \eqref{eq:r_hybrid_strel} with $L = 5$}
    \label{fig:superimposed_nonconformities_5}
    \end{subfigure}
    \begin{subfigure}[t]{0.32\textwidth}
    \includegraphics[width=\textwidth]{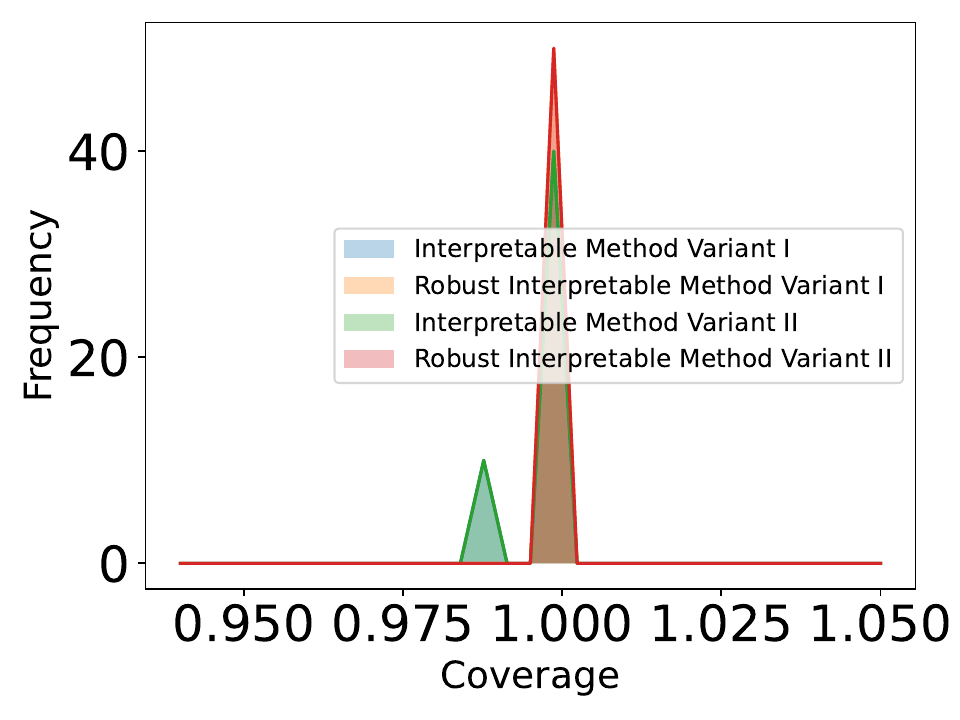}
    \caption{Histogram of coverage: interpretable methods with $L = 5$.}
    \label{fig:indirect_coverages_5}
    \end{subfigure}
    \hspace{1mm}
     \begin{subfigure}[t]{0.32\textwidth}
    \includegraphics[width=\textwidth]{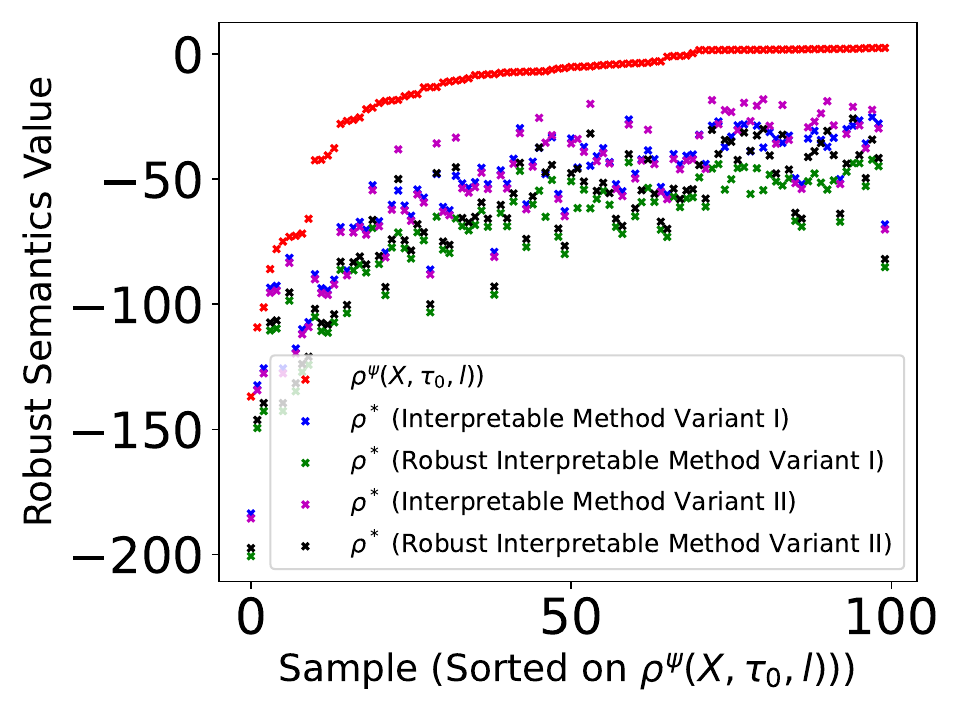}
    \caption{$\rho^\psi(X, \tau_0, l)$ and $\rho^*$ with $L = 5$ For the interpretable methods.}
    \label{fig:indirect_robustness_5}
    \end{subfigure}
    \caption{Results for STREL RPRV Case Study with $L := 5$.}
    \label{fig:results_strel_5}
\end{figure*}
We would like to verify at runtime that agent 1 can safely navigate through the cluster of obstacles, reach the goal configuration, and  maintain stable and prompt connection to the central observer. Formally, we consider the specification $\psi := \psi_1 \wedge G_{[0, T]}(\mathcal{M}_{[0, 6]}\psi_2)$ where $\psi_1 := G_{[0, T]}(X[l][2] \ge 10 \wedge \psi_3) \wedge \psi_4$, $\psi_2 := X[l][2] \le 50$, $\psi_3 := \min_{oc \in \text{obstacle centers}}\|X[l][0, 1] - oc[0, 1]\|_\infty \ge 18.75$, and $\psi_4 := F_{[0, T]}(X[l][0] \ge 600)$. Here, $X[l][n]$ extracts the $n$-th dimension of $X[l]$ and $X[l][0, 1]$ extracts the 1st and 2nd dimension of $X[l]$. We  seek to monitor $\psi$ on agent $l \coloneqq 1$ with $\tau_0 := 0$. The formula $G_{[0, T]}(X_t[l][2] \ge 10 \wedge \psi_3)$, similar to the single-agent STL specification, is a safety requirement. The formula $\psi_4$ is  a task completion requirement. The formula $\psi_2$ specifies that an agent is below the height of 50 (and thus maintains communication with the centralized monitor) and $G_{[0, T]}(M_{[0, 6]}\psi_2)$ thus specifies that at all time through the journey, there exists an agent (which can communicate with the observer) within the communication time no more than $6$ from $l$.

\mypara{Validation and Comparison of Accurate Method.} For this case study, we seek to find $\rho^*$ from Problem \ref{prob2} for a failure probability of $\delta \coloneqq 0.2$, where we first consider $L \coloneqq 5$. Again, we compare with a baseline model where we consider non-robust conformal prediction from \eqref{eq:vanilla_quantile}. We run the following experiment 50 times: we sample $K := 500$ calibration data from $Z_0$ and $100$ test data from $Z$. For one of these experiments, we show the histogram of nonconformity scores $R^{(i)}$ from \eqref{eq:nonc_multi_direct} for the calibration data and the robust prediction region $\tilde{C}$ from \eqref{eq:C_tilde} in Figure \ref{fig:direct_nonconformities_5}. For comparison, we show the prediction region $C$ from the non-robust accurate method, which is smaller than $\tilde{C}$ and cannot deal with the distribution shifts. In Figure \ref{fig:direct_coverages_5}, we plot the empirical coverages over the 50 experiments. For each experiment, we compute the ratio of test trajectories satisfying $\rho^\psi(X^{(i)}, \tau_0, l) \ge \rho^\psi(\hat{X}^{(i)}, \tau_0, l) - C$ and $\rho^\psi(X^{(i)}, \tau_0, l) \ge \rho^\psi(\hat{X}^{(i)}, \tau_0, l) - \tilde{C}$ respectively for the non-robust and robust methods. As demonstrated, the non-robust method undercovers. For one experiment, we show in Figure \ref{fig:direct_robustness_5} the true robust semantics $\rho^\psi(X, \tau_0, l)$ for the 100 ground truth test data and the predicted worst-case robust semantics $\rho^*$ for the non-robust and the robust methods. We show the results for $L := 7$ and $L := 10$ in Appendix \ref{sec:supp_results} in the supplementary material.

\mypara{Validation and Comparison of Interpretable Methods.} We perform the same validation experiments as for the accurate method again for 50 times with $\delta \coloneqq 0.2$. We first consider $L \coloneqq 5$. For validation purpose, we compare our RPRV methods to the non-robust interpretable methods, where we consider the prediction region $C$ from \eqref{eq:vanilla_quantile} instead of $\tilde{C}$ from \eqref{eq:C_tilde}. We run the following experiments $50$ times: we again sample $K \coloneqq 500$ calibration trajectories from $Z_0$ and $100$ test trajectories from Z. \emph{Variant I.} For one of these experiments, we show in Figure \ref{fig:superimposed_nonconformities_5} the histogram of $R^{(i)}$ from equation \eqref{eq:r_indirect_strel}. For comparison, we also show $C$ and $\tilde{C}$ following our result in Lemma \ref{lemma:variant1} where $C$ and $\tilde{C}$ are computed from equations \eqref{eq:vanilla_quantile} and \eqref{eq:C_tilde}. \emph{Variant II.} In Figure \ref{fig:superimposed_nonconformities_5}, we show the histogram of $R^{(i)}$ in equation \eqref{eq:r_hybrid_strel} over calibration data along with $\tilde{C}$ from \eqref{eq:C_tilde} and $C$ from \eqref{eq:vanilla_quantile}. As expected, the interpretable algorithms are more conservative than the accurate algorithm and all achieve an empirical coverage for $\rho^\psi(X^{(i)}, \tau_0, l) \ge \rho^*$ greater than the desired coverage of $1 - \delta = 0.8$. This is demonstrated in Figure \ref{fig:indirect_coverages_5} where we plot the coverage over the $50$ experiments. For one experiment, we show the ground truth robust semantics $\rho^\psi(X, \tau_0, l)$ for the $100$ ground truth test data and the predicted worst-case robust semantics $\rho^*$ for Variants I and II in Figure \ref{fig:indirect_robustness_5}. Again, we show the experimental results for $L := 7$ and $L := 10$ in Appendix \ref{sec:supp_results}.

\mypara{Computational Efficiency and Scalability.} Following the analysis in Section \ref{sec:data_complexity}, we show empirically now that given the system and specification setup considered in this case study, the accurate method requires more offline computation time (considering a complex specification $\psi$) but less online computation time as compared to the interpretable methods. Between the interpretable methods, Variant I suffers less from the offline computation but needs high online computation time. We further show empirically the scalability of the MAS RPRV algorithms.

We empirically validate the complexity by recording the computational time over the 50 experiments for each case of $L := 5, 7, 10$. For each of the accurate and interpretable methods (with 2 variants), we record the calibration time for each experiment for computing $C$ and $\tilde{C}$, including the time for computing $R^{(i)}$ over the 500 calibration data, and show the average calibration times over the $50$ experiments for the accurate and the interpretable methods in Table \ref{table:calibration}. For both the accurate and the interpretable methods, we see the offline computational time scale reasonably with increasing number of agents. Notably, as expected, the accurate methods take the longest, and the interpretable method (variant I) takes a shorter time than variant II. For each experiment, we also record the online computation time for computing the coverages (with both robust and non-robust conformal prediction) over the $100$ test data, including the time for computing state-level and predicate-level prediction regions for the indirect methods for each experiment. We show the average test time over the $50$ experiments and over $100$ test trajectories from each experiment in Table \ref{table:test} for both the accurate and interpretable methods. We remark that the interpretable method (variant I) takes the longest as expected at test time due to the optimization involved. We also emphasize that for both timings of calibrations and testings, the time for generating the predicted trajectories is ignored. For the 2 variants of interpretable methods, we time the computation of $\alpha_{\tau, l'}$, which are 0.040, 0.061, and 0.096 seconds for $L := 5, 7, 10$, and for Variant I and $\alpha_{\pi, \tau, l'}$ on Variant II, which are 4.565, 6.457, and 9.202 seconds for $L := 5, 7, 10$, all of which are conducted offline.
\begin{table}
\small
\begin{tabular}{||c c c c||} 
 \hline
& Average Time ($L = 5$) & Average Time ($L = 7) $ & Average Time ($L = 10)$ \\ [0.5ex] 
 \hline\hline
 Accurate Method & 22.646 & 32.214 & 47.036 \\ 
 \hline
 Interpretable Method (Variant I) & 0.093 & 0.129 & 0.181 \\
 \hline
 Interpretable Method (Variant II) & 11.414 & 15.987 & 22.863 \\
 \hline
\end{tabular}
\caption{Average Calibration Time (seconds)}
\label{table:calibration}
\end{table}
\begin{table}
\small
\begin{tabular}{||c c c c||} 
 \hline
& Average Time ($L = 5$) & Average Time ($L = 7) $ & Average Time ($L = 10)$ \\ [0.5ex] 
 \hline\hline
 Accurate Method & 0.04533 & 0.06484 & 0.09383 \\ 
 \hline
 Interpretable Method (Variant I) & 3.13622 & 4.9504 & 7.82622 \\
 \hline
 Interpretable Method (Variant II) & 0.09069 & 0.12926 & 0.18639 \\
 \hline
\end{tabular}
\caption{Average Test Time (seconds)}
\label{table:test}
\end{table}
\begin{figure*}
    \centering
    \begin{subfigure}[t]{0.32\textwidth}
    \includegraphics[width=\textwidth]{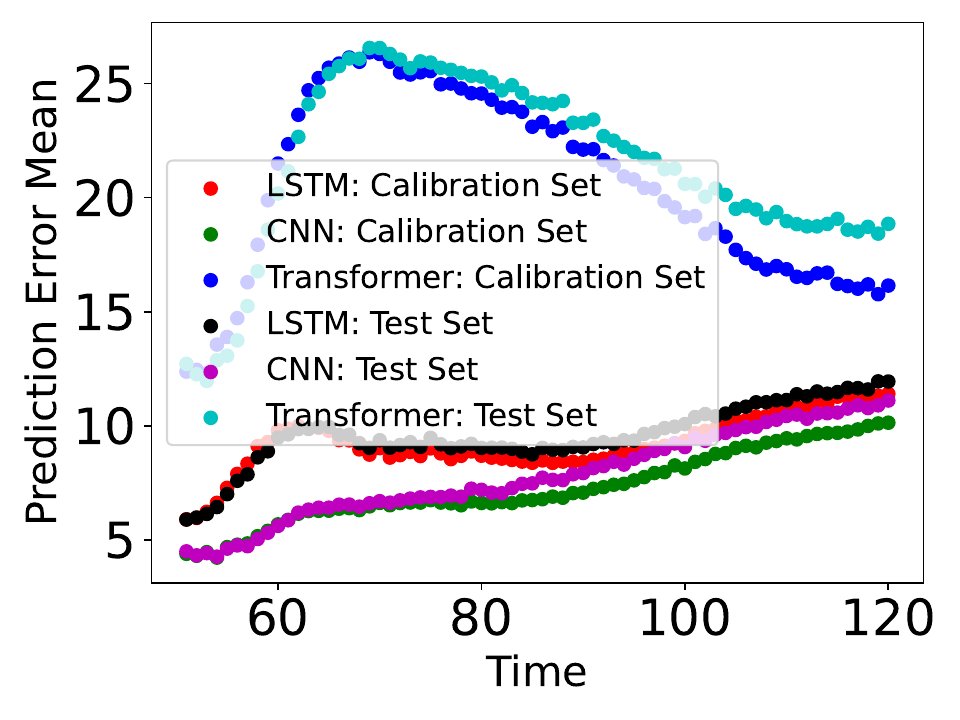}
    \caption{Mean prediction Errors.}
    \label{fig:prediction_errors_comparison_mean}
    \end{subfigure}
    \hspace{1mm}
    \begin{subfigure}[t]{0.32\textwidth}
    \includegraphics[width=\textwidth]{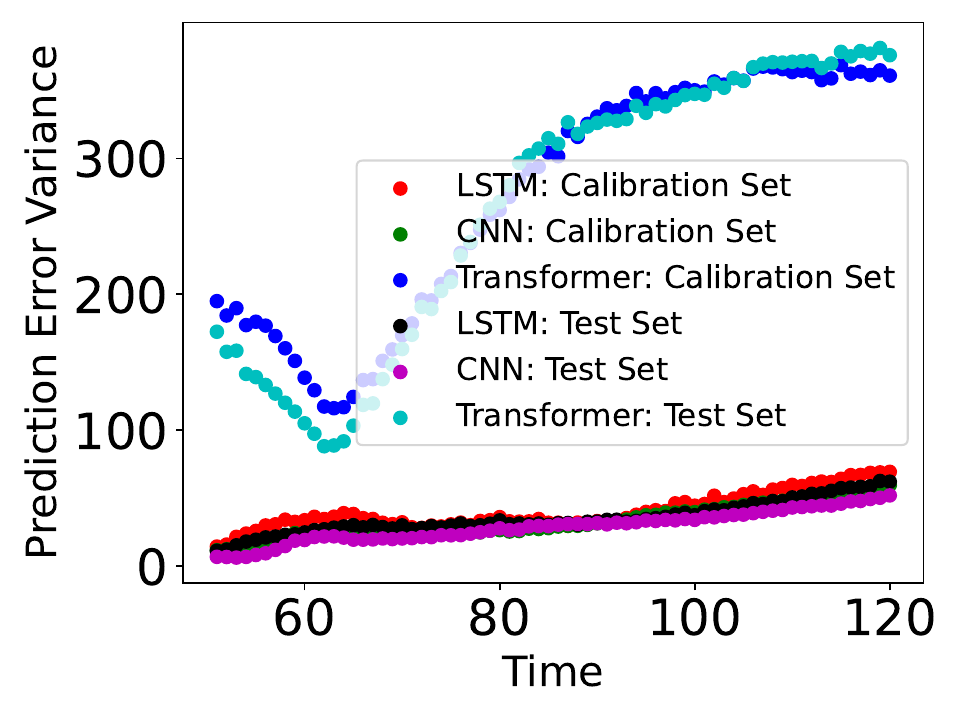}
    \caption{Prediction error variances.}
    \label{fig:prediction_errors_comparison_variance}
    \end{subfigure}
    \hspace{1mm}
    \begin{subfigure}[t]{0.32\textwidth}
    \includegraphics[width=\textwidth]{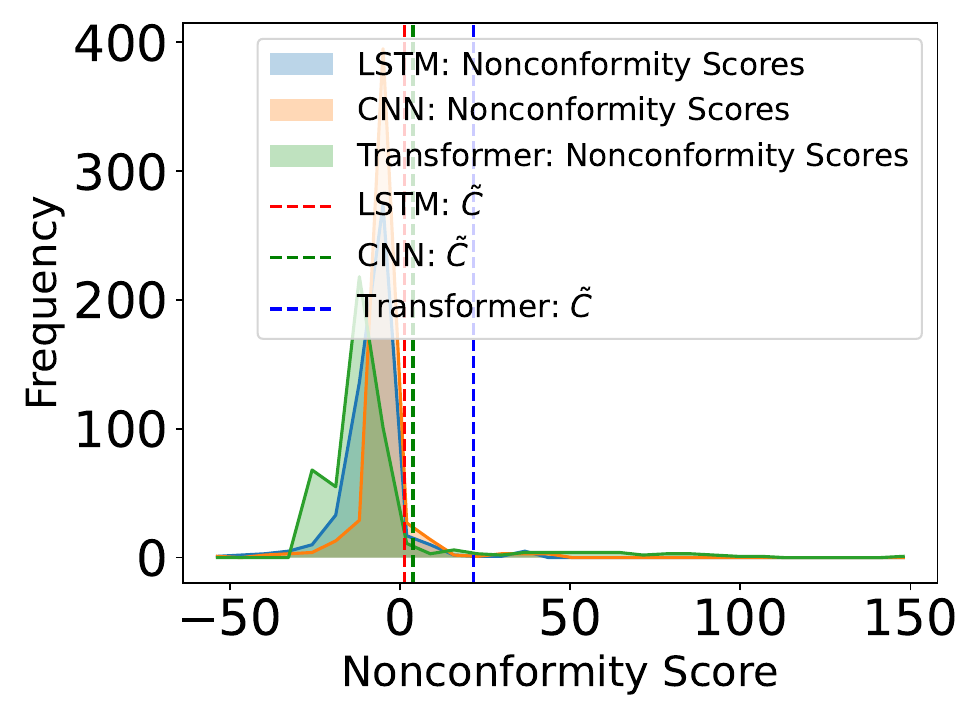}
    \caption{Histogram of $R^{(i)}$ from \eqref{eq:nonc_multi_direct}.}
    \label{fig:comparison_direct_nonconformities}
    \end{subfigure}
    \hspace{3mm}
    \begin{subfigure}[t]{0.32\textwidth}
    \includegraphics[width=\textwidth]{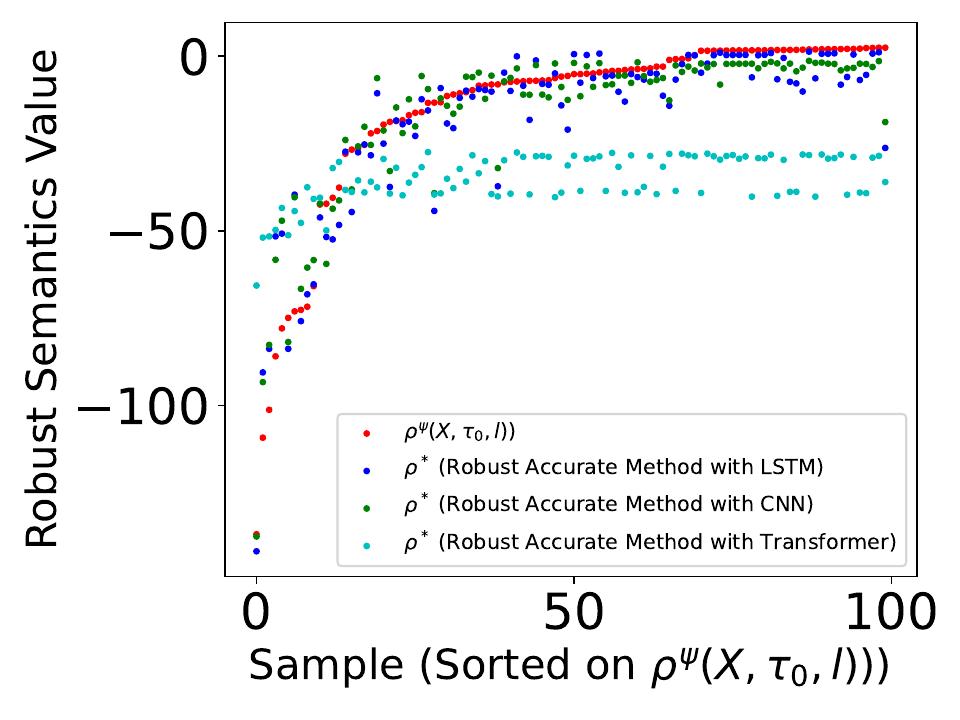}
    \caption{$\rho^\phi(X, \tau_0, l)$ and $\rho^*$ for the accurate methods.}
    \label{fig:comparison_direct_robustness}
    \end{subfigure}
    \hspace{1mm}
    \begin{subfigure}[t]{0.32\textwidth}
    \includegraphics[width=\textwidth]{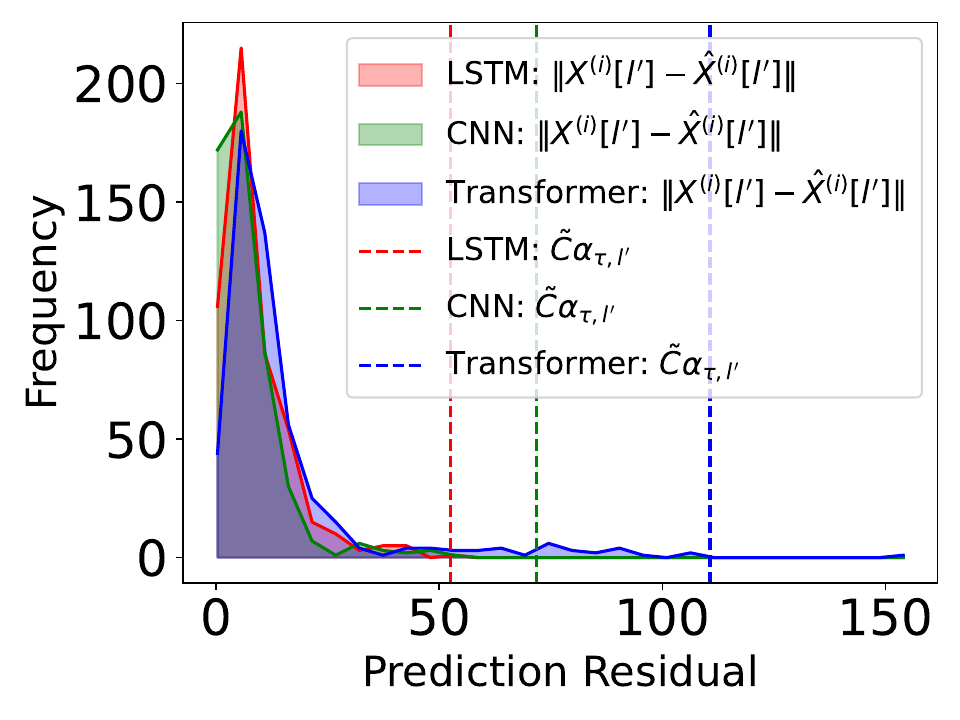}
    \caption{Histogram of $\|\hat{X}^{(i)}[l'] - X^{(i)}[l']\|$.}
    \label{fig:comparison_indirect_prediction_regions}
    \end{subfigure}
    \hspace{1mm}
    \begin{subfigure}[t]{0.32\textwidth}
    \includegraphics[width=\textwidth]{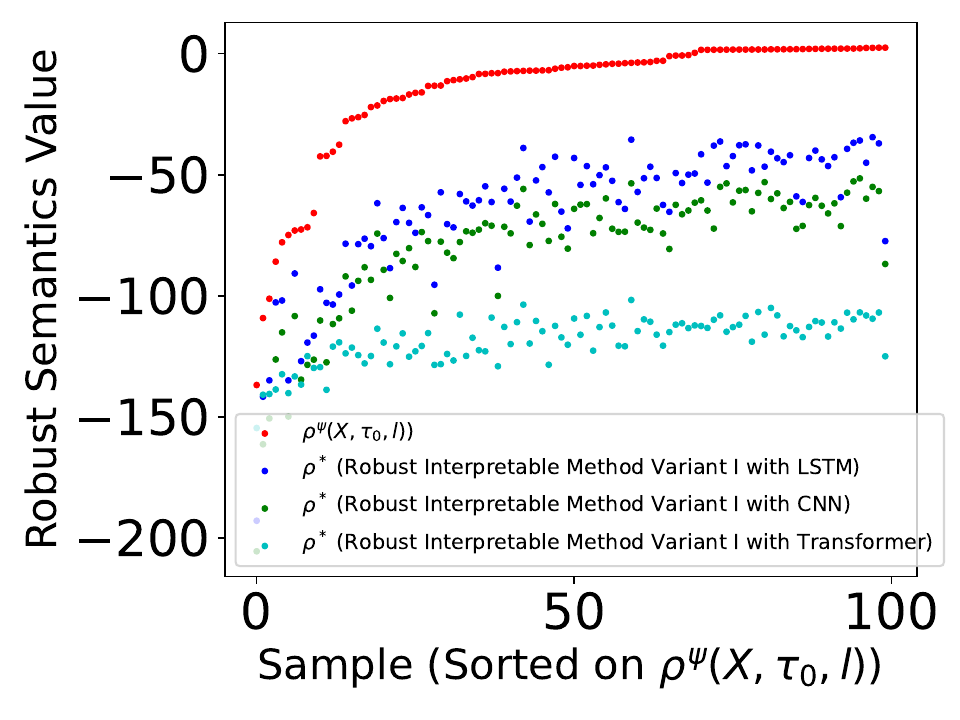}
    \caption{ $\rho^\phi(X, \tau_0, l)$ and $\rho^*$ with $L = 5$ for the interpretable methods (variant I).}
    \label{fig:comparison_indirect_robustness}
    \end{subfigure}
    \hspace{1mm}
    \begin{subfigure}[t]{0.35\textwidth}
    \includegraphics[width=\textwidth]{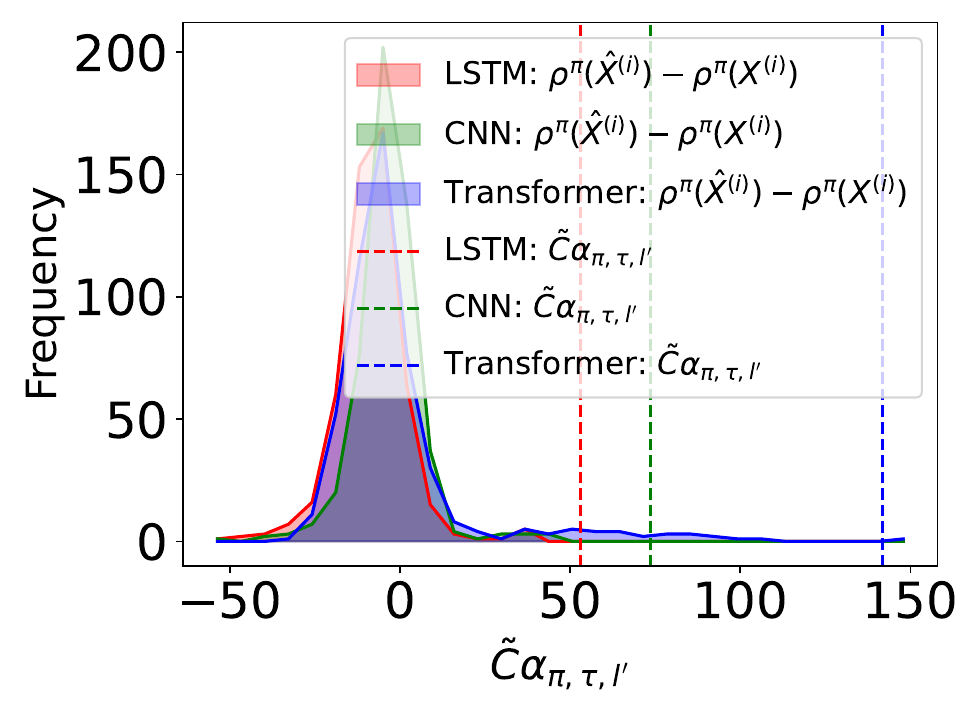}
    \caption{Histogram of $\rho^\pi(\hat{X}^{(i)}, \tau_0, l') - \rho^\pi(X^{(i)}, \tau_0, l')$.}
    \label{fig:comparison_hybrid_prediction_regions}
    \end{subfigure}
    \hspace{1mm}
    \begin{subfigure}[t]{0.35\textwidth}
    \includegraphics[width=\textwidth]{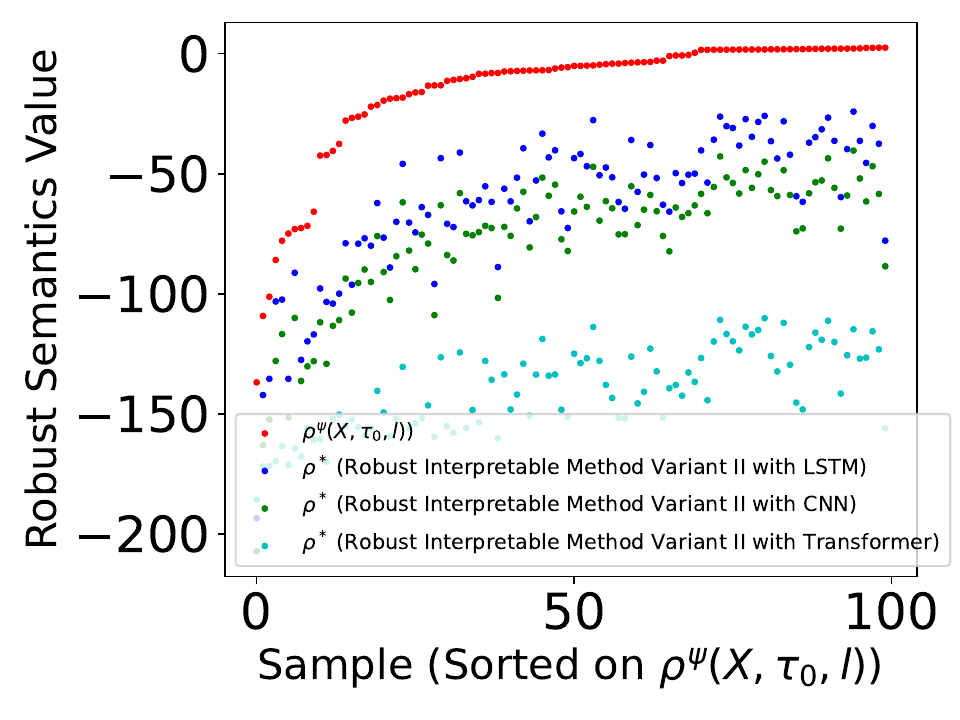}
    \caption{$\rho^\phi(X, \tau_0, l)$ and $\rho^*$ for the interpretable methods (variant II).}
    \label{fig:comparison_hybrid_robustness}
    \end{subfigure}
    \caption{Comparisons between the Verification Results with Different Predictors}
    \label{fig:results_comparison}
\end{figure*}
\subsection{Validation of STREL RPRV Methods with Other Trajectory Predictors}\label{sec:model_comparison}
We now empirically evaluate the effect of predictors in the verification results and show that better predictors lead to tighter verification results. We conduct the following experiment. Consider three trajectory predictors: the LSTM predictor that we use in Section \ref{sec:validation_rprv}, a CNN predictor \cite{lecun1998gradient}, and a transformer \cite{vaswani2017attention}. For illustration purposes, we show an example trajectory from $\mathcal{D}_0$ and from $\mathcal{D}$ with prediction for each of CNN and Transformer predictors in Figure \ref{fig:multiagent_example_trajectories_cnn} and \ref{fig:multiagent_example_trajectories_transformer}, respectively, in Appendix \ref{sec:supp_results} with $L \coloneqq 5$. As shown, the transformer is the most inaccurate from a visual inspection. We use a similar procedure for validation of accurate and interpretable methods from Section \ref{sec:validation_rprv} with $L := 5$ but now each with one experiment only for easy illustration of the effect of the predictor. In this experiment, we consider $\delta \coloneqq 0.3$ and we again sample $K := 500$ calibration data and $100$ test data. For both the calibration set and the test set, for each timestamp $\tau \in \{t + 1, \hdots, T\}$ and each agent $l' \in \{1, \hdots, L\}$, we collect the prediction error $\|\hat{X}^{(i)}_\tau[l'] - X^{(i)}_\tau[l']\|$. In Figure \ref{fig:prediction_errors_comparison_mean}, we plot the mean prediction errors over all trajectories in the calibration set and over all agents, $l'$, with respect to the time $\tau$. We do the same for the test set in Figure \ref{fig:prediction_errors_comparison_mean}. In Figure \ref{fig:prediction_errors_comparison_variance}, we plot the variance of the prediction errors over all trajectories in the calibration set and over all agents with respect to the time $\tau$, which we do the same for the test set. As illustrated, the transformer performs the worst in both accuracy and precision, whereas the performance of CNN and LSTM are roughly comparable with CNN achieving a slightly higher accuracy. We conduct the same procedure as the one for computing of $\epsilon$ in Section \ref{sec:validation_rprv} for the LSTM, the CNN and the transformer, and select the largest among the three, $\epsilon := 0.217$ as our tuning parameter for the experiment.

\mypara{Effect of Predictor on the Accurate Method.} We show in Figure \ref{fig:comparison_direct_nonconformities} the histogram of $R^{(i)}$ from equation \eqref{eq:nonc_multi_direct} together with $\tilde{C}$ from robust conformal prediction for all three predictors. We remark that the nonconformity scores from the transformer model in general, as expected, are more spread out. We show in Figure \ref{fig:comparison_direct_robustness} the ground truth robust semantics $\rho^\psi(X, \tau_0, l)$ for the test data and the corresponding $\rho^*$ from the robust accurate method. As expected, we see that the use of a transformer model leads to conservative verification results with $\rho^*$ that do not well reflect the ground truth test robust semantics for individual samples. We also note that the empirical coverage for robust accurate method with the LSTM predictor is $0.78$, with the CNN predictor is $0.69$, and with the transformer is $0.87$, which is within our expectation. The reason why we see a 0.69 for the CNN result (which is below 0.8) is that empirical coverage is not ensured to be above 0.8.

\mypara{Effect of Predictor on the Interpretable Method.} \emph{Variant I}: For the interpretable method (Variant I), we show in Figure \ref{fig:comparison_indirect_prediction_regions} the histogram of prediction errors $\|\hat{X}^{(i)}_\tau[l'] - X^{(i)}_\tau[l']\|$ over the calibration set with $\tau \coloneqq 120$ and $l' \coloneqq 1$ together with $\tilde{C}\alpha_{\tau, l'}$ with $\tilde{C}$ from \eqref{eq:C_tilde} for all three predictors. As expected, the transformer has the largest normalized prediction region $\tilde{C}\alpha_{\tau, l'}$. We remark that the reason why $\tilde{C}\alpha_{\tau, l'}$ associated with the LSTM model is smaller than that with the CNN model despite CNN having a slightly better accuracy is that the normalization constant $\tilde{C}\alpha_{\tau, l'}$ affects the tightness of verification results (i.e., we would expect the opposite if $\alpha_{\tau, l'}$ are set to $1$). We also show in Figure \ref{fig:comparison_indirect_robustness} the ground truth robust semantics $\rho^\psi(X, \tau_0, l)$ for the test data and the corresponding $\rho^*$ from the robust interpretable method (Variant I), where we see that the transformer introduces the most conservative verification results. The empirical coverage for the robust interpretable method (Variant I) with LSTM, with CNN, and with transformer are all $1$. \emph{Variant II} For the interpretable method (Variant II), we show in Figure \ref{fig:comparison_hybrid_prediction_regions} the histogram of $\rho^\pi(\hat{X}^{(i)}, \tau_0, l') - \rho^\pi(X^{(i)}, \tau_0, l')$ where $\tau \coloneqq 120$, $l' \coloneqq 1$, and $\pi := X[l][0] \ge 600$ together with $\tilde{C}\alpha_{\pi, \tau, l'}$ with $\tilde{C}$ from \eqref{eq:C_tilde} for all three predictors. We show in Figure \ref{fig:comparison_hybrid_robustness} the ground truth robust semantics $\rho^\psi(X, \tau_0, 1)$ for the test data and the corresponding $\rho^*$ from the robust interpretable method (Variant II). We again see that the transformer produces the most conservative verification results. We notice the same pattern as we observe for the interpretable method (Variant I). The empirical coverage for the robust interpretable method (Variant II) with LSTM, with CNN, and with Transformer are again all 1.
\section{Conclusion}
\label{sec:conclusion}
In this paper, we discussed the predictive runtime STL verification algorithms, proposed in \cite{zhao2024robust} for general CPS that can predict system failures even when the test time system differs from the design time system. Specifically, our algorithms are robust against distribution shifts measured in terms of the $f$-divergence of their system trajectories. We first use trajectory predictors to predict the future motion of the system, and we robustly quantify prediction uncertainty with respect to signal temporal logic (STL) system specifications using robust conformal prediction and calibration data from the design time system. Our first algorithm (called accurate algorithm) provides tight verification guarantees, while our second algorithm (called interpretable algorithm), which we present in two variants, provides more interpretable runtime information. Building on \cite{zhao2024robust}, we propose the RPRV methods for MAS under STREL specifications, where we apply robust conformal prediction over STREL robust semantics. We analyze the  relationship between calibration data, desired confidence,
and permissible distribution shift. We also provide an exhaustive case study in a drone swarm simulator where we validate the guarantees from the STL and STREL RPRV methods. We analyze the scalability of the STREL RPRV methods and discuss the impact of trajectory predictors in the verification results.

\section{Acknowledgements}
This work was partially supported by the National Science Foundation through the following grants: CAREER award (SHF-2048094), CNS-1932620, CNS-2039087, FMitF-1837131, CCF-SHF-1932620, IIS-SLES-2417075, funding by Toyota R\&D and Siemens Corporate Research through the USC Center for Autonomy and AI, an Amazon Faculty Research Award, and the Airbus Institute for Engineering Research. This work does not reflect the views or positions of any organization listed.

\bibliographystyle{IEEEtran}
\bibliography{main}

\appendix
\newpage
\begin{center}
\section*{Appendix}
\end{center}
\section{Semantics of Signal Temporal Logic}
\label{app:STL}
For a trajectory $x:=(x_0,x_1,\hdots)$, the semantics of an STL formula $\phi$ that is enabled at time $\tau_0$, denoted by $(x,\tau_0)\models \phi$, can be recursively computed based on the structure of $\phi$ using the following rules:
	\begin{align*}
	(x,\tau)\models \text{True} & \hspace{0.5cm} \text{iff} \hspace{0.5cm} \text{True},\\
	(x,\tau)\models \pi & \hspace{0.5cm} \text{iff} \hspace{0.5cm} h(x_\tau)\ge 0,\\
	(x,\tau)\models \neg\phi & \hspace{0.5cm} \text{iff} \hspace{0.5cm} (x,\tau)\not\models \phi,\\
	(x,\tau)\models \phi' \wedge \phi'' & \hspace{0.5cm} \text{iff} \hspace{0.5cm} (x,\tau)\models\phi' \text{ and } (x,\tau)\models\phi'',\\
	(x,\tau)\models \phi' U_I \phi'' & \hspace{0.5cm} \text{iff} \hspace{0.5cm} \exists \tau''\in (\tau\oplus I)\cap \mathbb{N} \text{ s.t. } (x,\tau'')\models\phi''\\
	&\hspace{1.2cm} \text{ and } \forall \tau'\in(\tau,\tau'')\cap \mathbb{N}, (x,\tau')\models\phi'.
	\end{align*}
The robust semantics $\rho^{\phi}(x,\tau_0)$ provide more information than the semantics $(x,\tau_0)\models \phi$, and indicate how robustly a specification is satisfied. We can again recursively calculate $\rho^{\phi}(x,\tau_0)$ based on the structure of $\phi$ using the following rules:
\begin{align*}
	\rho^\text{True}(x,\tau)& := \infty,\\
	\rho^{\pi}(x,\tau)& := h(x_\tau) \\
	\rho^{\neg\phi}(x,\tau) &:= 	-\rho^{\phi}(x,\tau),\\
	\rho^{\phi' \wedge \phi''}(x,\tau) &:= 	\min(\rho^{\phi'}(x,\tau),\rho^{\phi''}(x,\tau)),\\
  \rho^{\phi' U_I \phi''}(x,\tau) &:= \underset{\tau''\in (\tau\oplus I)\cap \mathbb{N}}{\text{sup}}  \Big(\min\big(\rho^{\phi''}(x,\tau''), \underset{\tau'\in (\tau,\tau'')\cap \mathbb{N}}{\text{inf}}\rho^{\phi'}(x,\tau') \big)\Big).
\end{align*}
	 The formula length $L^{\phi}$ of a bounded STL formula $\phi$ can be recursively calculated based on the structure of $\phi$ using the following rules:
 \begin{align*}
     L^\text{True}&=L^\pi:=0\\
     L^{\neg\phi}&:=L^\phi\\
     L^{\phi'\wedge\phi''}&:=\max(L^{\phi'},L^{\phi''})\\
     L^{\phi' U_I \phi''}&:=\max \{I\cap \mathbb{N}\}+\max(L^{\phi'},L^{\phi''}). \\
 \end{align*}
 The probabilistic robust semantics $\bar{\rho}^{\phi}(X,\tau_0)$, which are defined over the random trajectory $X\sim \mathcal{D}$ and the predicted trajectory $\hat{X} := (X_\text{obs}, \hat{X}_{t + 1 | t} , \hdots, \hat{X}_{t + H | t})$, are recursively defined based on the structure of $\phi$ using the following rules:
\begin{align*}
    \bar{\rho}^\text{True}(\hat{X},\tau)& := \infty,\\
    \bar{\rho}^{\pi}(\hat{X},\tau)& := 
     \begin{cases}
     h(X_\tau) &\text{ if } \tau\le t\\
     \rho_{\pi,\tau}^* &\text{ otherwise }
     \end{cases}\\
	\bar{\rho}^{\phi' \wedge \phi''}(\hat{X},\tau) &:= 	\min(\bar{\rho}^{\phi'}(\hat{X},\tau),\bar{\rho}^{\phi''}(\hat{X},\tau)),\\
	\bar{\rho}^{\phi' U_I \phi''}(\hat{X},\tau) &:= \underset{\tau''\in (\tau\oplus I)\cap \mathbb{N}}{\text{sup}}  \Big(\min\big(\bar{\rho}^{\phi''}(\hat{X},\tau''), \underset{\tau'\in (\tau,\tau'')\cap \mathbb{N}}{\text{inf}}\bar{\rho}^{\phi'}(\hat{X},\tau') \big)\Big).
	\end{align*}
 where the constant $\rho_{\pi,\tau}^*$ defines probabilistic prediction regions that can be computed as explained in detail in Section \ref{sec:general_cps_rprv}.

\section{Semantics of Signal Reach and Escape Temporal Logic} \label{app:STREL}
Recall that in Section \ref{sec:rnn}, we describe an MAS with a concatenated trajectory $x:=(x_0,x_1,\hdots)$. We let $x_\tau$ denote the state of $x$ at time $\tau$ and $x_\tau[l]$ denote the state of agent $l$ at that time instance. For a concatenated trajectory $x$ and an agent labeled $l$, the semantics of an STREL formula $\psi$ that is enabled at time $\tau_0$, denoted by $(x,\tau_0, l)\models \psi$, can be recursively computed based on the structure of $\psi$ using the following rules:
	\begin{align*}
	(x,\tau, l)\models \text{True} & \hspace{0.5cm} \text{iff} \hspace{0.5cm} \text{True},\\
	(x,\tau, l)\models \pi & \hspace{0.5cm} \text{iff} \hspace{0.5cm} h(x_\tau[l])\ge 0,\\
	(x,\tau, l)\models \neg\psi & \hspace{0.5cm} \text{iff} \hspace{0.5cm} (x,\tau, l)\not\models \psi,\\
	(x,\tau, l)\models \psi' \wedge \psi'' & \hspace{0.5cm} \text{iff} \hspace{0.5cm} (x,\tau, l)\models\psi' \text{ and } (x,\tau, l)\models\psi'',\\
	(x,\tau, l)\models \psi' U_I \psi'' & \hspace{0.5cm} \text{iff} \hspace{0.5cm} \exists \tau''\in (\tau\oplus I)\cap \mathbb{N} \text{ s.t. } (x,\tau'', l)\models\psi''\\
	&\hspace{1.2cm} \text{ and } \forall \tau'\in(\tau,\tau'')\cap \mathbb{N}, (x,\tau', l)\models\psi', \\
 (x, \tau, l) \models \psi'R_{[d_1, d_2]}\psi''  & \hspace{0.5cm} \text{iff} \hspace{0.5cm} \exists r \in Routes(\tau, l) \text{ and } \exists i \in \mathbb{N}^\infty \\ & \hspace{1.2cm} \text{ with } d(i, r, \tau) \in [d_1, d_2] \\
  &\hspace{1.2cm}  \text{ s.t. } (x, \tau, r[i]) \models \psi'', \text{ and } \forall j < i, \\ & \hspace{1.2cm} (x, \tau, r[j]) \models \psi', \\
   (x, \tau, l) \models \mathcal{E}_{[d_1, d_2]}\psi' & \hspace{0.5cm} \text{iff} \hspace{0.5cm} \exists r \in Routes(\tau, l) \text{ s.t. } \exists l' \in r \\ &\hspace{1.2cm}  \text{ with } \tilde{d}_{\min}(l, l', \tau) \in [d_1, d_2] \\  &\hspace{1.2cm}  \text{ s.t. } \forall i \le r(l'), (x, \tau, r[i])\models \psi'.
	\end{align*}
The robust semantics $\rho^\psi(x, \tau, l)$ is recursively defined below:
\begin{align*}
	\rho^\text{True}(x,\tau, l)& := \infty,\\
	\rho^{\pi}(x,\tau, l)& := h(x_\tau[l]) \\
	\rho^{\neg\psi}(x,\tau, l) &:= 	-\rho^{\psi}(x,\tau, l),\\
	\rho^{\psi' \wedge \psi''}(x,\tau, l) &:= 	\min(\rho^{\psi'}(x,\tau, l),\rho^{\psi''}(x,\tau, l)),\\
  \rho^{\psi' U_I \psi''}(x,\tau, l) &:= \underset{\tau''\in (\tau\oplus I)\cap \mathbb{N}}{\text{sup}} \Big(\min\big(\rho^{\psi''}(x,\tau'', l), \underset{\tau'\in (\tau,\tau'')\cap \mathbb{N}}{\text{inf}}\rho^{\psi'}(x,\tau', l) \big)\Big),\\
  \rho^{\psi' R_{[d_1, d_2]} \psi''}(x,\tau, l) &:= \underset{r \in Routes(\tau, l)}{\text{sup}}\Big(\underset{i \in \{i' \mid d(r, i', \tau) \in [d_1, d_2]\}}{\text{sup}}\big(\min(\rho^{\psi''}(x, \tau, r[i]), \underset{j < i}{\text{inf}}\rho^{\psi'}(x, \tau, r[j]))\big)\Big), \\
  \rho^{\mathcal{E}_{[d_1, d_2]} \psi'}(x,\tau, l) &:=\underset{r \in Routes(\tau, l)}{\text{sup}}\Big( \underset{l' \in \{l'' \in r\mid \tilde{d}_{\min}(l, l'', \tau) \in [d_1, d_2]\}}{\text{sup}}\big(\underset{i \le r(l')}{\text{inf}}\rho^{\psi'}(x, \tau, r[i])\big)\Big).
\end{align*}
The formula length $L^\psi$ is computed recursively as follows:
 \begin{align*}
     L^\text{True}&=L^\pi = L^{\psi' R_{[d_1, d_2]} \psi''} = L^{\mathcal{E}_{[d_1, d_2]}\psi'}:=0\\
     L^{\neg\psi}&:=L^\psi\\
     L^{\psi'\wedge\psi''}&:=\max(L^{\psi'},L^{\psi''})\\
     L^{\psi' U_I \psi''}&:=\max \{I\cap \mathbb{N}\}+\max(L^{\psi'},L^{\psi''}).
 \end{align*}

We define the probabilistic robust semantics of STREL as follows:
\begin{align*}
    \bar{\rho}^\text{True}(\hat{X},\tau, l)& := \infty,\\
    \bar{\rho}^{\pi}(\hat{X},\tau, l)& := 
     \begin{cases}
     h(X_\tau[l]) &\text{ if } \tau\le t\\
     \rho_{\pi,\tau, l}^* &\text{ otherwise }
     \end{cases}\\
	\bar{\rho}^{\psi' \wedge \psi''}(\hat{X},\tau, l) &:= 	\min(\bar{\rho}^{\psi'}(\hat{X},\tau, l),\bar{\rho}^{\psi''}(\hat{X},\tau, l)),\\
	\bar{\rho}^{\psi' U_I \psi''}(\hat{X},\tau, l) &:= \underset{\tau''\in (\tau\oplus I)\cap \mathbb{N}}{\text{sup}}  \Big(\min\big(\bar{\rho}^{\psi''}(\hat{X},\tau'', l), \underset{\tau'\in (\tau,\tau'')\cap \mathbb{N}}{\text{inf}}\bar{\rho}^{\psi'}(\hat{X},\tau', l) \big)\Big), \\
  \bar{\rho}^{\psi' R_{[d_1, d_2]} \psi''}(\hat{X},\tau, l) &:= \underset{r \in Routes(\tau, l)}{\text{sup}}\Big(\underset{i \in \{i' \mid d(r, i', \tau) \in [d_1, d_2]\}}{\text{sup}}\big(\min(\bar{\rho}^{\psi''}(\hat{X}, \tau, r[i]), \underset{j < i}{\text{inf}}\bar{\rho}^{\psi'}(\hat{X}, \tau, r[j]))\big)\Big) \\
  \bar{\rho}^{\mathcal{E}_{[d_1, d_2]} \psi'}(\hat{X},\tau, l) &:=\underset{r \in Routes(\tau, l)}{\text{sup}}\Big( \underset{l' \in \{l'' \in r \mid \tilde{d}_{\min}(l, l'', \tau) \in [d_1, d_2]\}}{\text{sup}}\big(\underset{i \le r(l')}{\text{inf}}\bar{\rho}^{\psi'}(\hat{X}, \tau, r[i])\big)\Big).
	\end{align*}
 where the constant $\rho_{\pi,\tau, l}^*$ defines probabilistic prediction regions.

\mypara{Computation of STREL Robust Semantics} The RPRV algorithms rely on efficient computation of the robust semantics in finite time. While this computation is fairly standard for STL with bounded temporal operators following the recursive definition, we recall the computation of the robust semantics for the reach operator in Algorithm \ref{alg:reach} (with spatially bounded reach operator in Algorithm \ref{alg:bounded_reach} and spatially unbounded reach operator in Algorithm \ref{alg:unbounded_reach}), where spatial boundedness refers to the finite length of $[d_1, d_2]$ in the STREL syntax. The computation for the escape operator can be found in Algorithm \ref{alg:escape}. The returned matrix $s$ from Algorithms \ref{alg:reach} and \ref{alg:escape} is such that $s[l] \coloneq \rho^\psi(x, \tau, l)$ where it holds that $\psi \coloneqq \psi' R_{[d_1, d_2]} \psi''$ for Algorithm \ref{alg:reach} and $\psi \coloneqq \mathcal{E}_{[d_1, d_2]}\psi'$ for Algorithm \ref{alg:escape}. The algorithms follow \cite{nenzi2022logic}, in which the computational complexity, the termination, and the correctness of the algorithms are analyzed.

\begin{algorithm}
    \caption{Computation of the Robust Semantics for the Reach Operator}\label{alg:reach}
    \begin{algorithmic}
    \Require $d_1, d_2 \in \mathbb{R}^\infty$, $s_1: \{1, \hdots, L\} \rightarrow \mathbb{R}^\infty \text{ where } s_1[l] = \rho^{\psi'}(x, \tau, l)$, $s_2: \{1, \hdots, L\} \rightarrow \mathbb{R}^\infty \text{ where } s_2[l] = \rho^{\psi''}(x, \tau, l).$
    \If{$d_2 \neq \infty$}
    \State \Return $BoundedReach(d_1, d_2, s_1, s_2)$
    \Else
    \State \Return $UnboundedReach(d_1, s_1, s_2)$
    \EndIf
    \end{algorithmic}
\end{algorithm}

\begin{algorithm}
    \caption{Computation of the Robust Semantics for the Bounded Reach Operator}\label{alg:bounded_reach}
    \begin{algorithmic}
    \Require $d_1, d_2 \in \mathbb{R}^\infty$, $s_1: \{1, \hdots, L\} \rightarrow \mathbb{R}^\infty \text{ where } s_1[l] = \rho^{\psi'}(x, \tau, l)$, $s_2: \{1, \hdots, L\} \rightarrow \mathbb{R}^\infty \text{ where } s_2[l] = \rho^{\psi''}(x, \tau, l).$
    \State $\forall l, s[l] = \begin{cases}
        s_2[l], \text{ if } d_1 = 0, \\
        -\infty, \text{ otherwise}
    \end{cases}$
    \State $Q = \{(l, s_2[l], 0) \mid l \in L\}$
    \While{$Q \neq \emptyset$}
    \State $Q' = \emptyset$
    \For{all $(l, v, d) \in Q$}
    \For{all $l':l' \xrightarrow{w} l$}
    \State $v' = \min(v, s_1[l'])$
    \State $d' = d + w$
    \If{$d_1 \le d' \le d_2$}
    \State $s[l'] = \max(s[l'], v')$
    \EndIf
    \If{$d' < d_2$}
    \If{$\exists(l',v'', d') \in Q'$}
    \State $Q' = (Q' - \{(l, v'', d')\}) \cup \{(l', \max(v', v''), d')\}$
    \Else
    \State $Q' = Q' \cup \{(l', v', d')\}$
    \EndIf
    \EndIf
    \EndFor
    \EndFor
    \State $Q = Q'$
    \EndWhile
    \State \Return $s[l]$
    \end{algorithmic}
\end{algorithm}

\begin{algorithm}
    \caption{Computation of the Robust Semantics for the Unbounded Reach Operator}\label{alg:unbounded_reach}
    \begin{algorithmic}
    \Require $d_1 \in \mathbb{R}^\infty$, $s_1: \{1, \hdots, L\} \rightarrow \mathbb{R}^\infty \text{ where } s_1[l] = \rho^{\psi'}(x, \tau, l)$, $s_2: \{1, \hdots, L\} \rightarrow \mathbb{R}^\infty \text{ where } s_2[l] = \rho^{\psi''}(x, \tau, l).$
    \If{$d_1 = 0$}
    \State $s = s_2$
    \Else
    \State $w_{\max} = \max\{w \mid \exists l, l' \in \{1, \hdots, L\} \text{ s.t. } l \xrightarrow{w} l'\}$
    \State $s = BoundedReach(d_1, d_1 + w_{\max}, s_1, s_2)$
    \EndIf
    \State $T = \{1, \hdots, L\}$
    \While{$T \neq \emptyset$}
    \State $T' = \emptyset$
    \For{all $l \in T$}
    \For{all $l' : l' \xrightarrow{w} l$}
    \State $v' = \max(\min(s[l], s_1[l']), s[l'])$
    \If{$v' \neq s[l']$}
    \State $s[l'] = v'$
    \State $T' = T' \cup \{l'\}$
    \EndIf
    \EndFor
    \EndFor
    \State $T = T'$
    \EndWhile
    \State \Return $s[l]$
    \end{algorithmic}
\end{algorithm}

\begin{algorithm}
    \caption{Computation of the Robust Semantics for the Escape Operator}\label{alg:escape}
    \begin{algorithmic}
    \Require $d_1, d_2\in \mathbb{R}^\infty$, $s_1: \{1, \hdots, L\} \rightarrow \mathbb{R}^\infty \text{ where } s_1[l] = \rho^{\psi'}(x, \tau, l)$
    \State $D = MinDistance()$ \Comment i.e., $D$ is an $L \times L$ matrix such that $D[l, l'] := \tilde{d}_{\min}(l, l', \tau), \forall l, l' \in \{1, \hdots, L\}$.
    \State $\forall l, l' \in \{1, \hdots, L\}, e[l, l'] = -\infty$
    \State $\forall l \in \{1, \hdots, L\}, e[l, l] = s_1[l]$
    \State $T = \{(l, l) \mid l \in \{1, \hdots, L\}\}$
    \While{$T \neq \emptyset$}
    \State $e' = e$
    \State $T' = \emptyset$
    \For{all $(l_1, l_2) \in T$}
    \For{all $l_1' : l_1' \xrightarrow{w} l_1$}
    \State $v = \max(e[l_1', l_2], \min(s_1[l_1'], e[l_1, l_2]))$
    \If{$v \neq e[l_1', l_2]$}
    \State $T' = T' \cup \{(l_1', l_2)\}$
    \State $e'[l_1', l_2] = v$
    \EndIf
    \EndFor
    \EndFor
    \State $T = T'$
    \State $e = e'$
    \EndWhile
    \State $s = []$
    \For{all $l \in L$}
    \State $s[l] = \max\{e[l, l'] \mid D[l, l'] \in [d_1, d_2]\}$
    \EndFor
    \State \Return $s[l]$
    \end{algorithmic}
\end{algorithm}

\section{Proofs for Technical Theorems and Lemmas}
\label{app:proof}

\subsection{Proof of Theorem \ref{thm:soundness_strel}}
\begin{proof}
    The proof is conducted by way of induction similar to the STL semantics in \cite{fainekos2009robustness}. We first note that by [\cite{fainekos2009robustness}, Proposition 16], the soundness property holds for the atomic truth value and predicates as well as for the temporal operators. It is, therefore, sufficient to show the soundness property for the two spatial operators from STREL that are not present in STL. We follow the same strategy from \cite{fainekos2009robustness} where we prove the contrapositives of the claims: $(x,\tau_0,l)\models \psi$ implies $\rho^\psi(x,\tau_0,l) \ge 0$ and $(x,\tau_0,l)\models \neg \psi$ implies $\rho^\psi(x,\tau_0,l) \le 0$.
    
    We first consider the reach operator: 1) Suppose $(x, \tau, l) \models \psi' R_{[d_1, d_2]}\psi''$. By definition, $\exists r \in Routes(\tau, l) \text{ s.t. } \exists i \in \mathbb{N}^\infty \text{ and } d(i, r, \tau) \in [d_1, d_2] \text{ s.t. } (x, \tau, l_i) \models \psi'', \text{ and } \forall j < i, (x, \tau, l_j) \models \psi'$. By the inductive hypothesis, there exists a route $r$ starting from $l$ and an index $i$ with $l_i \in r$ such that $\rho^{\psi''}(x, \tau, l_i) \ge 0$ and $\rho^{\psi'}(x, \tau, l_j) \ge 0$ for all $0 \le j < i$. By the robust semantics, for our selected route $r$ and $i$ that satisfies the requirement above $\min(\rho^{\psi''}(x, \tau, r[i]), \min_{j < i}\rho^{\psi'}(x, \tau, r[j])) \ge 0$. Then, for the maximum evaluation over all possible routes $r$ and index $i$ satisfying the distance constraint, the robust semantics is no less than $0$, and thus $\rho^{\psi' R_{[d_1, d_2]}\psi''}(x, \tau, l) \ge 0$. 2) Suppose now $(x, \tau, l) \not\models \psi' R_{[d_1, d_2]}\psi''$. Again by definition, for all route $r$ in $Routes(\tau, l)$ either there does not exist an $i$ such that $d(i, r, \tau) \in [d_1, d_2]$ or for all $i$ with $d(i, r, \tau) \in [d_1, d_2]$ it does not satisfy that $(x, \tau, l_i) \models \psi'', \text{ and } \forall j < i, (x, \tau, l_j) \models \psi'$. The former case leads to $\rho^{\psi' R_{[d_1, d_2]}\psi''}(x, \tau, l) = -\infty$. For the latter case, either $(x, \tau, l_i) \not\models \psi''$ for all $i$ with $d(i, r, \tau) \in [d_1, d_2]$ or there exists $j < i$ such that $(x, \tau, l_j) \not\models \psi'$. By the inductive hypothesis and the robust semantics, either $\rho^{\psi''}(x, \tau, r[i]) \le 0$ for all $i$ with $d(i, r, \tau) \in [d_1, d_2]$ or $\rho^{\psi''}(x, \tau, r[j]) \le 0$ for some $j < i$. It has to hold that $\min(\rho^{\psi''}(x, \tau, r[i]), \min_{j < i}\rho^{\psi'}(x, \tau, r[j])) \le 0$ for all permissible options of $r$ and $i$, and thus $\rho^{\psi' R_{[d_1, d_2]}\psi''}(x, \tau, l) \le 0$. The soundness property for reach operator is proven by contrapositive.

    We now consider the escape operator: 1) Suppose $(x, \tau, l) \models \mathcal{E}_{[d_1, d_2]}\psi'$. By definition, $\exists r \in Routes(\tau, l) \text{ s.t. } \exists l' \in r \text{ and } \tilde{d}_{\min}(l, l', \tau) \in [d_1, d_2] \text{ s.t. } \forall i \le r(l'), (x, \tau, r[i])\models \psi'.$ Consider the selected $r$ and the index $i$ that satisfies the previous sentence. By the inductive hypothesis, $\rho^{\psi'}(x, \tau, r[i]) \ge 0$ for all $i \le r(l')$, and thus $\min_{i \le r(l')}\rho^{\psi'}(x, \tau, r[i]) \ge 0$. Since this holds for the selected $r$ and $i$, it also holds for the maximum evaluation, which implies that $\rho^{\mathcal{E}_{[d_1, d_2]}\psi'} \ge 0$. 2) Suppose now $(x, \tau, l) \not\models \mathcal{E}_{[d_1, d_2]}\psi'$. Again by definition, for all routes $r$ in $Routes(\tau, l)$ and for all $l' \in r$, either $\tilde{d}_{\min}(l, l', \tau) \notin [d_1, d_2]$ or $\tilde{d}_{\min}(l, l', \tau) \in [d_1, d_2]$ but there exists an $i \le r(l')$ such that $(x, \tau, r[i]) \not\models \psi'$. For the former case, $\rho^{\mathcal{E}_{[d_1, d_2]}\psi'} = -\infty$. For the latter case, consider the selected $i$ such that $(x, \tau, r[i]) \not\models \psi'$. By the inductive hypothesis, $\rho^{\psi'}(x, \tau, r[i]) \le 0$. With the robust semantics, we see that $\min_{j \le r(l')}\rho^{\psi'}(x, \tau, r[j]) \le 0$ for all permissible $\tau$ and $l'$, and thus $\rho^{\mathcal{E}_{[d_1, d_2]}\psi'} \le 0$. The soundness property for escape operator is proven by contrapositive.
\end{proof}

\subsection{Proof of Theorem \ref{theorem:1}}
\begin{proof}
 By the data processing inequality in Lemma \ref{lemma:2} and since $D_f(\mathcal{D},\mathcal{D}_0)\le \epsilon$ holds by Assumption \ref{ass2}, we know that $D(\mathcal{R},\mathcal{R}_0)\le \epsilon$. We can thus apply Lemma \ref{lemma:1} and construct $\tilde{C}$ according to equation \eqref{eq:C_tilde} with $R^{(i)}$ as in \eqref{eq:R_direct}. We then know that $Prob(\rho^\phi(\hat{X}, \tau_0) - \rho^\phi(X, \tau_0) \le \tilde{C}) \ge 1 - \delta$. Therefore, it follows that $Prob(\rho^\phi(X, \tau_0) \ge \rho^\phi(\hat{X}, \tau_0) - \tilde{C}) \ge 1 - \delta$.
\end{proof}

\subsection{Proof of Theorem \ref{theorem:2}}

\begin{proof}
By assumption we know that equation \eqref{eq:guarantee_prediate} holds, i.e., that $\text{Prob}(\rho^{\pi}(X, \tau) \ge \rho_{\pi,\tau}^*, \forall (\pi, \tau) \in \mathcal{P}) \ge 1 - \delta$. Since we define $\bar{\rho}^{\pi}(\hat{X},\tau) :=h(X_\tau)$ if $\tau\le t$ and $\bar{\rho}^{\pi}(\hat{X},\tau) :=\rho_{\pi,\tau}^*$ otherwise, we know that $\text{Prob}(\rho^{\pi}(X, \tau) \ge \bar{\rho}^{\pi}(\hat{X},\tau), \forall (\pi, \tau) \in \mathcal{P}) \ge 1 - \delta$. Since $\phi$ is in positive normal form, we further know that, for any two signals $x,x'$ with the same prefix $x[\tau] = x'[\tau]$ if $\tau \le t$, it holds that ${\rho}^{\pi}(x, \tau) \ge {\rho}^{\pi}(x', \tau)$ for all $(\pi, \tau) \in \mathcal{P}$ implies ${\rho}^{\phi}(x, \tau_0) \ge {\rho}^{\phi}(x', \tau_0)$ \cite[Corollary 1]{lindemann2019robust}. Consequently, we conclude that $\text{Prob}(\rho^\phi(X, \tau_0) \ge \bar{\rho}^\phi(\hat{X}, \tau_0)) \ge 1 - \delta$ since the Boolean and temporal operators for $\bar{\rho}^\phi$ follow the same semantics as for ${\rho}^\phi$.
\end{proof}

\subsection{Proof of Lemma \ref{lemma:variant1}}

\begin{proof}
     By the data processing inequality in Lemma \ref{lemma:2} and since $D_f(\mathcal{D},\mathcal{D}_0)\le \epsilon$ holds by Assumption \ref{ass2}, we know that $D(\mathcal{R},\mathcal{R}_0)\le \epsilon$. We can thus apply Lemma \ref{lemma:1} and construct $\tilde{C}$ according to equation \eqref{eq:C_tilde} with $R^{(i)}$ as in \eqref{eq:R_indirect}. We then know that $\text{Prob}(\max_{\tau \in \{t+1, \hdots, t+H\}} \frac{\|X_\tau - \hat{X}_{\tau|t}\|}{\alpha_\tau} \le \tilde{C}) \ge 1 - \delta$, which implies that $\text{Prob}(\frac{\|X_\tau - \hat{X}_{\tau|t}\|}{\alpha_\tau} \le \tilde{C}, \forall \tau \in \{t + 1, \hdots, t + H\}) \ge 1 - \delta$. Since  $\alpha_\tau > 0$, this is equivalent to  $\text{Prob}(\|X_\tau - \hat{X}_{\tau|t}\|\le \tilde{C}\alpha_\tau,  \forall \tau \in \{t+1, \hdots, t+H\}) \ge 1 - \delta$. Finally, by the definition of $\rho^*_{\pi, \tau}$ in equation \eqref{eq:worst_case_robustness}, which considers the worst case value of $h(\zeta)$ over $\zeta\in \mathcal{B}_{\tau}$, it follows that  $\text{Prob}(\rho^{\pi}(X, \tau) \ge \rho_{\pi,\tau}^*, \forall (\pi, \tau) \in \mathcal{P}) \ge 1 - \delta$ holds.
\end{proof}

\subsection{Proof of Lemma \ref{lemma:variant2}}

\begin{proof}
 By the data processing inequality in Lemma \ref{lemma:2} and since $D_f(\mathcal{D},\mathcal{D}_0)\le \epsilon$ holds by Assumption \ref{ass2}, we know that $D(\mathcal{R},\mathcal{R}_0)\le \epsilon$. We can thus apply Lemma \ref{lemma:1} and construct $\tilde{C}$ according to equation \eqref{eq:C_tilde} with $R^{(i)}$ as in \eqref{eq:R_hybrid}. We then know that $\text{Prob}(\max_{(\pi, \tau) \in \mathcal{P}}\frac{\rho^{\pi}(\hat{X}, \tau) - \rho^{\pi}(X, \tau)}{\alpha_{\pi, \tau}} \le \tilde{C}) \ge 1 - \delta$, which implies that  $\text{Prob}(\frac{\rho^{\pi}(\hat{X}, \tau) - \rho^{\pi}(X, \tau)}{\alpha_{\pi, \tau}} \le \tilde{C}, \forall (\pi, \tau) \in \mathcal{P}) \ge 1 - \delta$. Since $\alpha_{\pi, \tau} > 0$, this is equivalent to $\text{Prob}(\rho^{\pi}(\hat{X}, \tau) - \rho^{\pi}(X, \tau) \le \tilde{C}\alpha_{\pi, \tau}, \forall (\pi, \tau) \in \mathcal{P}) \ge 1 - \delta$. From here, it follows that $\text{Prob}(\rho^{\pi}(X, \tau) \ge \rho_{\pi,\tau}^*, \forall (\pi, \tau) \in \mathcal{P}) \ge 1 - \delta$  with $\rho_{\pi,\tau}^*:= \rho^{\pi}(\hat{X}, \tau)-\tilde{C}\alpha_{\pi, \tau}$.
\end{proof}

\begin{figure*}
    \centering
    \begin{subfigure}[t]{0.35\textwidth}
    \includegraphics[width=\textwidth]{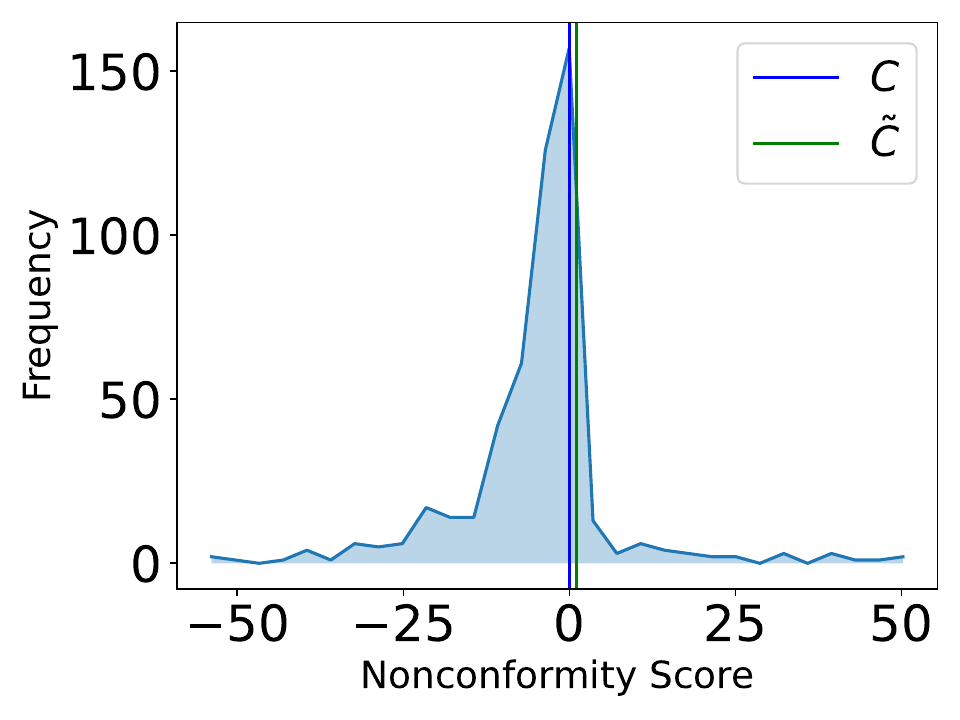}
    \caption{Histogram of $R^{(i)}$ from \eqref{eq:nonc_multi_direct} with $L = 7$.}
    \label{fig:direct_nonconformities_7}
    \end{subfigure}
     \hspace{1mm}
     \begin{subfigure}[t]{0.35\textwidth}
    \includegraphics[width=\textwidth]{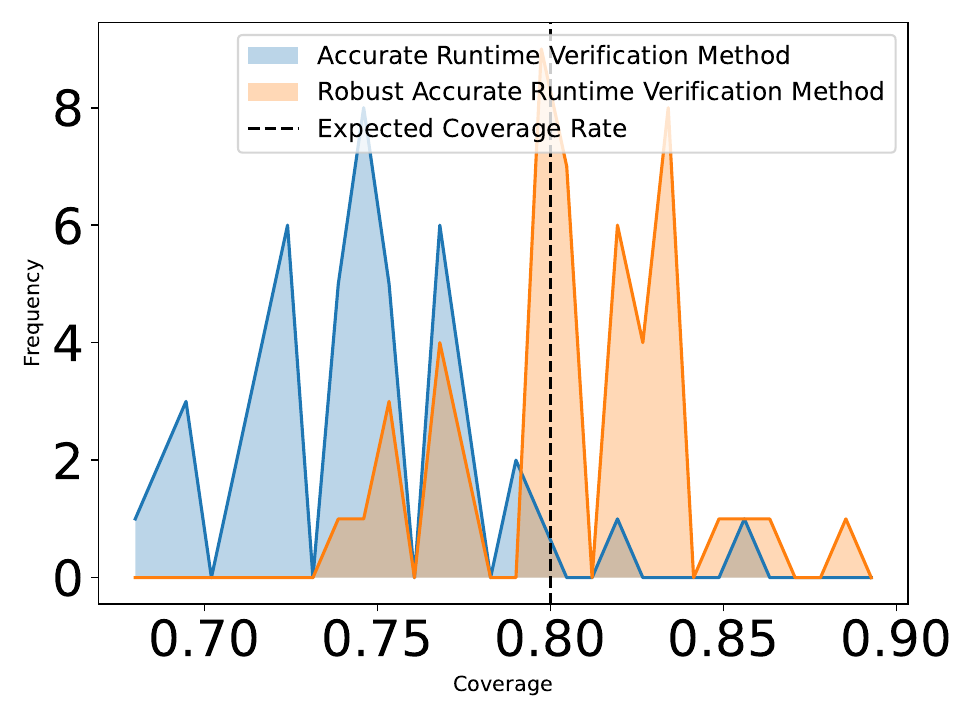}
    \caption{Histogram of empirical coverage from robust and non-robust accurate methods with $L = 7$.}
    \label{fig:direct_coverages_7}
    \end{subfigure}
    \hspace{1mm}
    \begin{subfigure}[t]{0.32\textwidth}
    \includegraphics[width=\textwidth]{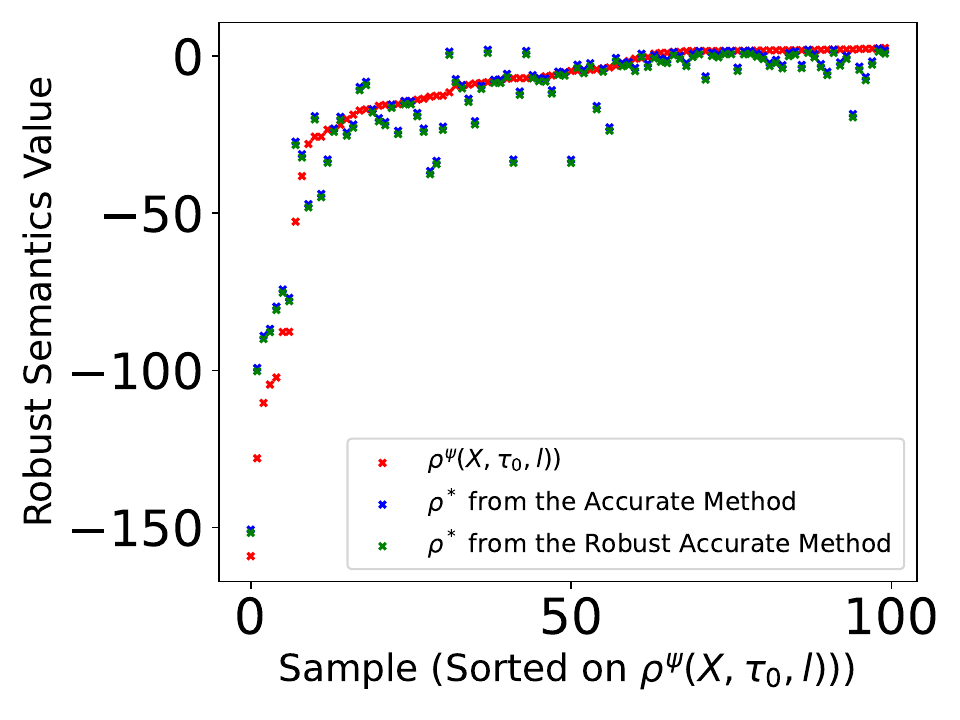}
    \caption{Comparison of $\rho^\psi(X, \tau_0, l)$ and $\rho^*$ with $L = 7$ for the accurate methods.}
    \label{fig:direct_robustness_7}
    \end{subfigure}
    \hspace{1mm}
    \begin{subfigure}[t]{0.32\textwidth}
    \includegraphics[width=\textwidth]{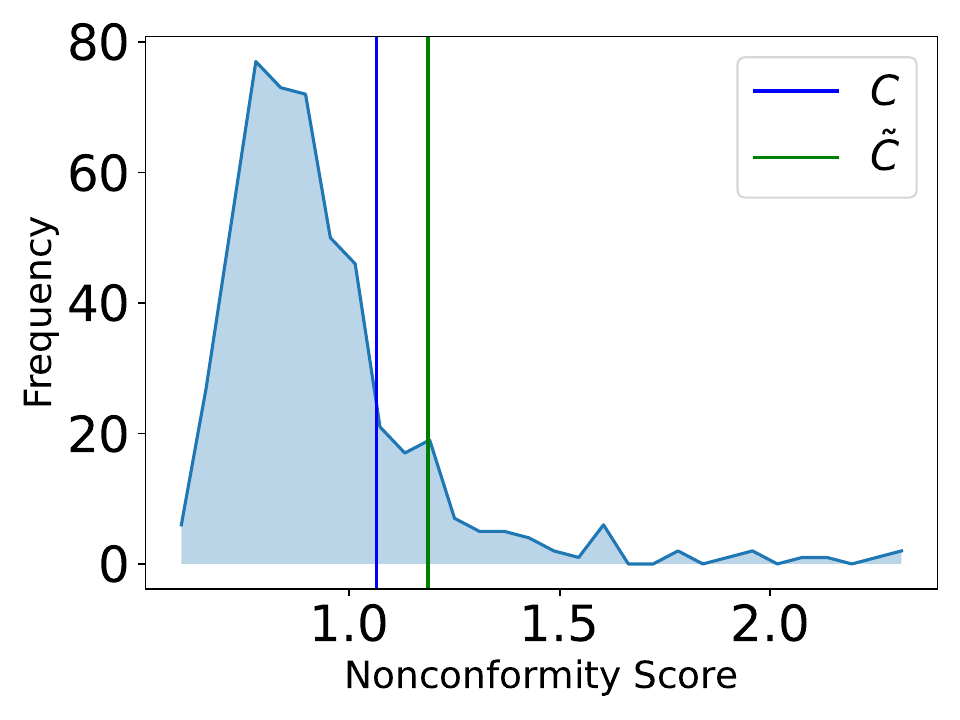}
    \caption{Histogram of $R^{(i)}$ from \eqref{eq:r_indirect_strel} with $L = 7$.}
    \label{fig:indirect_nonconformities_7}
    \end{subfigure}
    \hspace{1mm}
    \begin{subfigure}[t]{0.32\textwidth}
    \includegraphics[width=\textwidth]{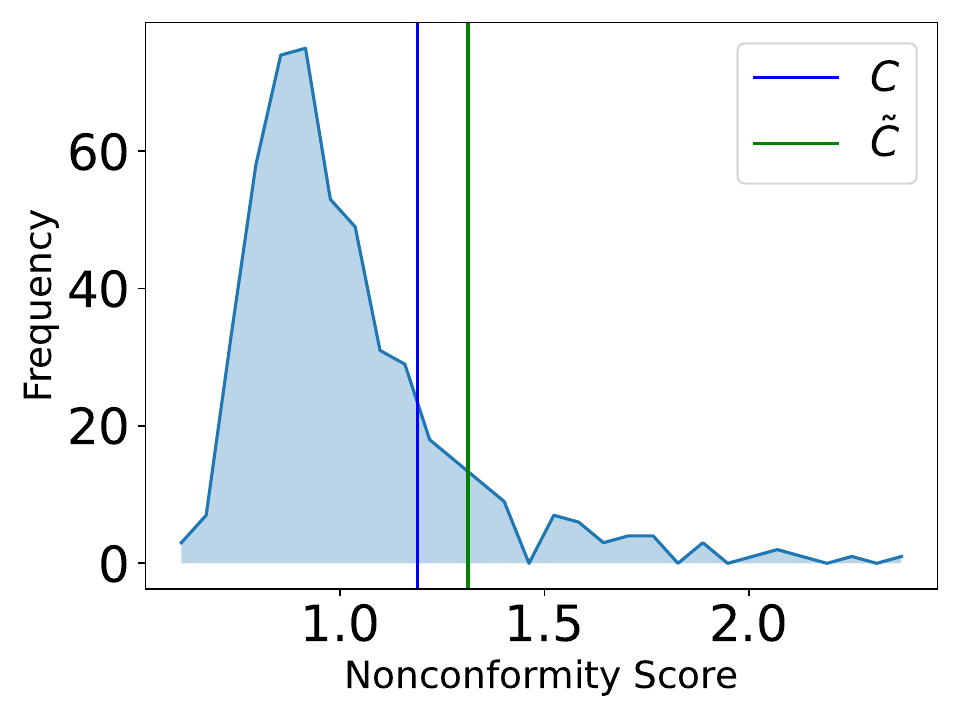}
    \caption{Histogram of $R^{(i)}$ from \eqref{eq:r_hybrid_strel} with $L = 7$.}
    \label{fig:hybrid_nonconformities_7}
    \end{subfigure}
     \hspace{1mm}
     \begin{subfigure}[t]{0.35\textwidth}
    \includegraphics[width=\textwidth]{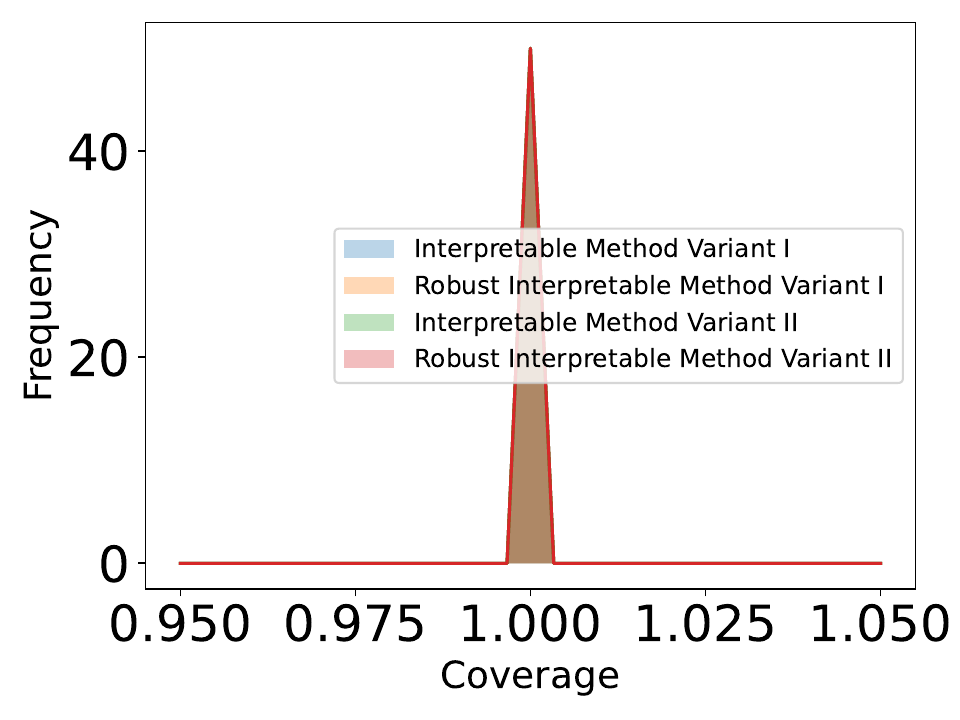}
    \caption{Histogram of empirical coverage from robust and non-robust interpretable methods with $L = 7$.}
    \label{fig:indirect_coverages_7}
    \end{subfigure}
    \hspace{1mm}
    \begin{subfigure}[t]{0.35\textwidth}
    \includegraphics[width=\textwidth]{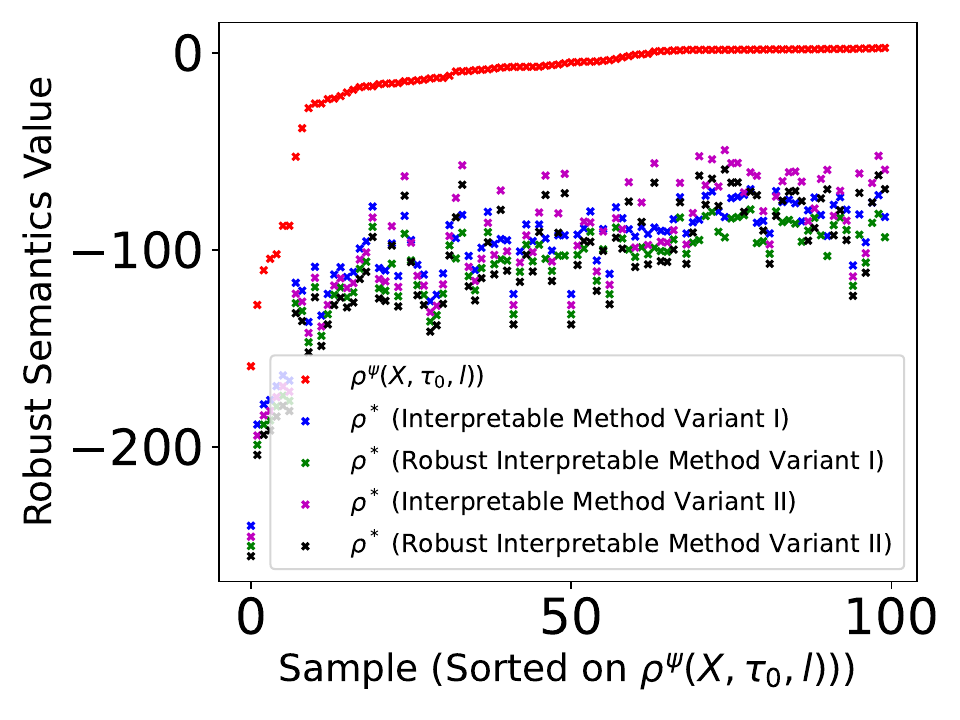}
    \caption{Comparison of $\rho^\psi(X, \tau_0, l)$ and $\rho^*$ with $L = 7$ for the interpretable methods.}
    \label{fig:indirect_robustness_7}
    \end{subfigure}
    \caption{Results for STREL RPRV Case Study with $L := 7$.}
    \label{fig:results_strel_7}
\end{figure*}
\begin{figure*}
    \centering
    \begin{subfigure}[t]{0.35\textwidth}
    \includegraphics[width=\textwidth]{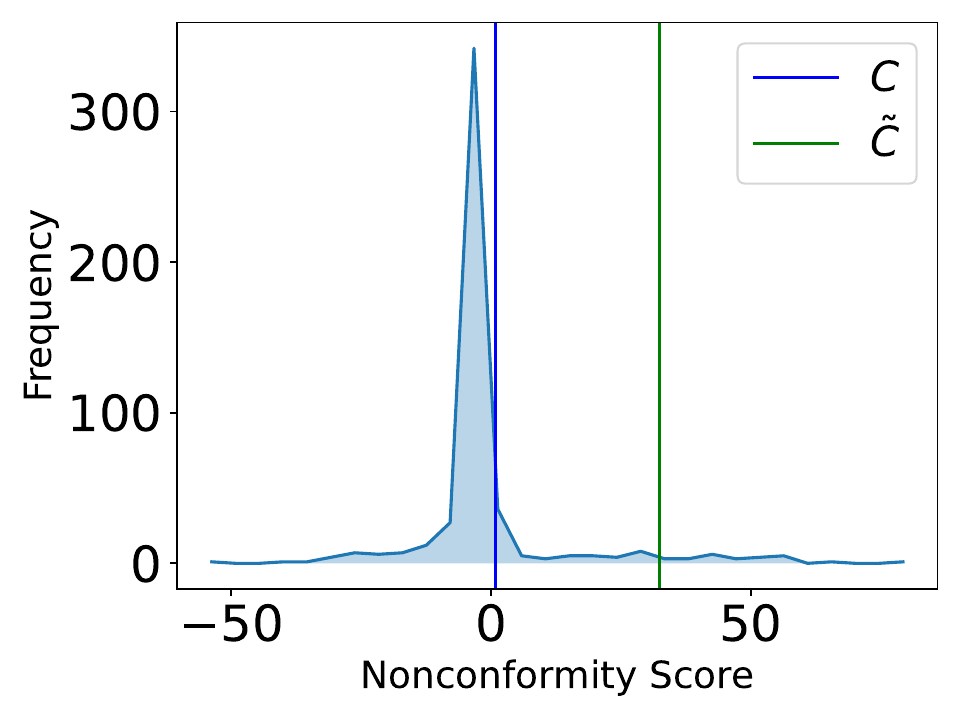}
    \caption{Histogram of $R^{(i)}$ from \eqref{eq:nonc_multi_direct} with $L = 10$.}
    \label{fig:direct_nonconformities_10}
    \end{subfigure}
    \hspace{1mm}
     \begin{subfigure}[t]{0.35\textwidth}
    \includegraphics[width=\textwidth]{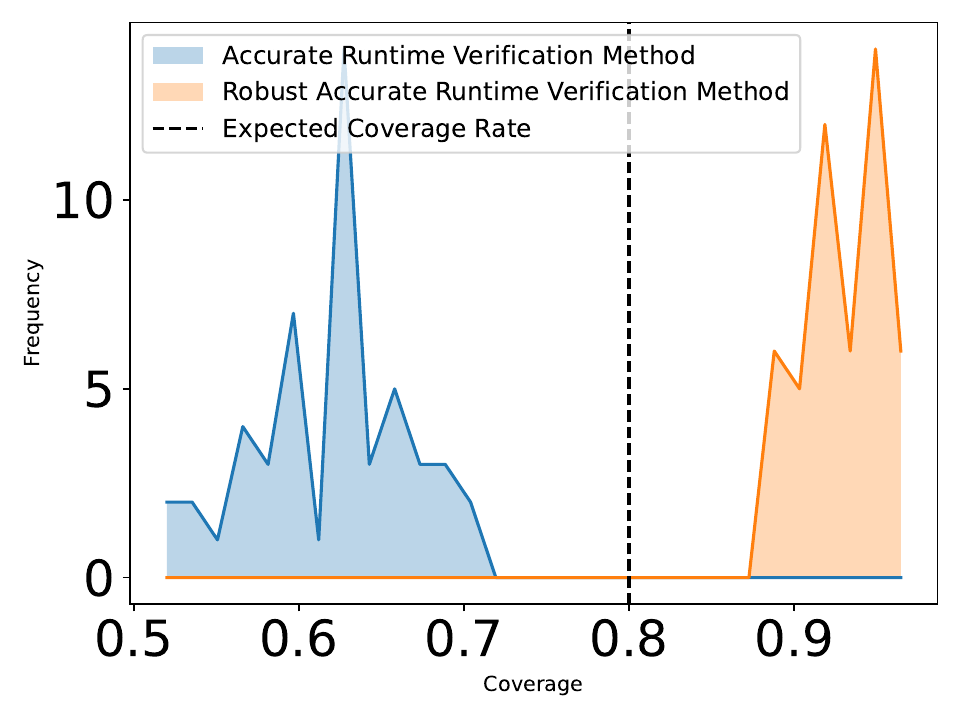}
    \caption{Histogram of coverage: accurate methods with $L = 10$.}
    \label{fig:direct_coverages_10}
    \end{subfigure}
    \hspace{1mm}
    \begin{subfigure}[t]{0.32\textwidth}
    \includegraphics[width=\textwidth]{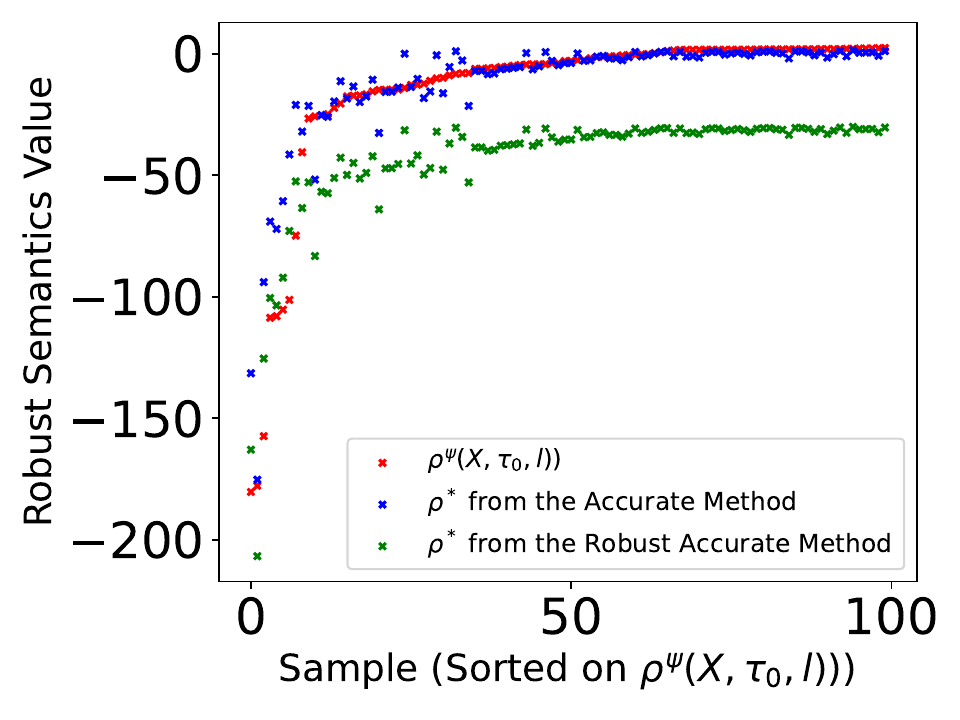}
    \caption{$\rho^\psi(X, \tau_0, l)$ and $\rho^*$ with $L = 10$ for the accurate methods.}
    \label{fig:direct_robustness_10}
    \end{subfigure}
    \hspace{1mm}
     \begin{subfigure}[t]{0.32\textwidth}
    \includegraphics[width=\textwidth]{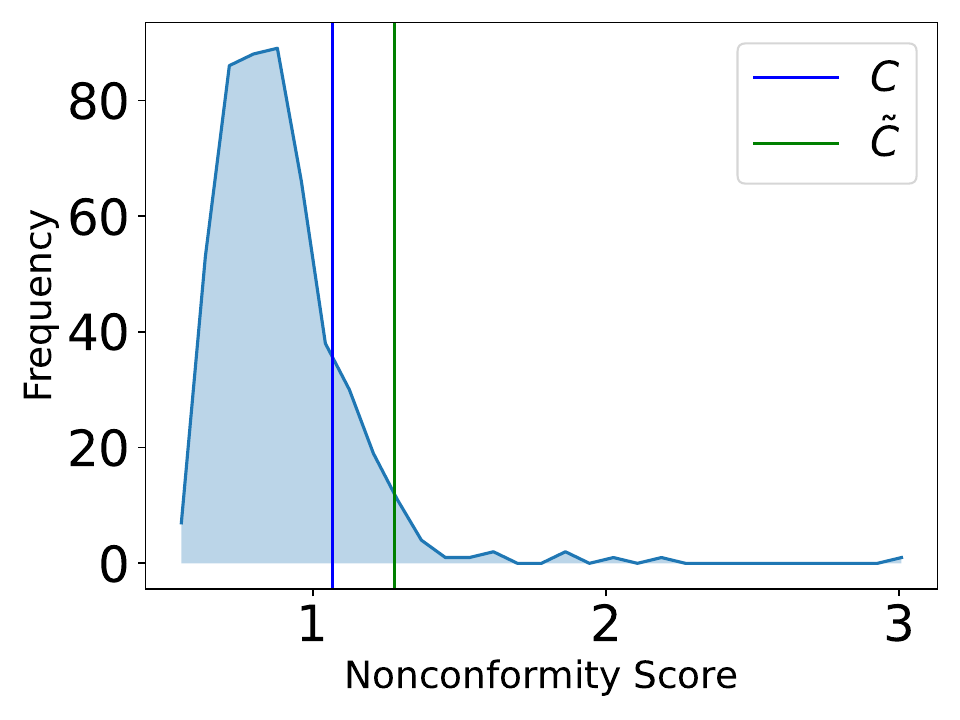}
    \caption{Histogram of $R^{(i)}$ from \eqref{eq:r_indirect_strel} with $L = 10$.}
    \label{fig:indirect_nonconformities_10}
    \end{subfigure}
    \hspace{1mm}
     \begin{subfigure}[t]{0.32\textwidth}
    \includegraphics[width=\textwidth]{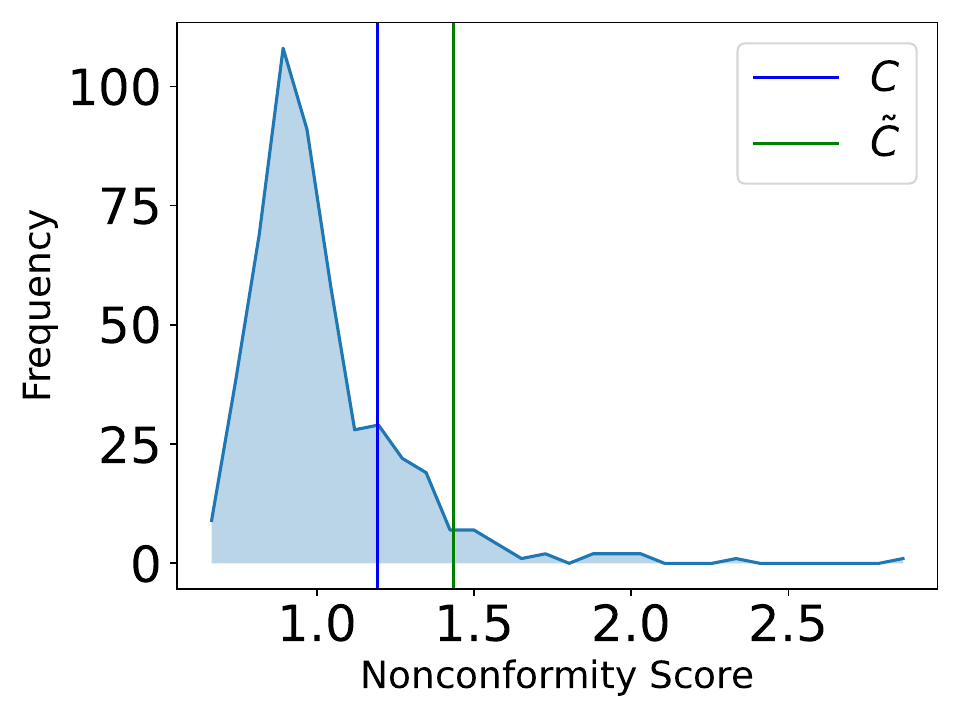}
    \caption{Histogram of $R^{(i)}$ from \eqref{eq:r_hybrid_strel} with $L = 10$.}
    \label{fig:hybrid_nonconformities_10}
    \end{subfigure}
    \hspace{1mm}
     \begin{subfigure}[t]{0.35\textwidth}
    \includegraphics[width=\textwidth]{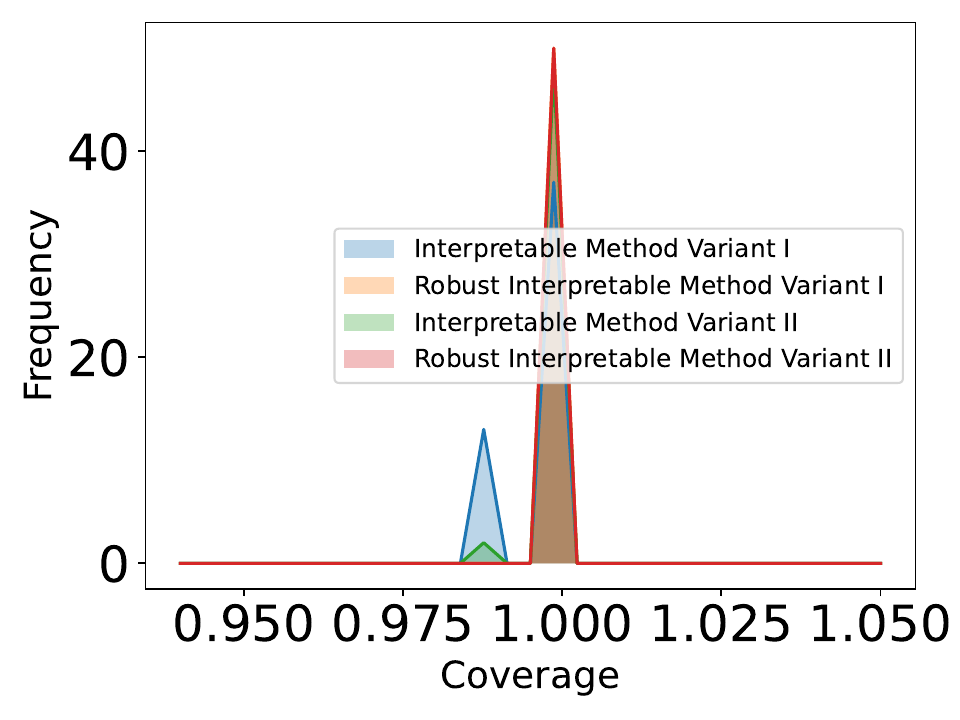}
    \caption{Histogram of coverage: interpretable methods with $L = 10$.}
    \label{fig:indirect_coverages_10}
    \end{subfigure}
    \hspace{1mm}
    \begin{subfigure}[t]{0.35\textwidth}
    \includegraphics[width=\textwidth]{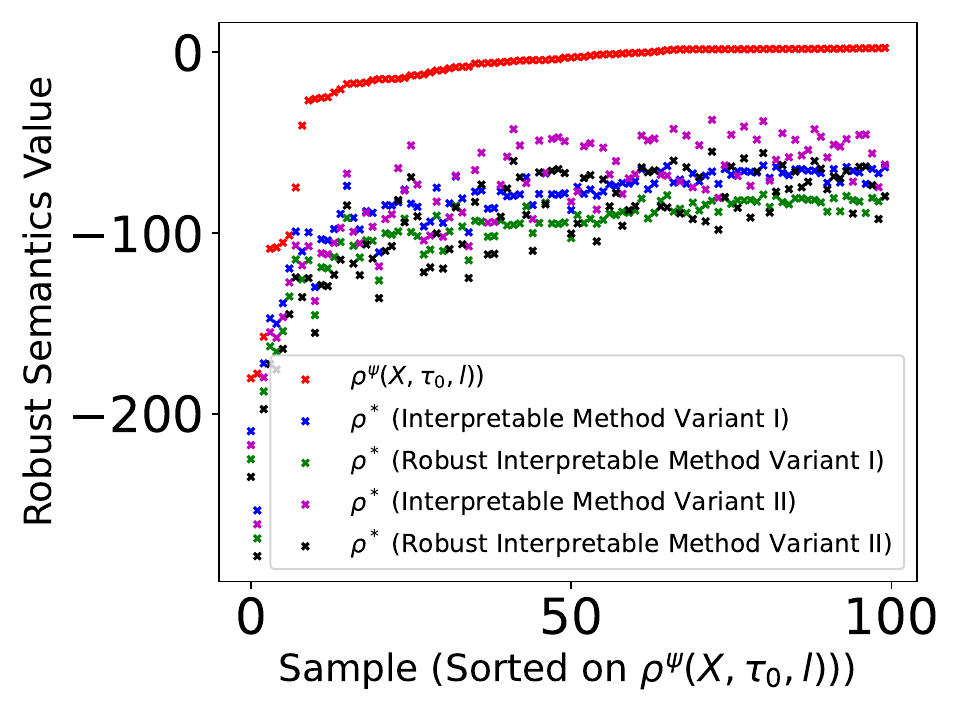}
    \caption{$\rho^\psi(X, \tau_0, l)$ and $\rho^*$ with $L = 10$ for the interpretable methods.}
    \label{fig:indirect_robustness_10}
    \end{subfigure}
    \caption{Results for STREL RPRV Case Study with $L := 10$.}
    \label{fig:results_strel_10}
\end{figure*}
\begin{figure*}
    \centering
    \begin{subfigure}[t]{0.3\textwidth}
    \includegraphics[width=\textwidth]{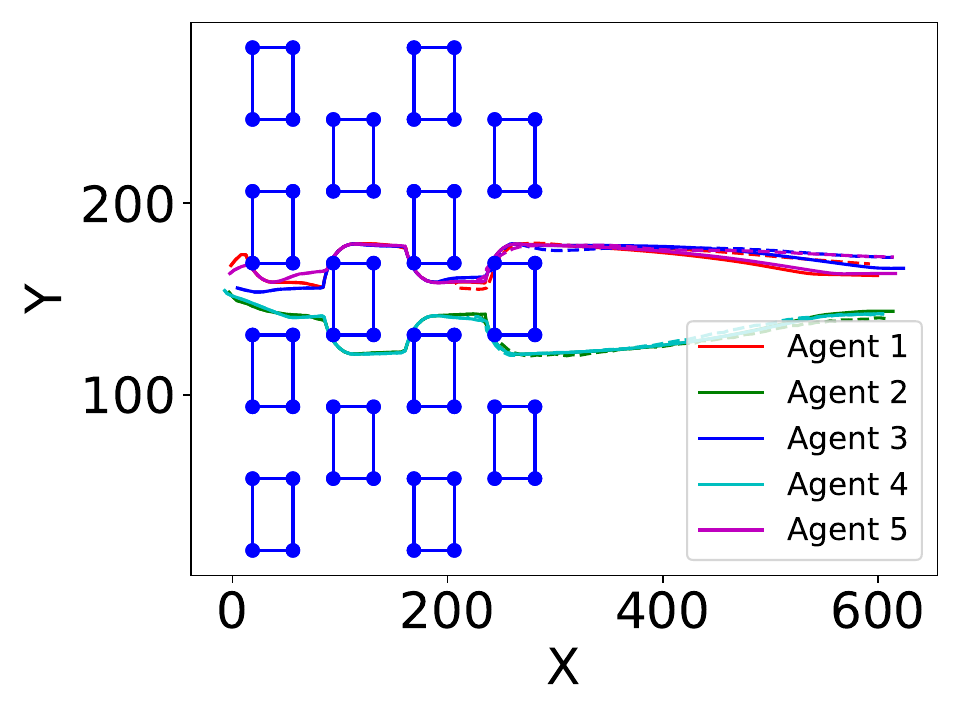}
    \caption{Example trajectory from $\mathcal{D}_0$ with prediction}
    \label{fig:example_trajectory_nominal_cnn}
    \end{subfigure}
    \hspace{3mm}
    \begin{subfigure}[t]{0.3\textwidth}
    \includegraphics[width=\textwidth]{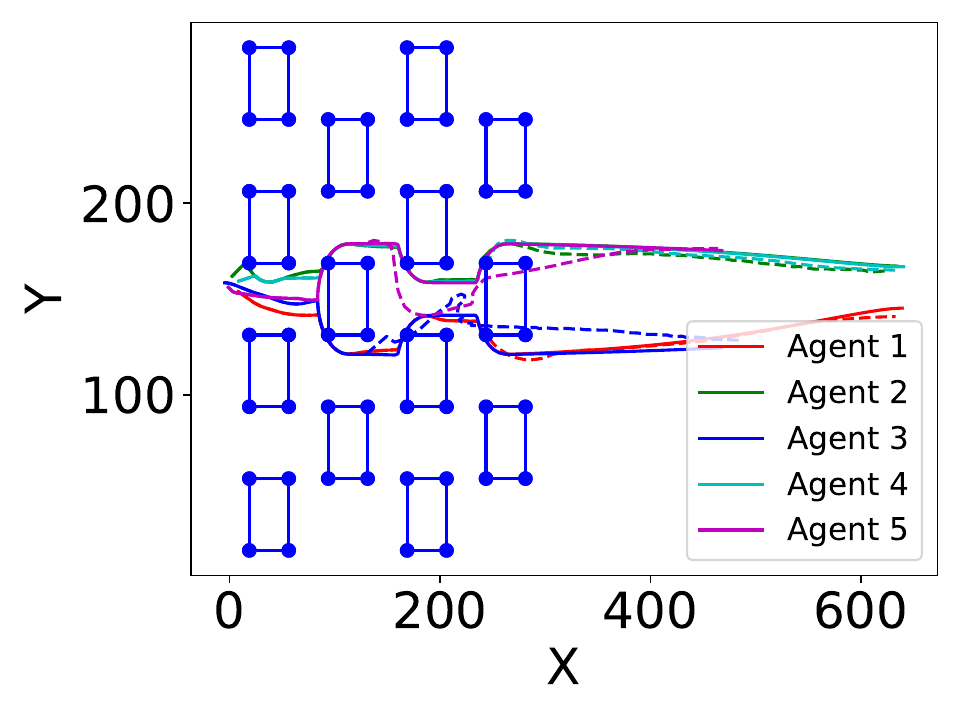}
    \caption{Example trajectory from $\mathcal{D}$ with prediction}
    \label{fig:example_trajectory_shifted_cnn}
    \end{subfigure}
    \caption{Example trajectories for the Drone Swarm Case Study with a CNN Predictor}
    \label{fig:multiagent_example_trajectories_cnn}
\end{figure*}
\begin{figure*}
    \centering
    \begin{subfigure}[t]{0.3\textwidth}
    \includegraphics[width=\textwidth]{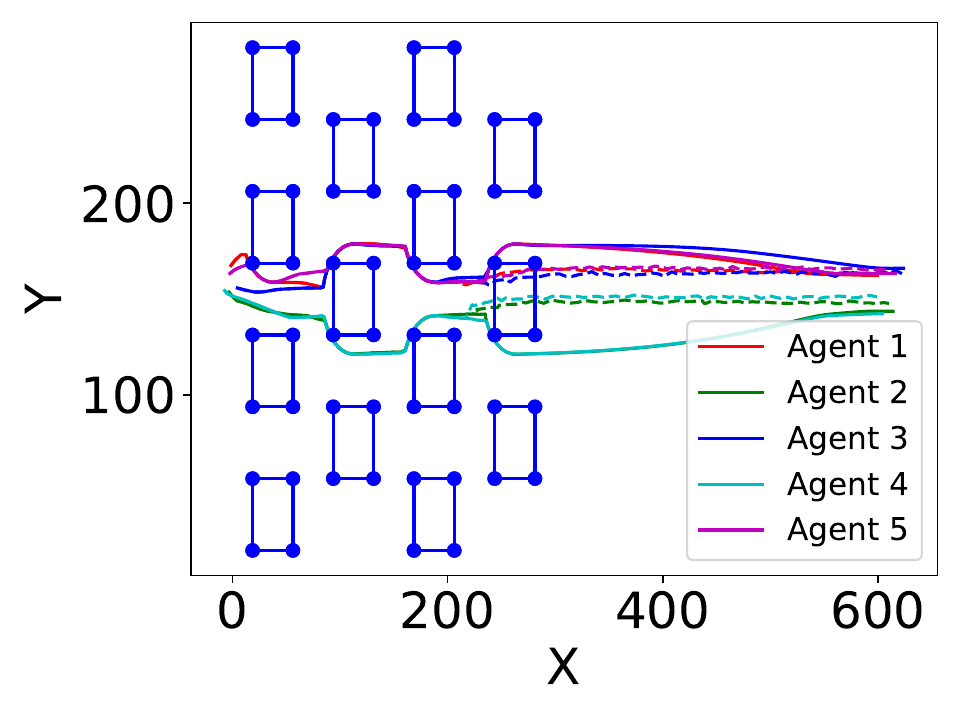}
    \caption{Example trajectory from $\mathcal{D}_0$ with prediction}
    \label{fig:example_trajectory_nominal_transformer}
    \end{subfigure}
    \hspace{3mm}
    \begin{subfigure}[t]{0.3\textwidth}
    \includegraphics[width=\textwidth]{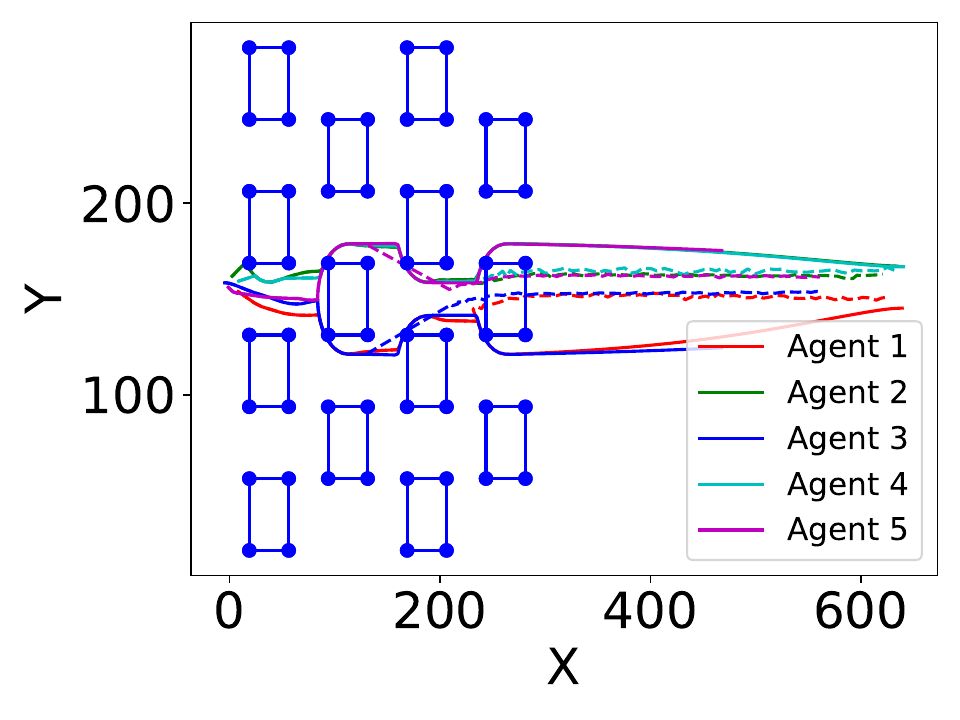}
    \caption{Example trajectory from $\mathcal{D}$ with prediction}
    \label{fig:example_trajectory_shifted_transformer}
    \end{subfigure}
    \caption{Example Trajectories for the Drone Swarm Case Study with a Tansformer Predictor}
    \label{fig:multiagent_example_trajectories_transformer}
\end{figure*}
\section{Supplementary Experimental Results}\label{sec:supp_results}
We show the results for the STREL RPRV case study in section \ref{sec:simulations} now with $L \coloneqq 7$ and $L \coloneqq 10$. We repeat the procedure for validation and comparison of accurate and interpretable methods as for $L \coloneqq 5$ in Section \ref{sec:simulations} but now with different values of $L$. We show the histogram of nonconformity scores from \eqref{eq:nonc_multi_direct} for $L \coloneqq 7$ in Figure \ref{fig:direct_nonconformities_7} and for $L \coloneqq 10$ in Figure \ref{fig:direct_nonconformities_10}. We show the empirical coverages over the 50 experiments for the non-robust and robust accurate methods in Figure \ref{fig:direct_coverages_7} and in Figure \ref{fig:direct_coverages_10} for $L \coloneqq 7$ and $L \coloneqq 10$. We demonstrate the robust semantics from one experiment, $\rho^\psi(X, \tau_0, l)$, for the $100$ ground truth test data together with the predicted worst-case robust semantics $\rho^*$ for the non-robust and robust accurate methods in Figure \ref{fig:direct_robustness_7} and in Figure \ref{fig:direct_robustness_10} for $L := 7$ and $L := 10$. For the interpretable methods, we show the histogram of $R^{(i)}$ from equation \eqref{eq:r_indirect_strel} for $L := 7$ in Figure \ref{fig:indirect_nonconformities_7} and for $L := 10$ in Figure \ref{fig:indirect_nonconformities_10} and the histogram of $R^{(i)}$ from equation \eqref{eq:r_hybrid_strel} for $L := 7$ in Figure \ref{fig:hybrid_nonconformities_7} and for $L := 10$ in Figure \ref{fig:hybrid_nonconformities_10}. We demonstrate the histograms of coverages for the interpretable methods with $L := 7$ in Figure \ref{fig:indirect_coverages_7} and with $L := 10$ in Figure \ref{fig:indirect_coverages_10}. Finally, we show the ground truth robust semantics as compared to the predicted worst-case robust semantics from the interpretable methods for $L := 7$ in Figure \ref{fig:indirect_robustness_7} and for $L := 10$ in Figure \ref{fig:indirect_robustness_10}.

We illustrate in Figure \ref{fig:multiagent_example_trajectories_cnn} and Figure \ref{fig:multiagent_example_trajectories_transformer} respectively the predictions from the CNN model and for the Transformer model as per Section \ref{sec:model_comparison} again with solid lines representing the ground truth trajectories and dashed lines as predictions.

\end{document}